\documentclass[twocolumn]{aastex7}
\usepackage{amsmath}
\usepackage{float}
\usepackage{graphicx} % Required for inserting images
\usepackage{booktabs}

\begin{document}

\title{JWST color sequence of transiting exoplanets and directly imaged substellar objects}
\author[0000-0002-3263-2251]{Guangwei Fu}\email{guangweifu@gmail.com}
\affiliation{Department of Physics and Astronomy, Johns Hopkins University, Baltimore, MD, USA}

\author[0000-0003-3667-8633]{Joshua D. Lothringer}\email{}
\affiliation{Space Telescope Science Institute, Baltimore, MD}

\author[0000-0003-1622-1302]{Sagnick Mukherjee}\email{}
\affiliation{Department of Astronomy and Astrophysics, University of California, Santa Cruz, CA 95064, USA \\ }
\affiliation{Department of Physics and Astronomy, Johns Hopkins University, Baltimore, MD, USA \\ }

\author[0000-0001-6050-7645]{David K. Sing}\email{}
\affiliation{Department of Physics and Astronomy, Johns Hopkins University, Baltimore, MD, USA}

\author[0000-0003-0192-6887]{Elena Manjavacas}\email{}
\affiliation{Space Telescope Science Institute, Baltimore, MD}

\author[0000-0002-7352-7941]{Kevin B. Stevenson}\email{}
\affiliation{JHU Applied Physics Laboratory, 11100 Johns Hopkins Rd, Laurel, MD 20723, USA}

\author[0000-0001-6396-8439]{William Balmer}\email{}
\affiliation{Department of Physics and Astronomy, Johns Hopkins University, Baltimore, MD, USA}

\begin{abstract}

Color–magnitude diagrams (CMDs) have been foundational across astrophysics, from galaxies and stars to brown dwarfs, and JWST now extends them to transiting exoplanets and self-luminous substellar objects at overlapping temperatures. We compile JWST dayside emission spectra for 13 transiting giant planets, 57 self-luminous objects, and the irradiated brown dwarf ZTF J0038+2030 B, spanning $\sim$350–2600~K and $\log g$ = 2.5--5.5, plus $\sim$2150 SPHEREx ultracool dwarfs, converted to synthetic photometry in 2MASS $J/K_s$ and five NIRCam medium bands isolating H$_2$O, CH$_4$, CO$_2$, and CO. Low gravity and inflated radii make transiting planets resemble young substellar objects, with shallower features and redder colors from high-altitude clouds that raise the photosphere. Transiting planets follow the L-dwarf sequence on the $J$ versus $J-K$ CMD, with WASP-80 b (T$_{eq}\sim$800~K) showing no T-dwarf-like blueward turn, suggesting a delayed or suppressed L/T transition in irradiated, low-gravity atmospheres. Methane onset is delayed from substellar to transiting-planet atmospheres, and cloudy radiative–convective models indicate that lower gravity and cloud back-warming shift atmospheres toward CO-dominated chemistry, suppressing photospheric CH$_4$. The irradiated but old, high-gravity brown dwarf ZTF J0038+2030 B ($T_{\rm day}$ = 1049 K, $\log g$ = 5.4) shows a deep, field-T-dwarf-like dayside methane band, implicating gravity rather than irradiation as the dominant control. CO$_2$-to-CO diagnostics place transiting and directly imaged planetary-mass companions at stronger relative CO$_2$ absorption than field brown dwarfs, consistent with metallicity enhancement from planetesimal accretion. These CMD sequences provide a unified, model-testable map for self-luminous and irradiated atmospheres.

\end{abstract}

\keywords{planets and satellites: atmospheres - techniques: spectroscopic}
\nopagebreak
\section{Introduction}

Measuring the color and brightness of objects and placing them on a log--log plot has been a long-standing technique in astronomy. The Hertzsprung-Russell diagram, which plots stellar luminosity against color is foundational to our understanding of stellar structure, evolution, and the physical processes governing stellar atmospheres. This simple yet profound approach of placing objects on color-magnitude diagrams (CMDs) has since been extended to diverse astronomical populations, from globular clusters to distant galaxies, providing fundamental insights into their formation and evolutionary stages.

Over the past three decades, color-magnitude diagrams have proven equally transformative for understanding substellar objects. The discovery of brown dwarfs \citep{nakajimaDiscoveryCoolBrown1995, reboloDiscoveryBrownDwarf1995} opened a new frontier in the study of cool, dense atmospheres intermediate between stars and planets. The subsequent identification of the L, T, and Y spectral classes revealed that brown dwarf atmospheres undergo dramatic transformations as they cool \citep{cushing_infrared_2005, kirkpatrick_discovery_2006, kirkpatrick_further_2012, burgasserUnifiedNearInfraredSpectral2006, beilerPreciseBolometricLuminosities2024}. Near-infrared CMDs, particularly the J versus J$-$K diagram, have been instrumental in characterizing these transitions at the population level \citep{saumonEvolutionDwarfsColorMagnitude2008, suarezUltracoolDwarfsObserved2022, bestUltracoolSheetPhotometryAstrometry2025}.

The L/T transition, occurring around $T_{\rm eff} \sim 1100-1400$ K, represents one of the most striking features in brown dwarf CMDs. As L dwarfs cool, they maintain red J$-$K colors due to thick silicate and iron cloud layers that redden the near-infrared spectrum. At the L/T boundary, these clouds rapidly dissipate or sink below the photosphere \citep{marleyPATCHYCLOUDMODEL2010}, revealing the underlying cloud-free atmosphere where strong methane absorption produces characteristically blue J$-$K colors in T dwarfs \citep{ackermanPrecipitatingCondensationClouds2001, burgasserUnifiedNearInfraredSpectral2006, saumonEvolutionDwarfsColorMagnitude2008}. This transition is not merely a spectral classification boundary but reflects fundamental changes in atmospheric chemistry, cloud physics, and vertical mixing.

Intriguingly, young brown dwarfs and directly imaged planetary companions do not follow the same CMD tracks as field objects. Low-gravity substellar objects exhibit redder colors \citep{faherty_population_2016} and delayed L/T transitions compared to their high-gravity counterparts \citep{marois_direct_2008, manjavacasMediumresolution09753Mm2024, milesJWSTEarlyreleaseScience2023}. This gravity dependence suggests that surface gravity, through its influence on atmospheric gaseous optical depth, cloud sedimentation, and chemical equilibrium, plays a critical role in shaping substellar atmospheres. Understanding these effects is essential for interpreting spectra of young planet mass objects and infer their atmospheric composition.

The launch of the James Webb Space Telescope (JWST) has inaugurated a new era in comparative atmospheric studies. For the first time, we can obtain high signal-to-noise infrared spectra of both transiting exoplanets and self-luminous substellar objects across overlapping temperature ranges with the same instrument. Eclipse spectroscopy of transiting hot Jupiters \citep{demingStrongInfraredEmission2006, charbonneauDetectionThermalEmission2005} provides dayside emission spectra from $\sim$1-5 $\mu$m, probing the same molecular features including H$_2$O, CH$_4$, CO, and CO$_2$ that define brown dwarf spectral sequences. By converting these spectra to synthetic photometry, we can place transiting exoplanets on the same CMDs as brown dwarfs and directly imaged companions, enabling direct population-level comparisons \citep{manjavacasCloudAtlasHubble2019}.

Critically, transiting exoplanets and directly imaged objects offer complementary strengths that, when combined, provide a more complete picture of substellar atmospheres. Transiting planets benefit from precisely measured masses (via radial velocities), radii (via transit depths), and equilibrium temperatures (via orbital parameters and stellar properties), providing empirically constrained surface gravities and irradiation environments. In contrast, directly imaged companions and free-floating brown dwarfs achieve much higher signal-to-noise ratios in their spectra, enabling direct spectral resolution of cloud features, identification of individual molecular species, and detailed characterization of atmospheric chemistry that remains challenging for stellar photon-noise dominated eclipse signals of transiting planets. By placing both populations on the same color-magnitude diagrams, we can leverage the well-characterized physical parameters of transiting systems to interpret the high-fidelity spectral information from directly imaged objects, and vice versa.

Transiting exoplanets also expand the parameter space relative to both field brown dwarfs and young directly imaged planets. Hot Jupiters have surface gravities ($\log g \sim 2.5-4$; all $\log g$ values in this paper are in cgs units), lower than some of the youngest planetary-mass companions ($\log g \sim 3.5-4$), which in turn have lower gravities than field brown dwarfs ($\log g \sim 4-5.5$). However, unlike self-luminous objects heated by residual formation energy, transiting planets receive intense external irradiation which alters their atmospheric thermal structure \citep{fortneyUnifiedTheoryAtmospheres2008} and composition \citep{tsaiPhotochemicallyProducedSO22023}. The tidal-locking induced day-night temperature gradient can also drive strong day-night atmospheric dynamics absent in isolated objects \citep{komacekAtmosphericCirculationHot2017}. As a result, an eclipse spectrum samples a single irradiated hemisphere rather than the rotation-averaged full disk measured for self-luminous objects \citep{cowanStatisticsAlbedoHeat2011}, a distinction we return to in Section \ref{sec:limitations}. Whether irradiated planets follow the same color-magnitude sequences as self-luminous objects, and if not, what physical processes drive the differences remains an open question.

\begin{figure*}[t]
    \centering
        \includegraphics[width=0.9\textwidth]{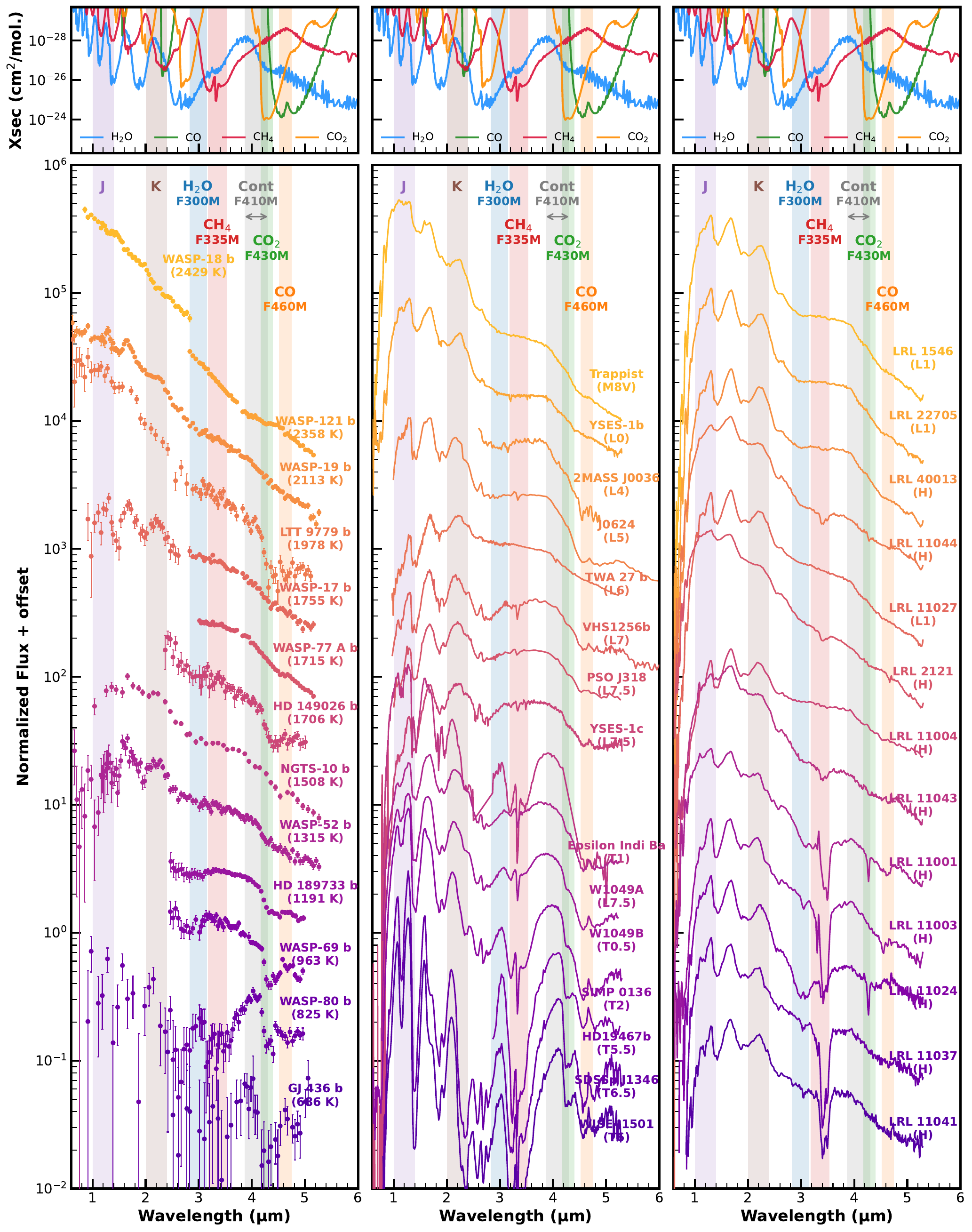}
        \caption{Emission spectra of a subset of objects with similar temperatures. The coolest late-T and Y dwarfs from \cite{beilerPreciseBolometricLuminosities2024} are omitted for clarity. The top row shows opacities of H$_2$O, CO, CH$_4$, and CO$_2$ at 1000K and 1 bar, representing the dominant molecular absorbers in H/He atmospheres. For the bottom row, the left panel displays the transiting exoplanets, while the middle panel shows the M, L and T dwarfs, and the right panel shows objects from the IC 348 star forming region \citep{luhmanNewSpectralClass2025}. All spectra are normalized and offset vertically; the transiting planets (left panel) are ordered by temperature from high to low.}
\label{spectra}   
\end{figure*}

In this paper, we present a systematic color-magnitude analysis of 13 transiting exoplanets, 57 substellar objects, and the irradiated brown dwarf ZTF J0038+2030 B spanning $T_{\rm eff} \sim 350-2600$ K. We construct CMDs using both traditional 2MASS J and K$_s$ bands and JWST NIRCam medium-band filters that isolate specific molecular absorption features. By comparing these observed sequences with atmospheric model grids, we constrain the roles of temperature, gravity, metallicity, clouds, and vertical mixing in shaping atmospheric properties across the substellar-to-planetary mass regime.

This paper is organized as follows. Section 2 describes our sample selection, data reduction, and synthetic photometry methods. Section 3 presents our results and their physical interpretation, including the J$-$K and NIRCam color-magnitude sequences, thermal inversions in ultra-hot Jupiters, the absence of L/T transitions in irradiated planets, the gravity-dependent onset of methane, and the comparison of CO$_2$ features. We summarize our conclusions in Section 4.

\section{Methods}

We collected 13 JWST transiting exoplanet emission spectra (Table \ref{tab:all_objects}), 57 ultracool dwarf and substellar object spectra, and the dayside spectrum of the irradiated brown dwarf ZTF J0038+2030 B (Section \ref{sec:ztf0038}). The transiting-planet sample spans $\sim$0.07--10 M$_J$ in mass and $\sim$680--2400 K in equilibrium temperature. The substellar objects range from Y1 (T$_{eff}\sim$350 K) to M8 (T$_{eff}\sim$2600 K) in spectral type. All datasets (Table \ref{tab:all_objects}) are from JWST/NIRSpec, except for two L dwarfs (2MASS J0036+1821 and 2MASS J1439+1929), which come from AKARI \citep{sorahanaAKARIOBSERVATIONSBROWN2012}, and $\epsilon$ Indi Ba and Bb, which come from VLT observations \citep{kingIndiBaBb2010}.

\subsection{Transiting exoplanets}

When a transiting exoplanet passes behind the star, one can obtain its disk-integrated dayside emission spectrum by measuring the relative combined flux changes during and outside the eclipse \citep{demingStrongInfraredEmission2006, charbonneauDetectionThermalEmission2005}. The resulting eclipse depth is a wavelength-dependent relative measurement of planet flux over stellar flux (F$_p$/F$_s$). To get the F$_p$ and enable direct comparison with spatially resolved objects, one can multiply the eclipse spectrum by F$_s$ from the data. 

Our transiting-planet sample comprises 13 JWST eclipse spectra from $\sim$1-5$\mu$m, of which 12 are previously published. We add WASP-52 b (GTO 1224, this work). We reanalyzed WASP-19 b, WASP-17 b, and WASP-80 b for this work to place their planet spectra on a more uniform wavelength grid and to flux-calibrate them against their stellar spectra, the resulting spectra are consistent with the previously published reductions.

\subsubsection{WASP-52 b}

The eclipse spectrum of WASP-52 b was taken as part of the GTO 1224 program on 2023 June 10th. The observation used the NIRSpec PRISM \citep{jakobsenNearInfraredSpectrographNIRSpec2022} with SUB512s subarray and 3 groups per integration. The data analysis started with the \texttt{uncal.fits} files and we used the default \texttt{jwst} stage1 pipeline \citep{bushouseJWSTCalibrationPipeline2025} to process the files and obtain the \texttt{darkcurrentstep.fits} files. Next, we calculated the column median with the first 3 and last 3 rows and subtracted them from each column to remove 1/f and background noise at the group level. Then we take the difference between the last read and the first read to obtain the 2D frame for each integration. We noticed columns 106 to 111 experienced partial saturation in the last read, and we only used the difference between the second and first read for those columns. Next, we used a 3-pixel width aperture to extract the spectral trace. All columns are first summed to create the whitelight curve, and then we fit it with mid-transit time, scaled semi-major axis (a/R$_s$), inclination, linear slope, constant, and eclipse depth using the \texttt{Batman} package \citep{kreidbergBatmanBAsicTransit2015}. During the fit, we noticed that the beginning and end segments of the eclipse light curve exhibit strong non-linear decay in flux. As a conservative approach, we decided to trim points at the beginning and the end of the light curve and fit a linear baseline slope. The detector settling at the start is non-linear which bias the linear baseline when included. As more of these points are trimmed, the non-linear component is removed and the eclipse depth stabilizes, and we trimmed the first 3000 and last 2096 points for the fit. The best-fit a/Rs, inclination, and mid-eclipse time are then fixed for the spectroscopic lightcurve fits to obtain the eclipse spectrum.

To get the flux-calibrated stellar spectrum, we ran the default \texttt{jwst} stage2 pipeline \citep{bushouseJWSTCalibrationPipeline2025} on the out-of-eclipse stellar spectrum to obtain the \texttt{x1dints.fits} file. The extracted stellar spectra fluxes are in units of MegaJansky (MJy). We then multiplied the eclipse spectrum by the stellar spectrum to get the planet flux (F$_p$) in MJy.

\subsubsection{Flux calibrated stellar spectrum}

For the other NIRSpec PRISM and G395H observation of WASP-19b and WASP-77 Ab, \texttt{x1dints.fits} files were also used to get the flux-calibrated stellar spectrum. For WASP-121 b, the spectral trace is not centered on the default spectral extraction aperture, and the resulting \texttt{x1dints.fits} stellar spectrum suffers from missing flux. To resolve this issue, we first extracted the WASP-121 b stellar spectrum from the stage 1 \texttt{rateints.fits} file. Then we used G395H observation on the standard CALSPEC star P330E calibration program (CAL/CROSS 6606) and the \texttt{p330e-mod-008.fits} stellar model spectrum from https://ssb.stsci.edu/cdbs/calspec/ to calibrate the conversion from counts from rateints to flux in Jansky. We used the same method as WASP-121 b to obtain flux-calibrated stellar spectra for all NIRCam observations \citep{schlawinMultipleCluesDayside2024}.

\subsection{Substellar objects}

The M-dwarf spectra for TRAPPIST-1 are from the \texttt{x1dints.fits} files downloaded from MAST, and other JWST spectra are obtained from the previously published works (Table \ref{tab:all_objects}). The two L-dwarf spectra (2MASS J0036+1821 and 2MASS J1439+1929) from AKARI are downloaded from the IRC Near-Infrared Point Source Spectral Catalogue \citep{usuiAKARIIRCNearInfrared2018}. They were added to fill the gap between M and late L in the currently published JWST spectra. 

\subsection{SPHEREx ultracool dwarf sample}

To place the small, curated JWST sample in the context of the full field population, we add synthetic photometry for known ultracool dwarfs from the SPHEREx SPIFF spectral library \citep{gagneSPHERExPipelineSpectral2026}, which provides $R\sim50$, 0.75--5~$\mu$m spectrophotometry that covers the five NIRCam medium bands used in this work. Starting from the 2304 known-ultracool-dwarf binned spectra, we compute synthetic photometry with the same photon-weighted band-averaged estimator and Vega zero points as for the JWST sample, and cross-match against the UltracoolSheet \citep{bestUltracoolSheetPhotometryAstrometry2025} to obtain parallax-based distances for 878 objects. Because the SPIFF spectra can retain uncleaned instrumental systematics, we apply a four-stage quality filter before plotting. Stages (i) and (ii) act on individual bands, so an object can survive in some colors and not others; stage (iii) removes an object entirely; stage (iv) affects only the color-magnitude diagrams. (i) \emph{Spikes.} A channel is flagged as a spike if it deviates from an exclude-self running median by more than 8 times the observed robust scatter of the residuals (computed separately blueward and redward of 2.5~$\mu$m) or by more than 7 times its combined quoted uncertainty (the channel error plus 10$\%$ of the local continuum, added in quadrature). The first criterion is the operative one, because the SPIFF per-channel uncertainties are overestimated by factors of $\sim$5--30 in the faint 3--5~$\mu$m region. We then determine which bands each spike corrupts by its photometric impact rather than by its wavelength: the flagged channels are repaired by interpolation, every band is recomputed through its real filter throughput, and a band is discarded if its magnitude shifts by more than 0.05 mag. This is necessary because a spike lying outside a band's half-power window can still corrupt that band through the filter wings. This stage removes 273 of the 11520 band measurements (64, 47, 73, 39 and 50 in F300M, F335M, F410M, F430M and F460M), affecting 225 objects. (ii) \emph{Band signal-to-noise.} Any band whose band-integrated flux signal-to-noise ratio is below 5 is discarded: 299 band measurements (94, 144, 4, 9 and 48 in the same order), affecting 202 objects. We test the band-integrated S/N, the throughput-weighted band-flux integral over its propagated uncertainty, rather than a single-channel value, because that integral is precisely the quantity the band magnitude measures; a band spans roughly ten SPHEREx channels, so this threshold corresponds to a median per-channel ratio of about 2. The plot-time cut below removes any survivors with larger errors. The counts are largest in F300M and F335M, dominated by T dwarfs whose deep water and methane absorption leaves them intrinsically faint in these 3~$\mu$m bands. (iii) \emph{Red-end dropout.} Because every L0 and later dwarf shows deep water absorption, an object whose F300M$-$F410M water color falls below 0.35 mag is a failed red-end extraction, in which the 3--5~$\mu$m continuum is undetected and all medium-band colors collapse toward zero; all of its photometry is discarded. Two objects are removed. (iv) \emph{Distance quality.} Absolute magnitudes require a reliable parallax, so the color-magnitude diagrams retain only objects with a fractional distance uncertainty below 0.33 (parallax signal-to-noise above 3), which excludes 7 of the objects that have UltracoolSheet distances. The color-color diagrams do not use distance and are unaffected. Every cut therefore targets an identifiable instrumental defect. Of the 2304 spectra we start from, 2153 appear in at least one panel: the color-magnitude diagrams retain 763 (H$_2$O), 746 (CH$_4$), 832 (CO$_2$) and 815 (CO) objects, all of which also require a parallax, and the color-color diagrams, which do not, retain 1956 (CH$_4$--CO) and 2108 (CO$_2$--CO). The full chain (synthetic photometry, UltracoolSheet cross-match, and the quality filter) is included in the code release with the paper, so the sample can be reproduced from the public SPIFF spectra. They are shown as a grey background population in Figures \ref{band_color_data_only} and \ref{H2O_color_mag}--\ref{CO2_CO_color_color}. While the SPHEREx sample does not reach the Y-dwarf regime, it densely samples the M/L/T sequence and the L/T transition in the 3--5~$\mu$m molecular bands, providing an empirical field locus for the CH$_4$ and CO$_2$ diagnostics.

\subsection{ZTF J0038+2030 B: an irradiated high-gravity brown dwarf}
\label{sec:ztf0038}

The eclipsing white dwarf--brown dwarf binary ZTF J0038+2030 was recently observed with JWST/NIRSpec PRISM spectroscopic phase curves (GO 4967; \citealt{broski-laingAsymmetricNightsideCO22026}). The system consists of a 0.505 M$_\odot$ white dwarf and a tidally locked brown dwarf (M $= 62.1 \pm 4.1$ M$_J$, R $= 0.759 \pm 0.011$ R$_J$, $\log g = 5.425^{+0.020}_{-0.030}$) on a 10.4 hr orbit at $d = 139 \pm 2$ pc, with a total system age of 7.5--8.8 Gyr and dayside and nightside brown dwarf temperatures of $1049 \pm 6$ K and $968 \pm 20$ K \citep{broski-laingAsymmetricNightsideCO22026}. This object provides a configuration absent from the rest of the sample: a dayside hemisphere that is eclipsing, irradiated, and tidally locked, exactly like a hot Jupiter, but with old, field brown-dwarf surface gravity. The irradiation is mild compared with a close-in planet, contributing roughly 25$\%$ of the dayside emitted power (Section \ref{sec:ztf_control}), but it is the coolest eclipsing brown dwarf--white dwarf system observed with JWST to date.

We performed an independent reduction of the public data, starting from the \texttt{uncal.fits} files with the \texttt{jwst} Detector1 pipeline plus a group-level 1/f correction, in which the per-column median of the off-trace background rows is subtracted from each group before ramp fitting. Spectra were extracted with a fixed box aperture and a local, wavelength-dependent background estimated from the rows between the aperture and the 1/f rows. The white dwarf spectrum is obtained empirically from the eclipse (out-of-eclipse minus totality flux), the nightside spectrum from totality, and the dayside spectrum from the median over phases 0.45--0.55 minus the white dwarf contribution. The absolute flux scale is transferred from the \texttt{jwst} pipeline \texttt{x1dints} products. We adopt the 5-row box with the adjacent rows as local background. The methane color is stable to $\pm 0.06$ mag across different extraction aperture and local-background configurations. The resulting dayside and nightside spectra are consistent with those of \citet{broski-laingAsymmetricNightsideCO22026}, including the Y/J/H/K band peaks and the asymmetric nightside CO$_2$ feature. Synthetic photometry is computed with the same estimator, filter curves, and zero points as the rest of the sample; the dayside methane color is F335M$-$F410M $= 2.21 \pm 0.02$ mag. In all figures, ZTF J0038+2030 B is plotted together with the transiting irradiated planets, with the same symbols and the same 1~R$_J$ radius normalization (its radius is directly measured from the eclipses). Its role as a high-gravity irradiated control data point is discussed in Section \ref{sec:ztf_control}.

\begin{figure*}[t]
    \centering
    \includegraphics[width=0.8\textwidth]{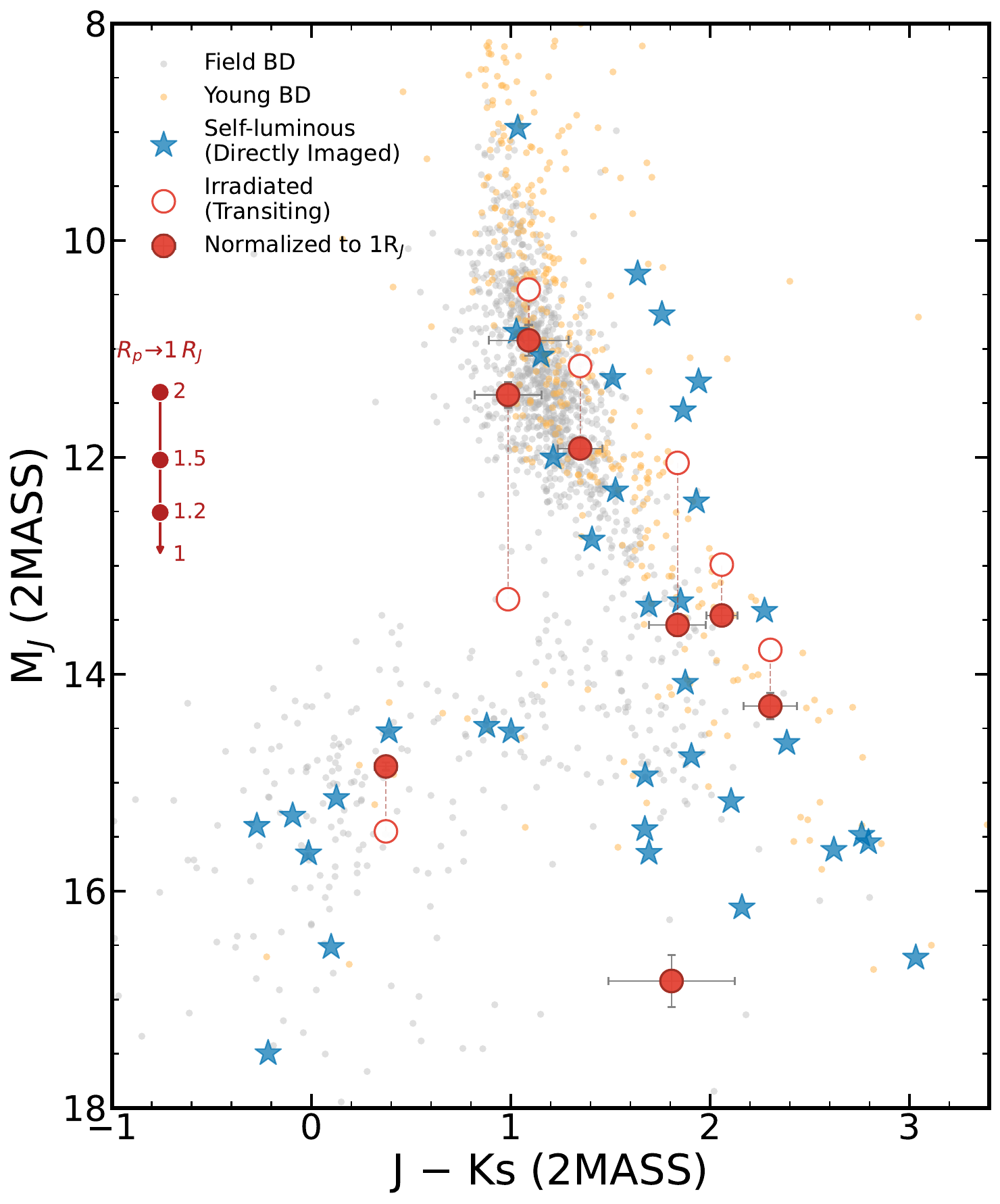}
    \caption{J vs. J-K color-magnitude diagram. The transiting exoplanets (red circles) and the self-luminous directly imaged objects of our sample (blue stars) are plotted alongside field (grey) and young (orange) brown dwarfs and substellar companions from the UltracoolSheet \citep{bestUltracoolSheetPhotometryAstrometry2025}. The transiting planets align with the L-dwarf sequence after normalizing their radii to 1 $R_J$ with no clear L/T transition down to M$_J \sim$17. Open and filled circles connected by dashed lines show each planet before and after the 1~$R_J$ normalization. The red arrow on the left shows the radius-normalization effect, $\Delta m = 5\log_{10}(R/R_J)$: an inflated object at the labeled radius ($2.0$, $1.5$, or $1.2\,R_J$, top to bottom) shifts down to the arrowhead, its 1~$R_J$ position ($+1.51$, $+0.88$, and $+0.40$ mag fainter, respectively). Field brown dwarfs are assumed to lie near 1~$R_J$ and receive no arrow. Because the shift is achromatic it is purely vertical and does not alter the $J-K$ color.}
\label{J_K_CMD}
\end{figure*} 

\subsection{Synthetic photometry}

To directly compare emission spectra from transiting exoplanets and substellar objects across the full wavelength range, we converted the observed flux-calibrated spectra (in units of F$_\nu$, Jy) to synthetic photometry using narrow-band filter transmission curves. For each object and filter, the synthetic flux density is calculated as a photon-weighted integral:

\begin{equation}
F_\nu = \frac{\int F_\nu(\lambda) \frac{T(\lambda)}{\lambda} d\lambda}{\int \frac{T(\lambda)}{\lambda} d\lambda}
\end{equation}

where $F_\nu(\lambda)$ is the observed spectral flux density in Jy, $T(\lambda)$ is the filter transmission function normalized to a maximum of 1, and the integral is over the wavelength range where the filter has significant throughput. Uncertainties are propagated using standard weighted-mean error propagation. We used five NIRCam filters (F300M, F335M, F410M, F430M, F460M) to capture molecular features in water (H$_2$O), methane (CH$_4$), continuum, carbon dioxide (CO$_2$), and carbon monoxide (CO). For comparison with the brown dwarf population, we also calculated 2MASS J and Ks band photometry using the corresponding filter transmission curves and the published 2MASS zero points. All synthetic photometry values are listed in Table \ref{tab:photometry_10pc}.

\subsection{Normalization}

We normalize all synthetic photometry to a standard distance of 10~pc. For transiting exoplanets (hot Jupiters) and the eclipsing irradiated brown dwarf ZTF J0038+2030 B, whose radii are directly measured, we divide the flux by $R_p^2$ to normalize to 1~$R_J$, enabling direct comparisons of intrinsic atmospheric properties across objects with different radii. We do not apply this radius normalization to brown dwarfs or directly imaged young substellar objects because their radii are inferred from evolutionary models and can carry large systematic uncertainties ($\sim$20$\%$, the typical disagreement between evolutionary-model radii and atmospheric-model-fit radii of substellar objects; e.g., \citealt{sanghiHawaiiInfraredParallax2023}), especially for young objects. On the other hand, transiting planets have more precisely measured transit radii ($\sim$few$\%$): the transit depth measures $(R_p/R_s)^2$ to well below a percent, and the stellar radius follows from the Gaia parallax and well-established stellar models to a few percent.

We emphasize that this radius normalization is achromatic: dividing the flux by $R_p^2$ scales every photometric band by the same factor, so it shifts an object only vertically (in absolute magnitude) and cancels identically in any color. All color-based results in this paper (the $J-K$ colors, the F335M$-$F410M methane color, and the CO$_2$/CO diagnostics) are therefore independent of the adopted radii; only the vertical placement on the CMDs carries a radius systematic. To quantify the residual vertical systematic from not normalizing the self-luminous objects, we note that radius enters as $\Delta m = 5\log_{10}(R/R_J)$: field-age brown dwarfs converge to the electron-degeneracy radius of $\sim$0.8--1.1~$R_J$ \citep{saumonEvolutionDwarfsColorMagnitude2008, marleySonoraBrownDwarf2021, phillipsNewSetAtmosphere2020}, whereas young ($\sim$10--200 Myr) planetary-mass objects can be inflated to $\sim$1.2--1.5~$R_J$, corresponding to shifts of $-0.48$ to $+0.88$ mag if normalized to 1~$R_J$. The arrows in Figures~\ref{J_K_CMD} and \ref{band_color_data_only} illustrate this shift for representative inflated radii of 1.2, 1.5, and 2.0~$R_J$ ($+0.40$, $+0.88$, and $+1.51$~mag fainter when normalized to 1~$R_J$). Figure~2 shows the $J$--$K$ color--magnitude diagram of all objects, while Figure~3 presents six panels comparing molecular feature strengths (delta magnitudes relative to the continuum) and molecular color--color ratios.

\subsection{J and Ks band}

The J and K bands are widely used to characterize and classify brown dwarfs originating from ground-based observations that need to limit the impact from Earth's atmospheric water absorption. This near-infrared color-magnitude diagram (CMD) provides population-level insights, specifically the L/T transition, explained by the onset of silicate clouds in L dwarfs and dissipation in T dwarfs \citep{saumonEvolutionDwarfsColorMagnitude2008, suarezUltracoolDwarfsObserved2022}. We have seven transiting exoplanets with NIR wavelength coverage with SOSS ($\sim$0.6-2.84$\mu$m) and PRISM ($\sim$0.6-5.3$\mu$m) in this study from $\sim2400$ to $\sim800$ K, spanning from M to T dwarfs. By converting their observed spectra to J and K band magnitudes, we can compare transiting exoplanets with self-luminous objects. First, we calculated the band integrated flux based on the 2MASS J and Ks bands in Jy and then used the 2MASS J and Ks zero points of 1594 and 666.7 Jy \citep{cohenSpectralIrradianceCalibration2003} to convert the flux into absolute magnitudes. 

Figure \ref{J_K_CMD} displays the J vs. J-K color-magnitude diagram. We compiled 2MASS J and Ks band photometry from UltracoolSheet \citep{bestUltracoolSheetPhotometryAstrometry2025} for field and young brown dwarfs and substellar companions. Our sample, which spans M, L, and T dwarfs, is overplotted on this comprehensive dataset. We also include all transiting exoplanets with J and K band wavelength coverage. With the radius normalization, the transiting exoplanets closely follow the L-dwarf sequence, with no clear evidence of an L/T transition down to M$_J \sim$17.

\subsection{Five infrared bands}

For objects with infrared coverage in the $\sim$3-5$\mu$m range, we calculate synthetic photometry using five NIRCam filters (F300M, F335M, F410M, F430M, and F460M) that capture key molecular features: H$_2$O, CH$_4$, continuum, CO$_2$, and CO, respectively. This choice of filters enables direct comparison with high-contrast directly imaged planetary companions \citep{balmerJWSTTSTHighContrast2025}.

The conversion from flux density to magnitudes requires filter-specific zeropoints. We adopt the NIRCam zeropoints from the JWST calibration reference file (jwst\_1126.pmap), which are derived using the CALSPEC Vega spectrum (alpha\_lyr\_stis\_011.ascii). For the NRCALONG detector, the Vega zeropoints are: F300M = 367.96 Jy, F335M = 299.65 Jy, F410M = 209.63 Jy, F430M = 190.67 Jy, and F460M = 163.78 Jy. These zeropoints correspond to the flux density of Vega in each filter bandpass and are used to compute magnitudes as $m = -2.5 \log_{10}(F_\nu / F_{\nu,\mathrm{Vega}})$.

\begin{figure*}[t]
    \centering
        \includegraphics[width=0.8\textwidth]{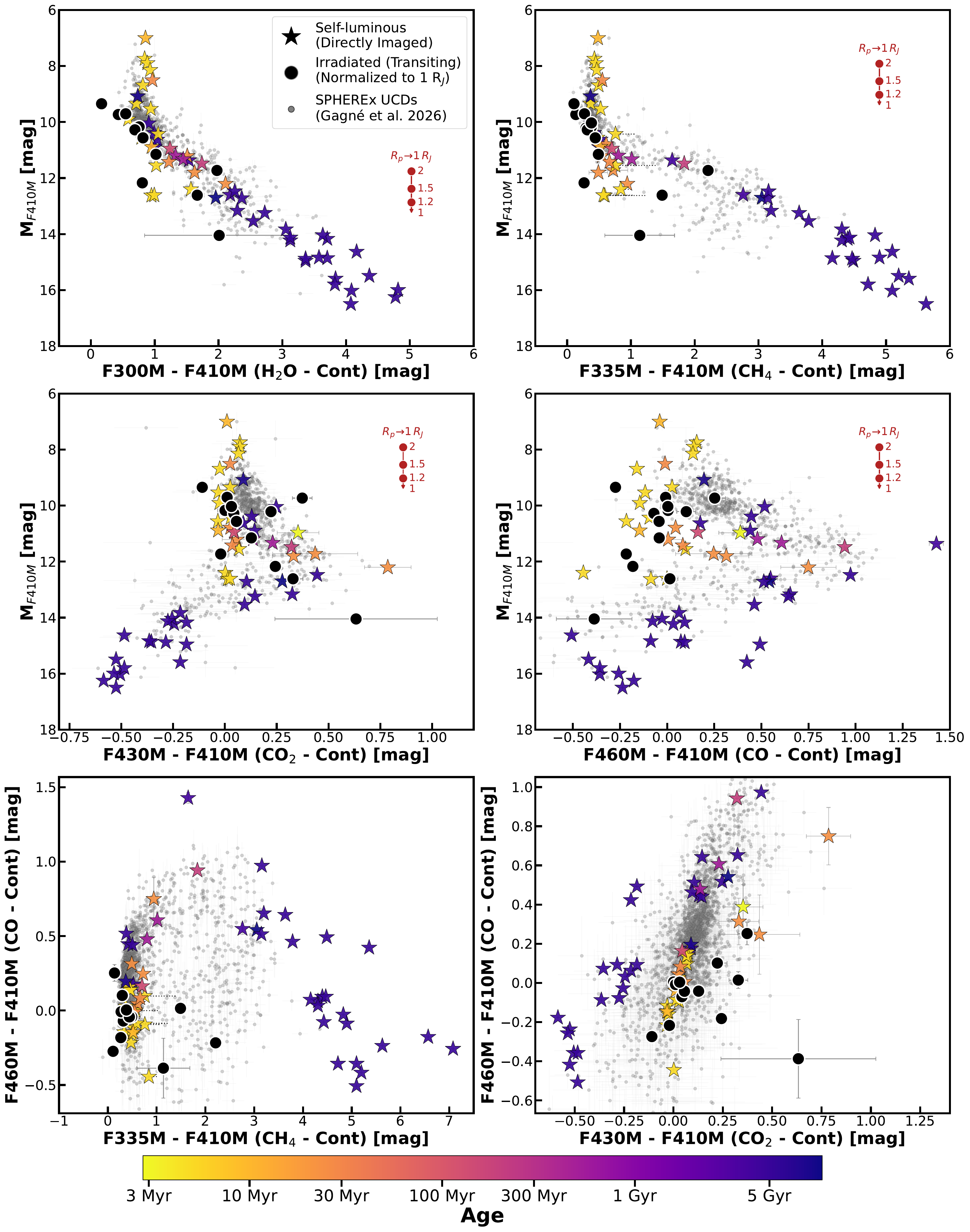}
            \caption{Six-panel analysis of molecular absorption features. The top four panels show color-magnitude diagrams of M$_{F410M}$ versus the absorption strength of H$_2$O, CH$_4$, CO$_2$, and CO relative to the F410M continuum (plotted on the x-axis; positive values indicate stronger absorption). The bottom two panels show color-color plots: CO-continuum versus CH$_4$-continuum (left) and CO-continuum versus CO$_2$-continuum (right). In each of the four color-magnitude panels, the red arrow shows the radius-normalization shift, $\Delta m = 5\log_{10}(R/R_J)$: an object at the labeled radius ($2.0$, $1.5$, or $1.2\,R_J$, top to bottom) moves down to the arrowhead, its 1~$R_J$ position; the two color-color panels carry no arrow because radius cancels in a color. Transiting irradiated planets are shown as circles, self-luminous objects as stars. Objects are color-coded by age: field objects (purple) and young objects (orange). Transiting planets (black) are all in mature systems ($\sim$Gyr); WASP-18 b lacks 3--5~$\mu$m coverage and appears only in the J vs. J$-$K diagram. In the top-right CH$_4$ CMD, unknown hydrocarbon features for objects from IC~348 are masked, and the colors derived from the unmasked spectra are indicated by horizontal dotted lines. Typical transiting planets show similar colors as very young self-luminous objects as both population share rather similar temperature and surface gravity. Small grey points show the quality-filtered SPHEREx known ultracool dwarfs with UltracoolSheet parallaxes \citep{gagneSPHERExPipelineSpectral2026, bestUltracoolSheetPhotometryAstrometry2025}, which trace the empirical field sequence.}
        \label{band_color_data_only}
\end{figure*}

Figure \ref{band_color_data_only} presents a six-panel analysis of these molecular features. The top four panels display color-magnitude diagrams of M$_{F410M}$ versus the absorption strength of each molecular feature (H$_2$O, CH$_4$, CO$_2$, and CO) relative to the F410M continuum, quantified as the magnitude difference between the molecular band and continuum on the x-axis (positive values indicate stronger absorption). The bottom two panels show color-color diagrams: CO-continuum versus CH$_4$-continuum (left) and CO-continuum versus CO$_2$-continuum (right), which serve as diagnostic tools for atmospheric composition and chemistry.

\subsection{Age}

For transiting exoplanets and field objects without known youth indicators, we assume a fiducial age of 5 Gyr, representative of typical field stars and consistent with the few-Gyr characteristic age of the field brown dwarf population \citep{dupuyIndividualDynamicalMasses2017}. For substellar companions and free-floating brown dwarfs with known moving group membership, ages are determined from the literature. The IC 348 objects are assigned an age of 5 Myr based on the age of the IC 348 star-forming region \citep{luhmanNewSpectralClass2025}. Other young objects (TWA 27 A/B, VHS 1256 b, W1049A/B, PSO J318, YSES-1b/c, HR 8799 b/c/d/e, 29 Cygni b, and SIMP 0136+0933) have ages ranging from 2.8 to 200 Myr based on their reported moving group membership or cluster age, with the corresponding references listed in Table \ref{tab:all_objects}. These age assignments are summarized in Table \ref{tab:all_objects}.

\subsection{Model tracks}
To contextualize our observations within theoretical frameworks, we compare our data with several atmospheric model grids. We compiled existing grids for self-luminous objects, and for transiting irradiated objects we generated new PICASO 1D self-consistent radiative-convective equilibrium models that include radiatively active clouds.

\subsubsection{Self-luminous models}
For self-luminous substellar objects, we employ three models from the Sonora suite as well as the Exo-REM models. The cloud-free Sonora Bobcat models \citep{marleySonoraBrownDwarf2021} span effective temperatures of $T_{\mathrm{eff}} = 200$–2400~K, surface gravities of $\log g = 3.0$–5.5 (cgs), and metallicities of $\mathrm{[M/H]} = -0.5, 0.0,$ and $+0.5$. The Sonora Diamondback models \citep{morleySonoraSubstellarAtmosphere2024} have self-consistent treatment of cloud physics. The Sonora Elf Owl models \citep{mukherjeeSonoraSubstellarAtmosphere2024} incorporate non-equilibrium chemistry relevant for cooler objects. The Exo-REM models \citep{charnaySelfconsistentCloudModel2018} have self-consistent treatment for clouds and mixing. 

\subsubsection{Irradiated models}

For irradiated objects, we generate a grid of one-dimensional radiative–convective equilibrium models using PICASO, including a self-consistent treatment of vertical mixing–induced disequilibrium chemistry \citep{mukherjeePICASO30Onedimensional2023, batalhaExoplanetReflectedlightSpectroscopy2019, mang_picaso_2026}, for Jupiter-sized planets. We adopt a fixed planetary radius of 1~$R_{\mathrm{J}}$ and vary planet masses to obtain surface gravity over $\log g=2.5$, 3.0, 3.5, and 4.0 (cgs). The stellar spectrum is set by a Sun-like host star, and we compute models over $T_{\mathrm{eq}}=600$--2000~K in 200~K increments. The internal temperature $T_{\mathrm{int}}$ is prescribed as a function of $T_{\mathrm{eq}}$ following Equation~3 of \citet{thorngrenIntrinsicTemperatureRadiative2019}, with a minimum floor of $T_{\mathrm{int}}=100$~K. For our grid points from $T_{\mathrm{eq}}=600$ to 2000~K, the corresponding $T_{\mathrm{int}}$ values are 100, 100, 222, 383, 526, 623, 666, and 664~K, respectively.

Current one-dimensional atmospheric model grids for irradiated exoplanets have largely focused on either cloud-free cases \citep{wiser_comparison_2026} or models with parameterized cloud treatments \citep{goyalLibraryATMOForward2018}. The radiative feedback effects of clouds have generally been neglected in 1D atmospheric models; however, they have been shown to significantly influence the thermal structure in both global circulation models \citep{kennedy_radiatively_2024} and brown dwarf atmospheres \citep{charnaySelfconsistentCloudModel2018}.

For this work, we use PICASO coupled with VIRGA \citep{batalha_condensation_2026, mang_picaso_2026} to include the effects of radiatively active clouds (absorption and scattering) with condensate species MgSiO$_3$, Mg$_2$SiO$_4$, Fe, Al$_2$O$_3$, MnS, Na$_2$S, KCl, and Cr. The grid spans three bulk metallicities (solar, 3$\times$ solar, and 10$\times$ solar). We also vary the C/O ratio over 0.5, 1.0, and 1.5 times the solar value (C/O$_\odot=0.458$) \citep{lodders44AbundancesElements2009}. We compute models for four cloud cases: clear, $f_{\rm sed}=2.0$, 1.0, and 0.5. High f$_{sed}$ values produce larger particles with optically thin clouds and
low values result in more vertically extended and optically thick
clouds.

\subsection{An empirical methane-onset separator}
\label{sec:svm_methods}

To summarize the gravity dependence of the 3.3~$\mu$m methane feature (Section 3.3), we construct an empirical separator in the temperature--$\log g$ plane that divides objects with strong and weak CH$_4$ absorption features. We label each object by its F335M$-$F410M color, adopting an operational threshold of 0.8 mag; this value is guided by the spectra of transitional objects (VHS 1256 b, W1049A/B, and YSES-1c), whose nascent 3.3~$\mu$m bands correspond to colors near 0.8 mag. This is an empirical decision boundary that best describes our sample, not a physical methane-onset law.

The separator is found with a linear support vector machine (SVM) classifier (\texttt{scikit-learn}, linear kernel). It is fit to all 56 objects with a temperature, a surface gravity and an F335M$-$F410M color. Because temperature (hundreds to thousands of Kelvin) and $\log g$ (a few dex) span very different numerical ranges, we first rescale each axis to zero mean and unit standard deviation so that neither one dominates the fit. The classifier then finds the straight line that leaves the widest possible gap between the methane-bearing and methane-free objects, and we convert that line back into physical units to quote it as $\log g = m \times T + b$. Boundary uncertainties are estimated with a 1000-iteration bootstrap in which each point is perturbed by Gaussian draws from its uncertainties in temperature, $\log g$, and F335M$-$F410M color; we construct 1$\sigma$, 2$\sigma$, and 3$\sigma$ envelopes, and verified that doubling the bootstrap to 2000 iterations changes the median coefficients and envelopes by less than 2$\%$ of their uncertainties.

Two caveats accompany this fit. First, the surface gravities are heterogeneous in origin: substellar gravities come from atmospheric model fits with systematic, model-dependent uncertainties whereas transiting-planet gravities follow nearly model-independently from radial-velocity masses and transit radii ($\lesssim$0.1 dex). Second, the bootstrap assumes Gaussian, uncorrelated errors, whereas model-derived $T_{\rm eff}$ and $\log g$ are in practice covariant. 

We also performed a sensitivity test of the 0.8 mag threshold choice by refitting the separator at color thresholds of 0.6, 0.7, 0.9, and 1.0 mag: the slope, which encodes the gravity--temperature trade-off and is the physically meaningful quantity, remains between $3.7\times10^{-3}$ and $5.0\times10^{-3}$ K$^{-1}$, consistent within the bootstrap uncertainties at every threshold, and varying the threshold mainly translates the boundary parallel to itself. The systematic delay of methane onset to lower temperatures at lower gravity is therefore robust to the threshold choice.

\section{Results and Discussion}

\subsection{Thermal Inversions in Ultra-hot Jupiters}

Two ultra-hot Jupiters in our sample, WASP-18 b ($T_{eq} = 2429$ K) and WASP-121 b ($T_{eq} = 2358$ K), exhibit clear thermal inversion signatures characterized by H$_2$O and CO emission features rather than absorption \citep{mansfieldHSTWFC3Thermal2018, coulombeBroadbandThermalEmission2023, evans-somaSiOSuperstellarRatio2025}. This behavior is consistent with the established theoretical framework that ultra-hot Jupiters ($T_{eq} > 2200$ K) develop stratospheric temperature inversions due to strong optical absorption by gas-phase TiO, VO, and atomic species such as Fe and Mg at high altitudes \citep{fortneyUnifiedTheoryAtmospheres2008, lothringerUVAbsorptionSilicate2022, lothringerExtremelyIrradiatedHot2018}. At these extreme temperatures, refractory species that would otherwise condense remain in the gas phase and absorb incident stellar radiation at pressures lower than where the thermal infrared emission originates, creating an inverted temperature-pressure profile.

This phenomenon is notably absent in the cooler hot Jupiters in our sample ($T_{eq} < 2000$ K), which uniformly display molecular absorption features indicative of temperature decreasing with altitude. The critical temperature threshold for thermal inversions appears to lie between $\sim$2100-2200 K, consistent with condensation curves for refractory-bearing species. 

Self-luminous brown dwarfs and substellar objects in our sample also lack prominent thermal inversions, as expected for objects without significant external irradiation. Their atmospheres are heated from below by internal luminosity, naturally producing temperature profiles that decrease monotonically with altitude in the radiative portions of their atmospheres. This fundamental difference in energy deposition geometry, external irradiation versus internal heat flux, produces the distinct spectral morphologies between ultra-hot Jupiters and self-luminous objects at similar temperatures.

\subsection{Irradiated Transiting Planets Resemble Young Brown Dwarfs}

In the J-K color-magnitude diagram (Figure \ref{J_K_CMD}), transiting exoplanets closely follow the L-dwarf sequence after normalizing for their inflated radii, maintaining red J-K colors. Critically, we observe no evidence for an L/T transition, the characteristic blueward color evolution seen in field brown dwarfs around $T_{eff} \sim 1200-1400$ K, down to $M_J \sim 17$. Even WASP-80 b, with $T_{eq} = 825$ K placing it in the expected T-dwarf temperature regime, retains L-dwarf-like red colors rather than exhibiting the blue J-K colors characteristic of cloud-free T-dwarf atmospheres.

This behavior mirrors that of young, low-gravity brown dwarfs and directly imaged planetary companions, which also show delayed or suppressed L/T transitions compared to field objects \citep{manjavacasMediumresolution09753Mm2024}. The L/T transition in field brown dwarfs is attributed to the breakup and sinking of silicate cloud decks, which reveals the underlying cloud-free photosphere and produces bluer near-infrared colors \citep{burgasserUnifiedNearInfraredSpectral2006, marleyPATCHYCLOUDMODEL2010}. The absence of this transition in both young brown dwarfs and irradiated planets suggests that similar physical mechanisms could operate in both populations. Delayed cloud clearing may reflect the combined effects of lower surface gravity, which slows sedimentation, and enhanced vertical mixing (higher $K_{zz}$), which lofts condensates to higher altitudes and sustains optically thick clouds to cooler temperatures.

The NIRCam color--magnitude analysis shows that transiting planets and young substellar objects have weaker molecular absorption features than field-age brown dwarfs at similar brightness (Figure \ref{band_color_data_only}). This muted spectral contrast likely reflects a combination of irradiation-driven, more isothermal temperature--pressure profiles and high-altitude clouds that raise the photosphere. Strong irradiation reduces the temperature contrast between in-band and out-of-band photospheric layers, diminishing band depths, while elevated clouds shift the effective photosphere to lower pressures with smaller molecular column densities, further weakening absorption. If young, low-gravity substellar objects are similarly cloudy, they should show comparably shallow molecular features for the similar reasons: higher clouds, a more elevated photosphere, and weaker temperature gradients.

\subsection{Delayed Onset of Methane in Low Surface Gravity Objects}

\begin{figure*}[t]
    \centering
    \includegraphics[width=\textwidth]{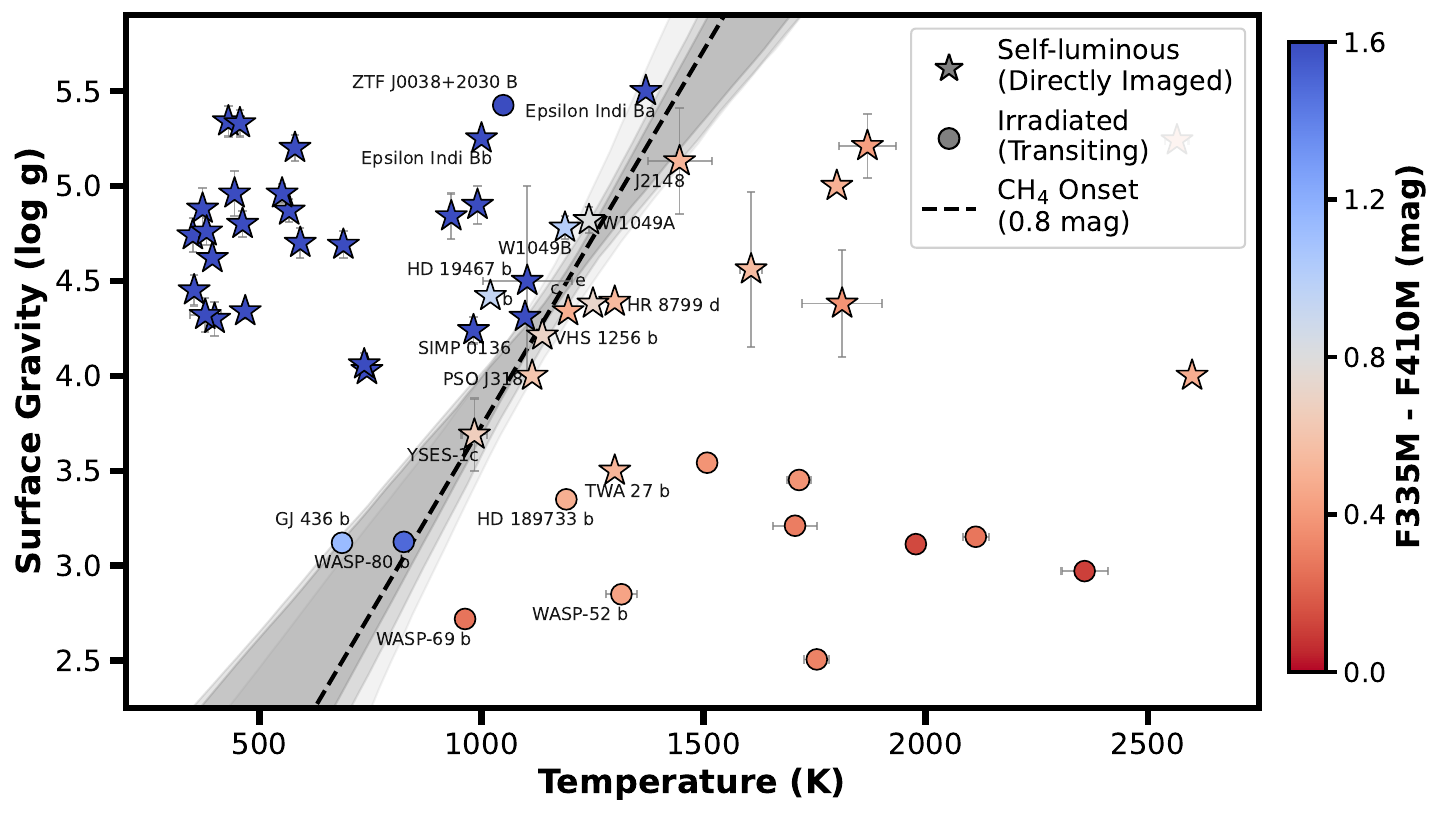}
    \caption{Temperature versus surface gravity diagram with the empirical CH$_4$ onset line. Substellar objects have model-derived surface gravities and effective temperatures from the literature, while transiting planets have measured equilibrium temperatures and surface gravities derived from radial velocity masses and transit radii. The data points are color-coded by the F335M$-$F410M color (methane absorption strength), where blue indicates stronger methane absorption (color $>$ 0.8 mag). The dashed black line shows an empirical decision boundary fit with a linear support vector machine classifier (Section \ref{sec:svm_methods}), with shaded regions indicating the 1$\sigma$, 2$\sigma$, and 3$\sigma$ bootstrap uncertainty envelopes. Objects above and to the left of the line (cooler, higher gravity) exhibit strong methane absorption (F335M$-$F410M $>$ 0.8 mag), while those below and to the right (hotter, lower gravity) show weak or absent CH$_4$ features. The onset of CH$_4$ absorption is systematically delayed to lower temperatures as surface gravity decreases. Irradiated brown dwarf ZTF J0038+2030 B (labeled) lands $\sim$1.5 dex inside the methane region (Section \ref{sec:ztf_control}), supporting surface gravity rather than irradiation as the dominant physical parameter determining the onset of CH$_4$ at a given temperature.}
\label{Teff_logg}
\end{figure*}

The F335M$-$F410M color traces CH$_4$ absorption at 3.3 $\mu$m relative to the continuum at 4.1 $\mu$m. In the color magnitude diagram (Figure \ref{band_color_data_only}), the field age L/T dwarfs have stronger CH$_4$ absorption compared to their younger counterparts and transiting planets. The $\sim$750 SPHEREx field ultracool dwarfs trace this transition continuously across the L/T boundary, confirming the onset of strong-CH$_4$ Q-branch is universal among field L/T dwarfs, from which the transiting planets and young companions clearly depart. To map out the onset of CH$_4$ and explore the underlying physical mechanisms, we plotted the points in the surface gravity and temperature space (Figure \ref{Teff_logg}). Surface gravities for brown dwarfs are derived from spectral model fits in the literature (with the late-T and Y dwarf parameters taken from the JWST retrieval analysis of \citealt{lueberCloudsChemistryAcross2026}), while transiting planet gravities are calculated from radial velocity masses and transit-derived radii (Table \ref{tab:all_objects}). Field brown dwarfs with $\log g \sim 5$ exhibit strong methane absorption beginning at $T_{eff} \sim 1400$ K, corresponding to spectral types near the L/T transition (T0.5--T1 in our sample). In contrast, transiting exoplanets with $\log g \sim 2.5-3.5$ show the onset of detectable CH$_4$ absorption only at $T_{eq} \lesssim 825$ K, a temperature shift of $\sim$600 K. Young substellar companions and low-gravity free-floating objects occupy intermediate positions in this trend, with methane appearing at progressively lower temperatures as gravity decreases. The irradiated high-gravity brown dwarf ZTF J0038+2030 B (labeled in Figure \ref{Teff_logg}) extends this sequence into the irradiated high-gravity regime, discussed in Section \ref{sec:ztf_control}.

\subsubsection{Methane-onset line}

To quantify this gravity-temperature dependence, we define the CH$_4$ onset threshold as F335M$-$F410M = 0.8 mag. This threshold value was determined through visual inspection of the spectra of objects where methane features are just beginning to emerge. Specifically, VHS 1256 b, W1049A/B, and YSES-1c all exhibit nascent CH$_4$ absorption features around 3.3 $\mu$m that correspond to F335M$-$F410M colors near 0.8 mag, making this a convenient operational threshold between objects with detectable versus undetectable methane signatures. We then fit a boundary line in the $T_{\rm eff}$--$\log g$ plane that optimally separates objects with strong methane absorption (F335M$-$F410M $>$ 0.8 mag) from those with weak or absent CH$_4$ features (F335M$-$F410M $<$ 0.8 mag).

The separator is fit with a linear support vector machine classifier, with uncertainties from a bootstrap analysis and robustness tests against the threshold choice; the full methodology and caveats are described in Section \ref{sec:svm_methods}. The resulting empirical decision boundary is:
\begin{equation}
\log g = m \times T_{eff} + b
\label{eq:ch4_onset}
\end{equation}
The best-fit CH$_4$ onset line is:
\begin{equation}
\log g = \left(3.95^{+0.34}_{-1.20}\right)\times10^{-3}\,T_{eff} - \left(0.218^{+1.459}_{-0.400}\right)
\label{eq:ch4_onset_bestfit}
\end{equation}
where $T_{eff}$ is in Kelvin and the uncertainties represent the 16th--84th percentile range from the bootstrap analysis.

A potential concern with this plane is that transiting planets are plotted at their equilibrium temperature $T_{eq}$ (computed for full heat redistribution and zero albedo) while self-luminous objects are plotted at their effective temperature. For an irradiated planet the dayside effective temperature can exceed $T_{eq}$ depending on the Bond albedo $A_B$ and the day--night recirculation efficiency $\epsilon$ (equation 4 of \citealt{cowanStatisticsAlbedoHeat2011}). However, $\epsilon$ is not a free parameter: both theory and phase-curve observations establish that heat redistribution becomes progressively less efficient (the day--night temperature contrast grows) with increasing irradiation \citep{cowanStatisticsAlbedoHeat2011, perezbeckerAtmosphericHeatRedistribution2013, komacekAtmosphericCirculationHot2016, komacekAtmosphericCirculationHot2017}. The ratio $T_{day}/T_{eq}$ is therefore an \emph{increasing} function of $T_{eq}$: the largest dayside excesses occur for the ultra-hot Jupiters (e.g. WASP-18 b, WASP-121 b), which sit far from the methane boundary, whereas the cool planets that define the boundary ($T_{eq}\lesssim1000$~K) lie in the efficient-redistribution regime where $T_{day}\approx T_{eq}$. Since this is  where the methane boundary lies, using dayside temperatures would not meaningfully shift the boundary. This expectation is confirmed by direct JWST measurements of the boundary planets, whose dayside temperatures are consistent with $T_{eq}$: $663\pm5$~K for GJ 436 b \citep{mukherjeeJWSTPanchromaticThermal2025}, $811^{+70}_{-69}$~K with a geometric albedo of 0.20 for WASP-80 b \citep{morelModerateAlbedoReflecting2025}, and a disk-integrated dayside consistent with $T_{eq}$ for WASP-69 b \citep{schlawinMultipleCluesDayside2024}. We therefore adopt $T_{eq}$ as a reproducible, model-independent abscissa available uniformly for the whole sample. The positive correlation between temperature and gravity for the onset of methane is consistent with a systematic delay: at lower gravities, objects must cool to significantly lower temperatures before CH$_4$ absorption feature starts to appear.

\begin{figure*}[t]
    \centering
    \includegraphics[width=\textwidth]{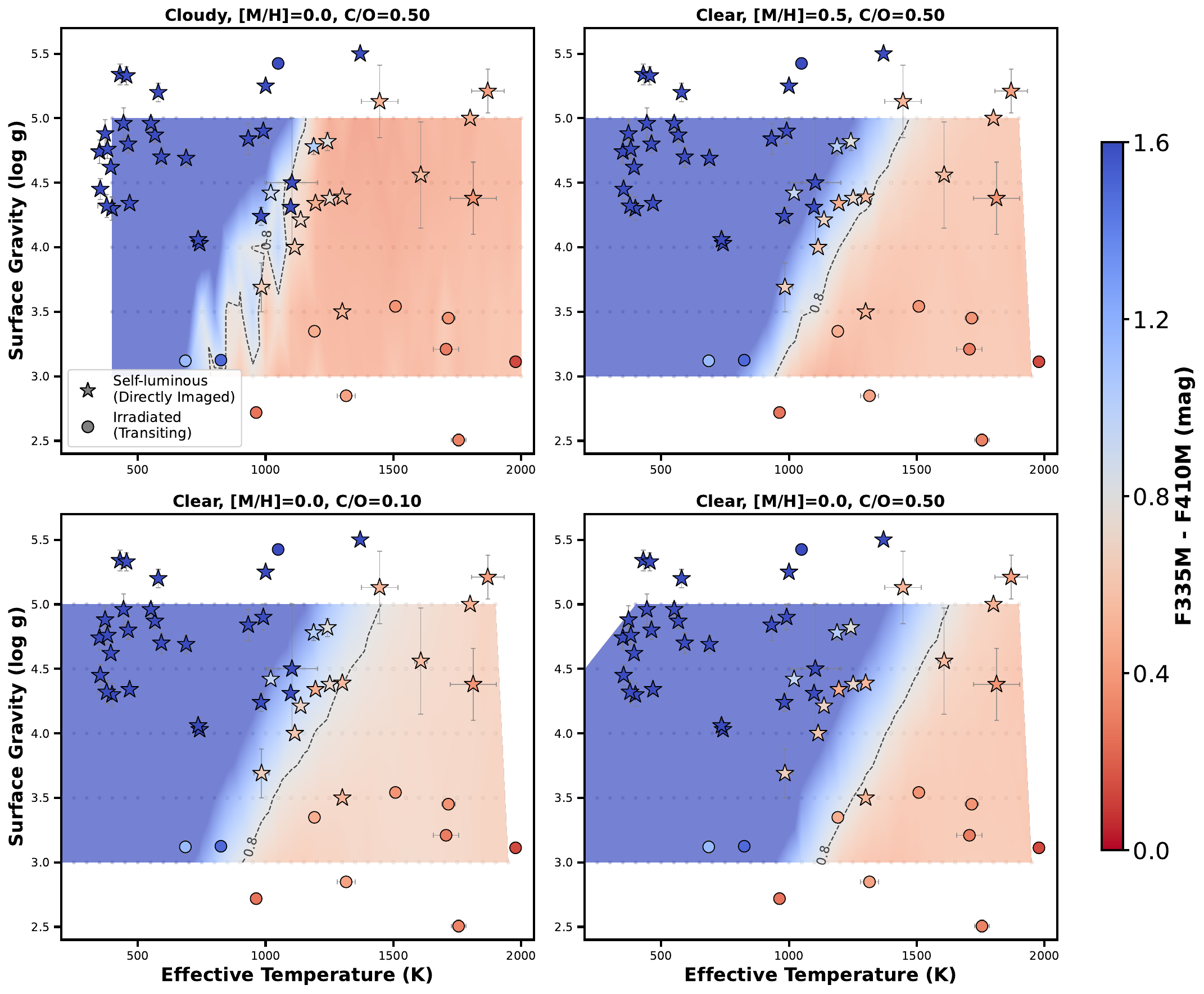}
    \caption{Comparison of the F335M - F410M color (CH$_4$ absorption) between the observed sample and Exo-REM model grids \citep{charnaySelfconsistentCloudModel2018} on the $T_{eff}$-$\log g$ plane. The four panels display model predictions for different atmospheric scenarios: cloudy (top left), enhanced metallicity ([M/H]=0.5, top right), low C/O ratio (C/O=0.1, bottom left), and a clear, solar-composition baseline ([M/H]=0.0, C/O=0.5, bottom right). The transiting exoplanets (circles) and substellar objects (stars) are overplotted, including the Late-T and Y dwarf sample from \citet{beilerPreciseBolometricLuminosities2024}.}
\label{exorem_2x2}
\end{figure*}

\subsubsection{Possible physical mechanisms}

This gravity-dependent methane onset arises from the interplay of several atmospheric processes:

\textit{Pressure-dependent chemical equilibrium:} The CO/CH$_4$ chemical transition is strongly pressure-dependent, with CO favored at lower pressures according to the net thermochemical reaction CO + 3H$_2$ $\rightleftharpoons$ CH$_4$ + H$_2$O \citep{visscherDeepWaterAbundance2010, zahnleMETHANECARBONMONOXIDE2014}. In low-gravity atmospheres, the photospheric pressure is reduced at a given temperature, shifting the equilibrium chemistry toward CO and delaying the appearance of CH$_4$ to cooler temperatures where the equilibrium eventually favors methane even at low pressures.

\textit{Cloud-induced greenhouse warming:} Higher and optically thicker cloud decks in low-gravity atmospheres produce a stronger greenhouse effect, trapping thermal radiation and heating the atmosphere below the cloud base \citep{charnaySelfconsistentCloudModel2018, morleySonoraSubstellarAtmosphere2024}. This elevated deep atmospheric temperature further suppresses CH$_4$ formation by pushing the local chemistry toward CO across a larger pressure range.

\textit{Vertical mixing and chemical quenching:} Enhanced vertical mixing in low-gravity atmospheres (parameterized by higher $K_{zz}$ values) can quench the CO-to-CH$_4$ conversion by transporting gas parcels upward faster than the chemical reaction timescale \citep{visscherQUENCHINGCARBONMONOXIDE2011}. This disequilibrium chemistry freezes in the CO-dominated composition from deeper, hotter layers, preventing methane formation even when thermochemical equilibrium would favor CH$_4$ at the photosphere \citep{mukherjeeProbingExtentVertical2022}.

\begin{figure*}[t]
    \centering
    \includegraphics[width=0.8\textwidth]{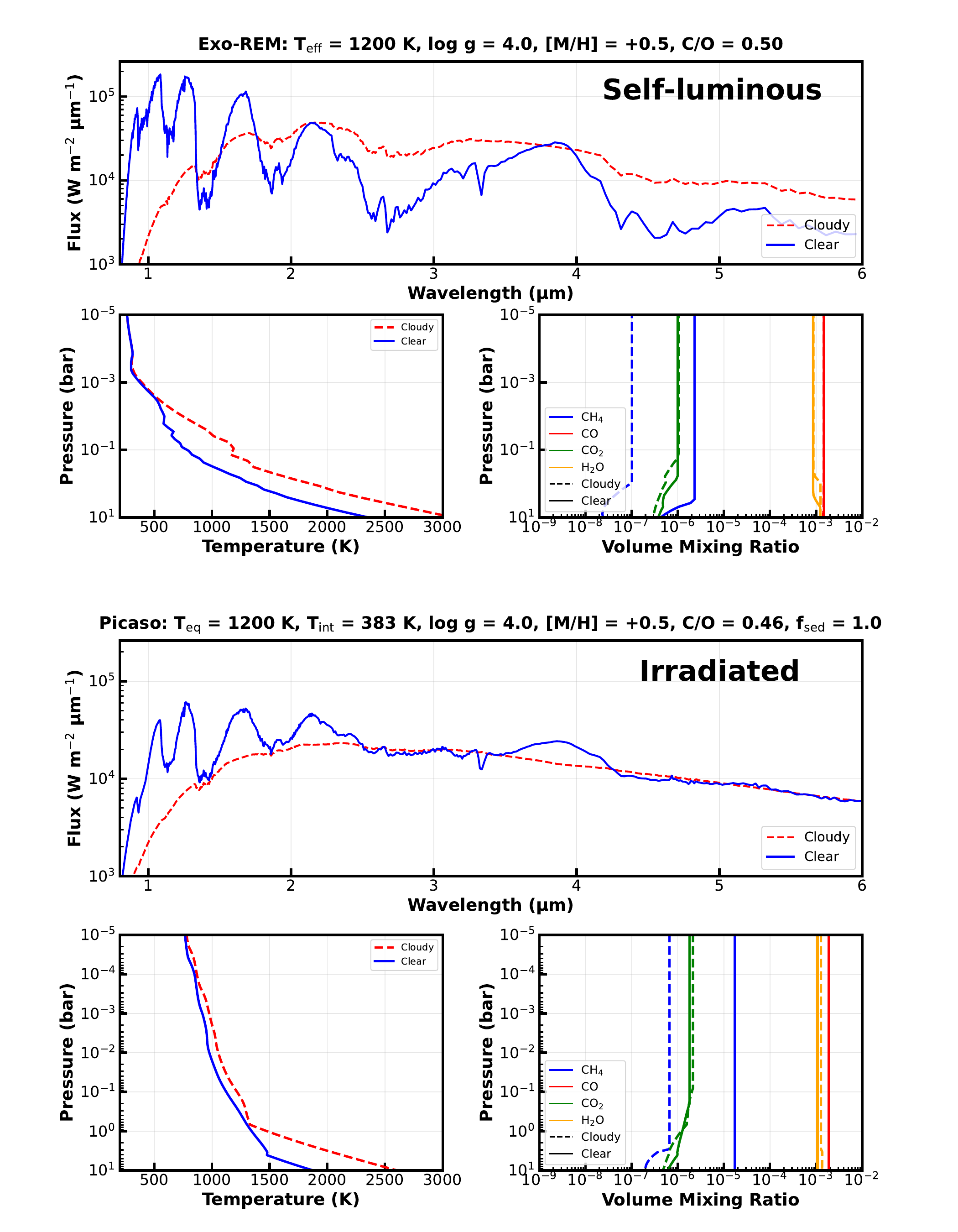}
    \caption{Comparison of Exo-REM and PICASO cloudy/clear predictions for a representative $T_{\rm eff}=1200$~K, $\log g=4.0$, [M/H]$=+0.5$ atmosphere. Dashed curves show cloudy models and solid curves show clear baselines. Cloud greenhouse back-warming heats the deep atmosphere ($P \gtrsim 0.1$~bar; left subpanels), shifting the chemistry toward CO and reducing the photospheric CH$_4$ abundance (right subpanels), which weakens the 3.3~$\mu$m CH$_4$ band in the emergent spectra (main panels). This effect can occur in both self-luminous and irradiated objects. For clear atmospheres, irradiated objects have more isothermal temperature--pressure profiles, leading to shallower spectral features. For cloudy atmospheres, the self-luminous and irradiated cases appear more similar as cloud opacity raises the photosphere.}
    \label{exorem_picaso_comparison}
\end{figure*}

\begin{figure*}[t]
    \centering
    \includegraphics[width=0.8\textwidth]{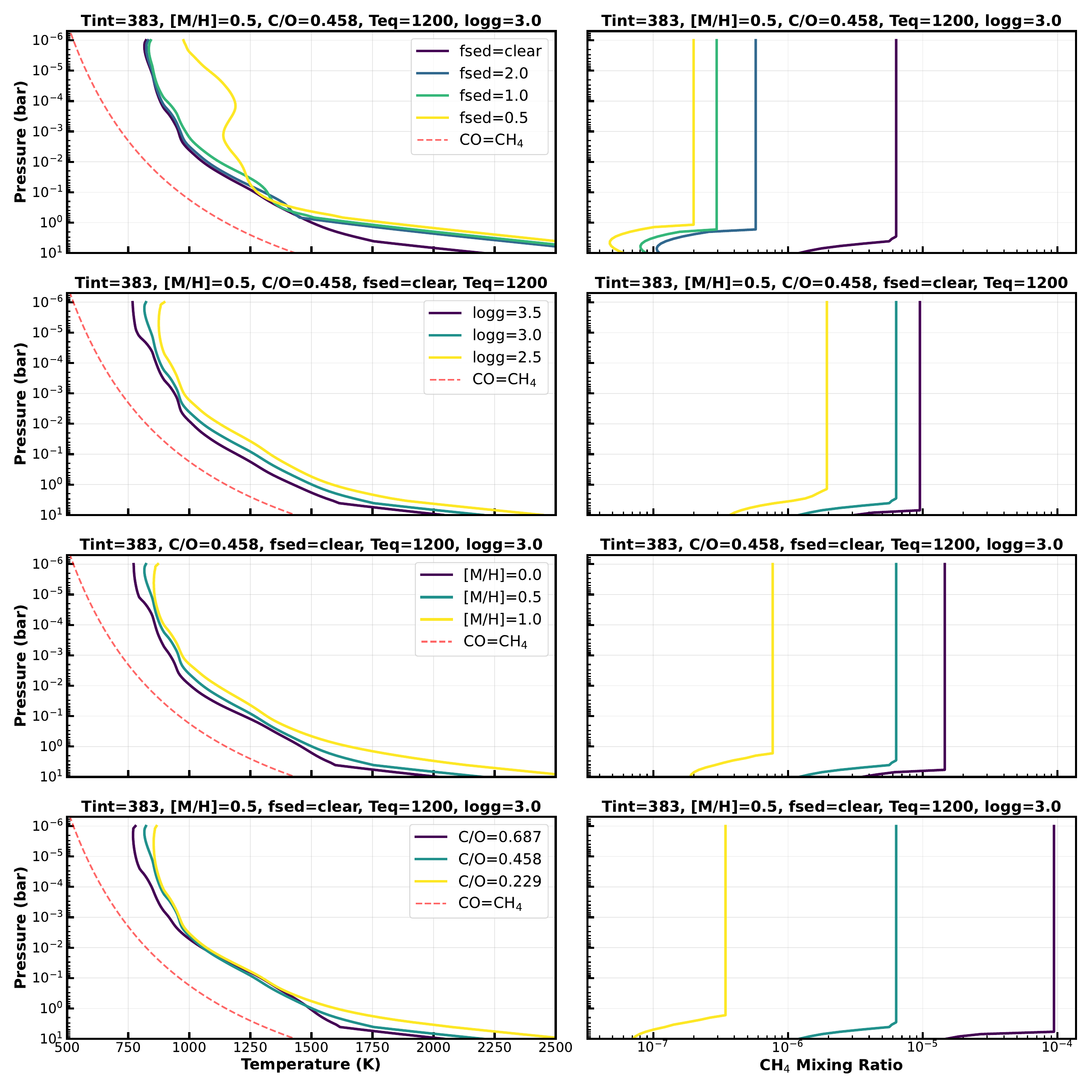}
    \caption{Thermal structure and CH$_4$ abundance for PICASO irradiated models at 1200~K, varying $f_{\mathrm{sed}}$, $\log g$, [M/H], and C/O. The dashed red line marks the CO$=$CH$_4$ VMR boundary at solar metallicity from \citep{fortneyEquilibriumTemperatureHow2020}. Radiative feedback from clouds can heat deeper atmospheric layers and lead to quenching at lower CH$_4$ VMR. Similarly, lower $\log g$, higher metallicity, and lower C/O can all reduce methane, delaying the onset of CH$_4$ to cooler temperatures. This highlights the fundamental degeneracies among different mechanisms that can explain the delayed appearance of CH$_4$ in warm H$_2$-dominated atmospheres.}
    \label{TP_profiles_CH4_VMR}
\end{figure*}

\subsubsection{Exo-REM (Self-luminous) and PICASO (Irradiated)}

The CH$_4$ onset line extends across both self-luminous and irradiated objects across 3 dex in surface gravity, indicating potential common underlying atmospheric mechanisms across both populations. Therefore, we use with both self-luminous Exo-REM \citep{charnaySelfconsistentCloudModel2018} and irradiated PICASO (This work) grids to explore the CH$_4$ onset. 

Figure \ref{exorem_2x2} compares the observed F335M$-$F410M colors with predictions from the Exo-REM atmospheric model grid \citep{charnaySelfconsistentCloudModel2018}, which self-consistently couples radiative transfer, convection, cloud formation, and equilibrium chemistry across a range of $T_{eff}$, $\log g$, metallicity, and C/O ratios. For clear, solar-composition atmospheres (bottom right panel), the models predict methane onset at temperatures $\sim$200-400 K hotter than observed in our low-gravity sample. Including clouds (top left panel) delays the predicted methane onset by heating the deep atmosphere through the greenhouse effect. Increasing metallicity (top right panel) shifts the CO=CH$_4$ transition toward colder temperatures. Decreasing the C/O ratio (bottom left panel) reduces the available carbon for CH$_4$ formation while maintaining oxygen for H$_2$O, also suppressing methane. 

We observe similar effects in our irradiated PICASO models (Figure \ref{exorem_picaso_comparison}, \ref{TP_profiles_CH4_VMR}). Radiatively active clouds increase the thermal infrared opacity for the outgoing planetary flux, driving up atmospheric temperatures below the clouds. This pushes the deep atmosphere to be hotter and more CO-dominated at the pressure the CH$_4$ is quenched, reducing the CH$_4$ feature observed in the emission spectrum. In figure \ref{TP_profiles_CH4_VMR}, varying f$_{sed}$, logg, [M/H] and C/O can all achieve similar effects of reducing CH$_4$ abundance similar to Exo-REM models. 

The multiple knobs that can produce similar levels of CH$_4$ depletion are a fundamental cause of modeling degeneracies when interpreting atmospheres in this CH$_4$ transition parameter space. While self-luminous objects typically have higher-S/N spectra with more precisely measured molecular abundances, their temperatures and surface gravities are often ambiguous. Irradiated transiting objects have independently constrained irradiation temperatures and surface gravities, but they have lower-S/N spectra, poorer precision on abundances, and limited information on the deep intrinsic temperature.

\begin{figure}[t]
    \centering
    \includegraphics[width=0.45\textwidth]{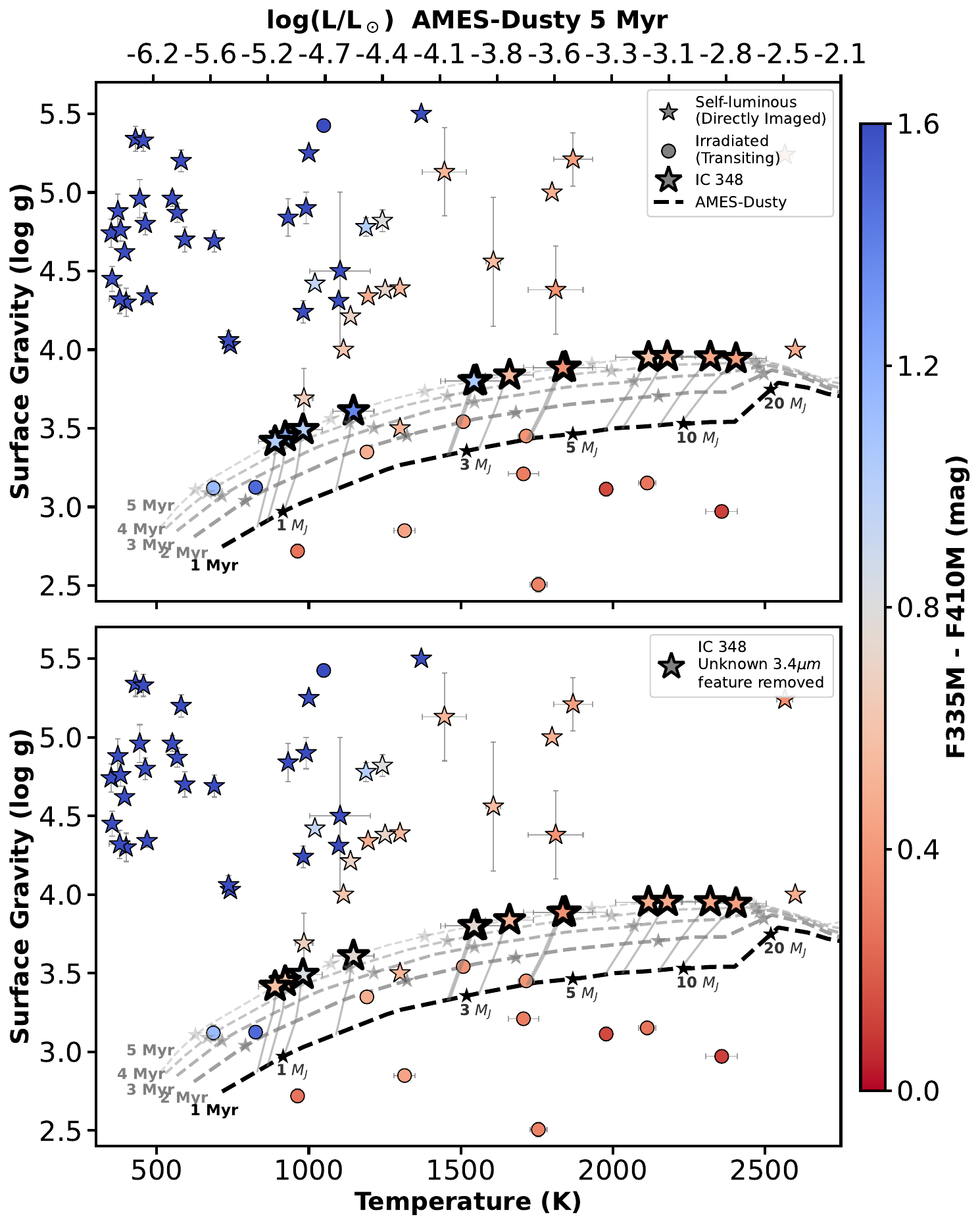}
    \caption{The objects from IC 348 star-forming region \citep{luhmanNewSpectralClass2025} are plotted alongside transiting exoplanets and other substellar objects. Given the unknown mass and radius of the IC 348 objects, we use AMES-Dusty evolutionary models to convert their measured bolometric luminosities to effective temperatures and surface gravities. We display isochrons from 1 to 5 Myr, representing the likely age range of the region. The points are color-coded by the F335M - F410M color, indicating methane absorption strength. The top panel uses the colors measured directly from the spectra, while the bottom panel uses the colors recomputed after masking the unknown 3.4 $\mu$m hydrocarbon feature; after masking, the IC 348 objects show a lack of CH$_4$ absorption, consistent with the delayed onset of methane seen in transiting hot Jupiters.}
\label{IC348_isochrones}
\end{figure}

\subsection{IC 348 star forming region objects}

The recently published JWST spectra of objects in the IC 348 star-forming region \citep{luhmanNewSpectralClass2025} provide a unique opportunity to study extremely young ($\sim$5 Myr), low-mass substellar objects with effective temperatures overlapping those of transiting exoplanets and brown dwarfs. Thirteen of these fifteen objects have usable flux-calibrated 1--5~$\mu$m spectra for synthetic photometry and are included in our sample; they span bolometric luminosities from $\log(L/L_\odot) \approx -2.8$ to $-5.0$, corresponding to inferred masses of $\sim$1-20 $M_J$ based on AMES-Dusty evolutionary models \citep{chabrierEvolutionaryModelsVery2000,allardLimitingEffectsDust2001}. At such young ages, these objects occupy the low-gravity regime ($\log g \sim 3-4$) similar to transiting hot Jupiters, making them valuable comparison targets.

Figure \ref{IC348_isochrones} places the IC 348 objects on the $T_{eff}$-$\log g$ plane alongside our transiting exoplanet and substellar object sample. Since the IC 348 objects lack direct mass and radius measurements, we convert their measured bolometric luminosities to effective temperatures and surface gravities using AMES-Dusty evolutionary isochrones, adopting the IAU 2015 nominal solar luminosity $L_\odot = 3.828\times10^{26}$ W \citep{prsaNominalValuesSelected2016}. We display isochrones spanning 1-5 Myr to represent the likely age range of the region, with lines connecting each object's position across the different age tracks to illustrate the systematic uncertainty in their derived atmospheric parameters. The youngest isochrones (1 Myr) predict higher temperatures and lower gravities for a given luminosity, while older isochrones (5 Myr) yield cooler temperatures and higher gravities. We adopt the 5 Myr track as our fiducial model for plotting positions.

The IC 348 objects are color-coded by their F335M$-$F410M colors, indicating methane absorption strength. However, \citet{luhmanNewSpectralClass2025} identified an unknown hydrocarbon absorption feature at 3.4 $\mu$m in these spectra that contaminates the F335M bandpass. This feature, whose origin remains unknown, can artificially enhances the apparent F335M$-$F410M color and could be mistaken for CH$_4$ absorption (Figure \ref{spectra}). To avoid potential bias from this feature, we recomputed the F335M synthetic photometry after masking the 3.2--3.6 $\mu$m wavelength region from the spectra. The masked region was replaced with flux values obtained by linear interpolation between the adjacent continuum regions at 3.0--3.2 $\mu$m and 3.6--3.8 $\mu$m. This approach effectively removes the unknown feature's contribution while preserving the underlying continuum shape. The bottom panel of Figure \ref{IC348_isochrones} shows the IC 348 objects with these corrected colors. After removing the unknown feature's contribution, the IC 348 objects exhibit a consistent lack of CH$_4$ absorption across the full temperature range sampled. This absence of methane in objects as cool as $T_{eff} \sim 890$ K is consistent with the gravity-dependent delayed methane onset observed in transiting hot Jupiters and young directly imaged companions, supporting the high entropy initial formation assumptions used in evolutionary models for these young objects. Given the overlapping opacity between the unknown hydrocarbon and the CH$_4$ feature, any potential CH$_4$ contribution would be heavily contaminated and challenging to isolate. Additional spectroscopic observations of low-luminosity star-forming regions will be highly informative on whether the hydrocarbon feature is common and if the delayed methane onset is universal in low surface gravity objects.

\subsection{Irradiation is not the first-order driver: the ZTF J0038+2030 B test}
\label{sec:ztf_control}
\label{sec:limitations}

A fundamental caveat in comparing the two populations is that transiting-planet eclipse spectra measure dayside emission from a single, externally irradiated hemisphere, whereas self-luminous spectra are rotation-averaged emission powered by internal heat. Day--night temperature contrasts, longitudinally non-uniform clouds, and stellar-driven photochemistry can all make the dayside unrepresentative of a global average, and the bias grows with incident stellar flux \citep{komacekAtmosphericCirculationHot2017}. While all of these processes operate at some level, the question is how they rank: are they first-order drivers, or second-order perturbations on sequences organized primarily by temperature and surface gravity?

The data contain two complementary control populations that address this. The first is the young, low-gravity self-luminous objects: they share the low gravities of the transiting planets but none of the irradiation-driven physics (internally heated, free of tidally locked day--night forcing, and viewed at random inclinations rather than edge-on), yet they track the transiting planets across color space and show a similar delayed L/T transition \citep{manjavacasMediumresolution09753Mm2024}. If viewing geometry, heat redistribution, or stellar-driven photochemistry were first-order drivers, these two differently forced low-gravity populations would separate; they do not. Several of the distinguishing observables are themselves downstream of gravity, which sets cloud sedimentation, vertical mixing and photosphere pressure levels.

The second control, from the opposite corner of parameter space, is the irradiated brown dwarf ZTF J0038+2030 B (Section \ref{sec:ztf0038}). It decouples irradiation from gravity: its dayside is irradiated and tidally locked like a hot Jupiter dayside, but its surface gravity ($\log g = 5.43$) is that of an old field brown dwarf.

\begin{figure}[t]
    \centering
    \includegraphics[width=0.48\textwidth]{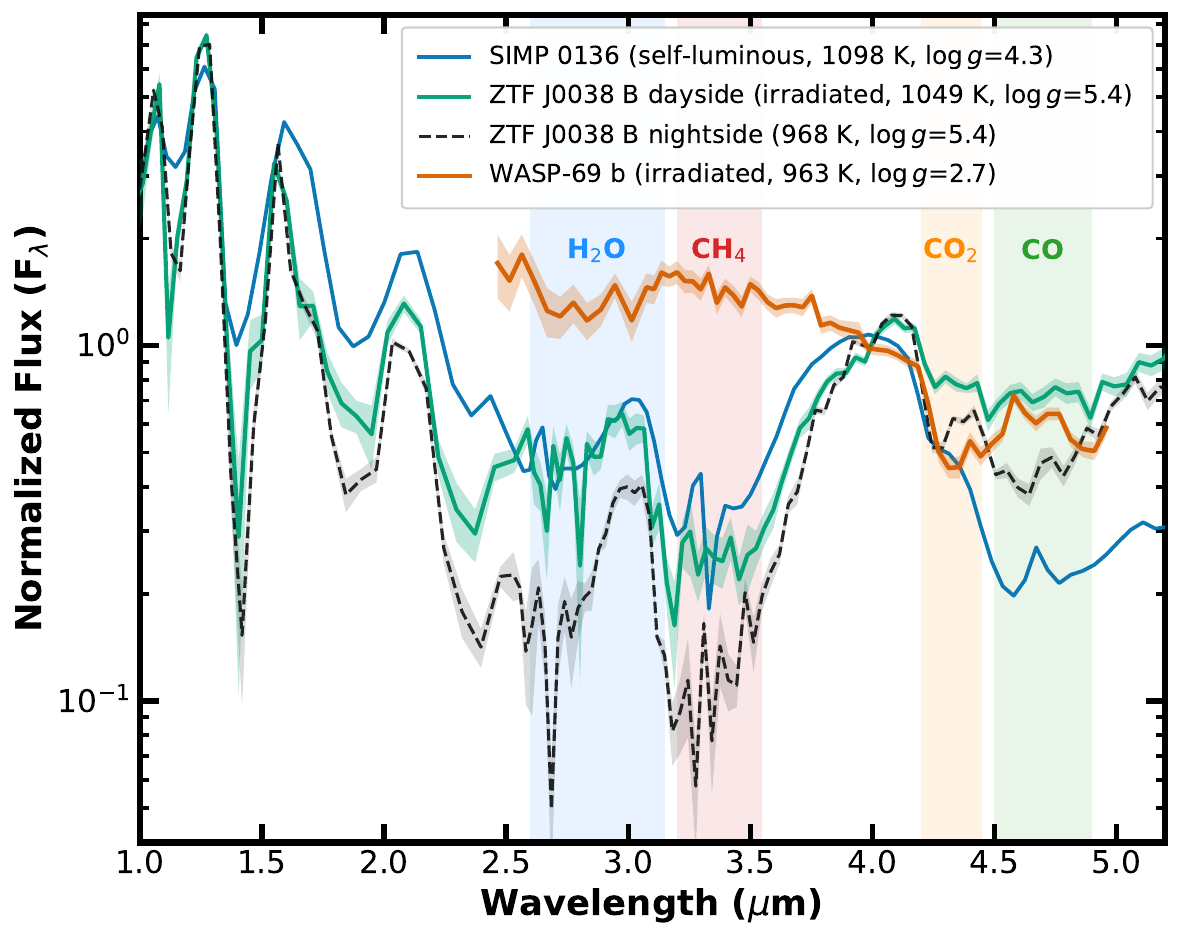}
    \caption{A same-temperature test of whether irradiation drives the spectral differences between transiting planets and isolated brown dwarfs, in the normalized F$_\lambda$ units of Figure \ref{spectra}: the self-luminous T2 dwarf SIMP 0136+0933 (1098 K, $\log g$ = 4.31, blue), the irradiated brown dwarf ZTF J0038+2030 B dayside (1049 K, $\log g$ = 5.43, green; Section \ref{sec:ztf0038}) and its effectively un-irradiated nightside (968 K, dashed black), and the transiting planet WASP-69 b dayside (963 K, $\log g$ = 2.72, orange). The two PRISM spectra (SIMP 0136+0933 and ZTF J0038+2030 B) are binned to $R = 100$ ($R = 30$ blueward of 2.5 $\mu$m, where the deep troughs approach zero flux); WASP-69 b is shown on its published grid. Shaded bands are 1$\sigma$. The two irradiated daysides do not resemble each other: ZTF J0038+2030 B tracks the self-luminous T dwarf band for band and is nearly identical to its own nightside, while the low-gravity planet is featureless. This comparison shows surface gravity rather irradiation is the first-order parameter shaping the observed methane onset.}
    \label{ztf_spectra}
\end{figure}

Figure \ref{ztf_spectra} shows the resulting controlled comparison among three objects at nearly the same temperature (960--1100 K). The self-luminous T2 dwarf SIMP 0136+0933 ($T_{\rm eff} = 1098$ K, $\log g = 4.31$) is not irradiated; WASP-69 b ($T_{\rm eq} = 963$ K, $\log g = 2.72$) and ZTF J0038+2030 B ($T_{\rm day} = 1049$ K, $\log g = 5.43$) are both irradiated, tidally locked daysides observed with the same eclipse-based technique. If the irradiation-specific physics (external heating, day--night circulation, stellar-driven photochemistry) were the first-order cause of the spectral differences between transiting planets and isolated brown dwarfs, the two irradiated daysides would resemble each other. They do not. The dayside of ZTF J0038+2030 B \citep{broski-laingAsymmetricNightsideCO22026} instead tracks the self-luminous T dwarf band for band, with deep H$_2$O absorption and a deep 3.3 $\mu$m CH$_4$ band, while WASP-69 b at the same temperature is nearly featureless. The object also carries its own internal control: its nightside, which faces permanently away from the white dwarf and is effectively un-irradiated, is nearly identical to the dayside apart from slightly deeper bands from its cooler temperature, so switching the irradiation on and off on the same high-gravity atmosphere leaves the spectral morphology largely unchanged. Quantitatively, the methane colors are F335M$-$F410M $= 0.27 \pm 0.03$ mag for WASP-69 b, $2.21 \pm 0.02$ mag for the ZTF J0038+2030 B dayside, and $1.84$ mag for SIMP 0136+0933: moving $\sim$2.7 dex in gravity at fixed irradiation state (WASP-69 b to ZTF J0038+2030 B) changes the methane color by $\sim$2 mag, whereas switching irradiation on at high gravity (SIMP 0136+0933 to ZTF J0038+2030 B) has limited impact on the deep, T-dwarf-like methane band.

The same conclusion follows from the empirical methane-onset separator (Figure \ref{Teff_logg}). At 1049 K the boundary sits at $\log g \approx 3.9$, so a high-gravity object is expected to be strongly methane-bearing regardless of its irradiation. ZTF J0038+2030 B is consistent with this: at $\log g = 5.43$ its measured color of 2.21 mag, far above the 0.8 mag threshold, places it $\sim$1.5 dex inside the methane region, and including it in the separator fit leaves the boundary essentially unchanged. On the color-magnitude diagrams (Figures \ref{J_K_CMD} and \ref{band_color_data_only}), ZTF J0038+2030 B likewise falls on the field T-dwarf locus rather than with the transiting planets.

There are two caveats. First, irradiation is a smaller fractional contribution to the dayside energy budget of ZTF J0038+2030 B than for a hot Jupiter, and we can quantify it: the object is strongly self-luminous. Its equilibrium temperature is $T_{\rm eq} = 745$ K, while the internal luminosity from the energy-balance analysis of \citet{broski-laingAsymmetricNightsideCO22026} corresponds to an internal effective temperature of $\approx$967 K, so the interior heat flux exceeds the irradiation. Comparing the absorbed irradiation with the dayside emission, $(T_{\rm eq}/T_{\rm day})^4$, irradiation supplies $\sim$25$\%$ of the dayside emitted power (lifting the dayside 81 K above the nightside, whose emission is $>$99$\%$ internal), compared with $\gtrsim$99$\%$ of the dayside power for WASP-69 b, whose internal temperature is $\lesssim$200 K. The dayside is nevertheless measurably irradiated, with inefficient day-to-night heat transport and a nightside-asymmetric CO$_2$ feature demonstrating that irradiation-driven physics is active \citep{broski-laingAsymmetricNightsideCO22026}. Second, this is only one irradiated high surface gravity object supporting our proposed hierarchy of surface gravity as the fundamental upstream parameter. Additional irradiated high-gravity brown dwarfs are needed to populate this currently sparse corner of the temperature--gravity plane and test whether this hierarchy holds generally.

\begin{figure*}[t]
    \centering
    \includegraphics[width=\textwidth]{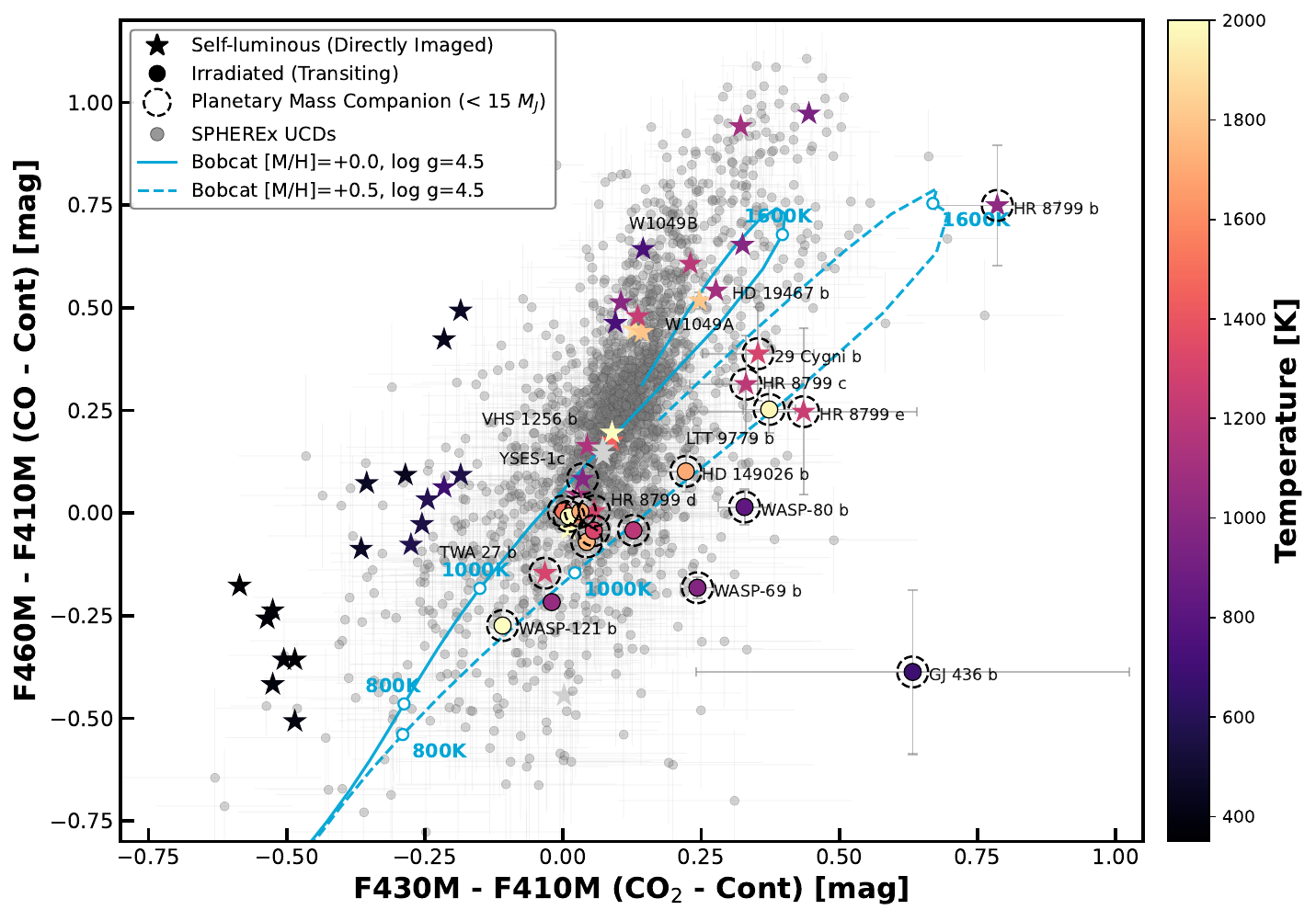}
    \caption{Sonora Bobcat predictions for the CO$_2$--CO color--color diagnostic, illustrating the effect of metallicity. The x-axis is the CO$_2$ absorption strength (F430M$-$F410M) and the y-axis is the CO absorption strength (F460M$-$F410M), where larger values indicate deeper molecular bands relative to the F410M continuum. Colored tracks show solar and 3$\times$ solar metallicities across the model temperature sequence. Higher metallicity preferentially strengthens CO$_2$ relative to CO, shifting models toward larger CO$_2$-continuum values. Points are color-coded by temperature, and grey stars mark the IC 348 objects, which are plotted without a temperature because they are free-floating and mostly carry the `H' spectral class of \citet{luhmanNewSpectralClass2025}. Dashed circles mark planetary-mass companions ($<15\,M_J$), and the small grey points are the quality-filtered SPHEREx ultracool dwarfs \citep{gagneSPHERExPipelineSpectral2026}. Planetary-mass companions appear to show stronger CO$_2$ features than isolated brown dwarfs, indicating possible metallicity enhancement in these objects.}
    \label{CO2_CO_metallicity}
\end{figure*}

\subsection{Metallicity enhancement in planets compared to isolated brown dwarfs}

The CO$_2$-continuum versus CO-continuum color-color diagram (Figure \ref{CO2_CO_color_color}) reveals a systematic offset between planetary objects and field brown dwarfs \citep{balmerJWSTTSTHighContrast2025}. The $\sim$2100 SPHEREx field ultracool dwarfs now define this field locus empirically as a tight diagonal sequence, against which the offset of the planetary population is measured rather than inferred from model tracks alone. Transiting planets exhibit stronger CO$_2$ absorption relative to CO compared to brown dwarfs at similar temperatures, occupying a distinct region of color-color space. Notably, directly imaged planetary-mass companions including HR 8799 b, c, d, e and 29 Cygni b also display enhanced CO$_2$ features, clustering near the transiting planet population rather than with field brown dwarfs \citep{balmerJWSTTSTHighContrast2025, xuanCompositionsHR87992026}; on the transiting side, CO$_2$ is likewise detected directly in dayside emission (e.g., WASP-69 b; \citealt{schlawinMultipleCluesDayside2024}). This separation provides a diagnostic for atmospheric composition differences between objects that formed in circumstellar disks versus those that formed via gravitational collapse like stars.

The enhanced CO$_2$/CO ratio in both transiting and directly imaged planets can be explained by elevated atmospheric metallicity. CO$_2$ abundance scales quadratically with metallicity ([CO$_2$] $\propto$ Z$^2$) because its formation requires both carbon and oxygen, while CO abundance scales linearly with metallicity \citep{loddersAtmosphericChemistryGiant2002}. Consequently, atmospheres with super-solar metallicities exhibit disproportionately enhanced CO$_2$ features relative to CO. Figure \ref{CO2_CO_metallicity} illustrates this with Sonora Bobcat tracks at solar and 3$\times$ solar metallicity, which shift toward larger CO$_2$-continuum values at fixed CO, in the same sense as the observed offset of the planetary-mass objects. This metallicity enhancement is consistent with predictions from core accretion formation, where heavy element enrichment occurs through planetesimal and pebble accretion during formation in the protoplanetary disk \citep{ikomaFormationGiantPlanets2025}. Metallicity enhancements of 3-10$\times$ solar have also been inferred from previous atmospheric retrieval studies of individual hot Jupiters \citep{fuHydrogenSulfideMetalenriched2024, wiserPreciseMetallicityCarbontoOxygen2025}, HR 8799 system \citep{xuanCompositionsHR87992026} and population-level meta-analysis \citep{lothringer_library_2025}. The fact that both close-in ($<$0.1 AU) transiting planets and wide-separation ($>$10 AU) directly imaged companions show similar CO$_2$ enhancement suggests that disk-based formation could result in metal-enriched atmospheres across orders of magnitude differences in orbital separation. 

For directly imaged objects, surface gravity introduces a degeneracy in interpreting the CO$_2$-CO color-color diagram. Lower surface gravity atmospheres have reduced photospheric pressures. The reduced pressure broadening can alter the apparent absorption depths in the F430M and F460M filters differently. Model comparisons (Figure \ref{CO2_CO_color_color}) show that decreasing $\log g$ from 5 to 3 produces a shift in CO$_2$-CO color space that partially overlaps with the effect of increasing metallicity. 

Breaking this gravity-metallicity degeneracy requires independent constraints on surface gravity (available for transiting planets through mass and radius measurements). The systematic offset of planets toward stronger relative CO$_2$ absorption supports the interpretation of enhanced metallicities in the planetary population compared to field brown dwarfs.

\subsection{Brown dwarf, planet-mass object, transiting planets synergies}

To first order, objects from $\sim$2~Earth-radius sub-Neptunes to main-sequence stars have H/He-dominated atmospheres, with temperature as the primary factor shaping their observed properties. Indeed, extremely irradiated planets like KELT-9~b have atmospheres remarkably similar to stars \citep{lothringerExtremelyIrradiatedHot2018, manjavacasCloudAtlasHubble2019}; evidence for silicate clouds is seen in both L dwarfs \citep{suarezUltracoolDwarfsObserved2022} and hot Jupiters \citep{fuStatisticalAnalysisHubble2017, gaoAerosolCompositionHot2020}; and CH$_4$ opacity dominates in both cool T dwarfs \citep{beilerPreciseBolometricLuminosities2024} and temperate sub-Neptunes \citep{bennekeJWSTRevealsCH$_4$2024}.

In the JWST era, transiting planets are beginning to overlap in temperature with self-luminous objects, enabling comparative studies that probe second-order effects beyond temperature, such as stellar irradiation, photochemistry, surface gravity, tidally locked day--night circulation, and imprints of different formation channels.

Directly imaged objects often have high-SNR spectra thanks to reduced photon noise, enabling unambiguous detection of molecular \citep{burgasserObservationUndepletedPhosphine2025} and cloud species \citep{milesJWSTEarlyreleaseScience2023} in their atmospheres. These spectra can serve as benchmarks for searching and interpreting lower-SNR spectra of transiting-planet counterparts.

Transiting planets, on the other hand, suffer from lower-SNR spectra due to stellar photon noise and often struggle to robustly identify spectral features, especially at low spectral resolution and with opacity confusion \citep{welbanksChallengesDetectingGases2025}. However, radial-velocity and transit measurements provide independently measured masses and radii, which directly imaged objects typically lack. This allows transiting planets to map atmospheric transitions as a function of surface gravity empirically. In particular, transiting planets provide unique access to the low surface-gravity regime ($\log g \sim 2$--3.5), which overlaps with very young ($\sim$few Myr) planetary-mass objects. If the onset of methane directly traces temperature and surface gravity, then methane features could provide a diagnostic tool for inferring the properties of young, directly imaged planetary-mass objects.

Metallicity is a direct tracer of planet-formation channels: core accretion predicts enhanced metal enrichment compared to gravitational collapse. As a sensitive metallicity tracer, comparative studies of CO$_2$ features can help distinguish populations with different formation mechanisms, especially across a range of masses and current orbital separations. Transiting planets are typically close-in ($<0.1$~AU) and lower mass ($<1\,M_{\mathrm{J}}$), whereas directly imaged objects are farther out (a few to tens of AU) and more massive (a few to tens of $M_{\mathrm{J}}$). Combining these populations expands the parameter space by 1--2 dex in both mass and orbital separation, offering more leverage to test planet-formation models.

\section{Summary}

We present a unified color--magnitude analysis for transiting exoplanets and directly imaged substellar objects with JWST. We compile spectra for 13 transiting giant planets, 57 self-luminous objects, and the irradiated brown dwarf ZTF J0038+2030 B spanning $\sim$350--2600~K and a wide range of surface gravities, then convert them to synthetic photometry in 2MASS $J/K_s$ and JWST NIRCam medium-band filters that isolate H$_2$O, CH$_4$, CO$_2$, and CO. The NIR CMDs show that transiting planets follow the L-dwarf sequence in $J$ versus $J\!-\!K$ but lack a clear T-dwarf blueward turn. The NIRCam CMD also shows delayed CH$_4$ onset in transiting planets. Together they suggest a delayed L/T transition in low-gravity, irradiated atmospheres similar to young self-luminous objects. Transiting planets also share the observed shallow molecular features with young, low-gravity objects, pointing to surface gravity as a shared driver for the observed colors similarities.

Methane absorption is systematically delayed in transiting planets relative to self-luminous objects, consistent with cloud greenhouse back-warming and CO-favored chemistry at low gravity. The irradiated but high-gravity brown dwarf ZTF J0038+2030 B decouples irradiation from gravity: its dayside shows a deep, field-T-dwarf-like methane band at a temperature where equally irradiated low-gravity planets show none, landing $\sim$1.5 dex inside the methane region predicted by our pre-existing temperature--gravity separator and demonstrating that irradiation alone does not reorganize the sequence. Comparisons to multiple self-consistent model grids indicate that clouds can explain the suppressed CH$_4$ in both low-gravity self-luminous and irradiated objects, though metallicity and C/O remain degenerate factors. CO$_2$-to-CO diagnostics show stronger relative CO$_2$ absorption in transiting and directly imaged planetary-mass companions than in field brown dwarfs, consistent with enhanced metallicities expected from formation in disks. Overall, temperature and surface gravity emerge as the dominant parameters shaping H/He-dominated atmospheres, with low gravity linked to cloudy spectra, delayed L/T transitions and methane onset.

Future work should test this picture with models that treat clouds, chemistry, and thermal structure self-consistently across the wide gravity--temperature space, and with observations that fill remaining gaps (e.g., L/T dwarfs with dynamical masses spanning $\log g \sim 5.5$ to 3.5, warm transiting planets, and brown dwarfs below 1000~K). If the methane-onset boundary remains sharp, temperature and gravity likely dominate; if it blurs, additional mechanisms must contribute. As Henry Norris Russell stated in his original work on stellar classification \citep{russell_relations_1914}, variations in a single physical condition lead to a series, but two or more independent variable conditions of comparable degree result in scattering. 

\begin{acknowledgments}
Some of the data presented in this article were obtained from the Mikulski
Archive for Space Telescopes (MAST) at the Space Telescope Science Institute.
The specific observations analyzed can be accessed via
\dataset[doi:10.17909/tn1m-p398]{https://doi.org/10.17909/tn1m-p398}.
This work has benefited from The UltracoolSheet, maintained by Will Best,
Trent Dupuy, Michael Liu, Aniket Sanghi, Rob Siverd, and Zhoujian Zhang.
The code and data needed to reproduce every figure in this paper are archived at
Zenodo:
\dataset[doi:10.5281/zenodo.22101390]{https://doi.org/10.5281/zenodo.22101390}.
An interactive browser for the spectra and color--magnitude diagrams presented
here is available at \url{https://spectra.guangweifu.com}.
\end{acknowledgments}

\begin{figure*}[t]
    \centering
    \includegraphics[width=0.8\textwidth]{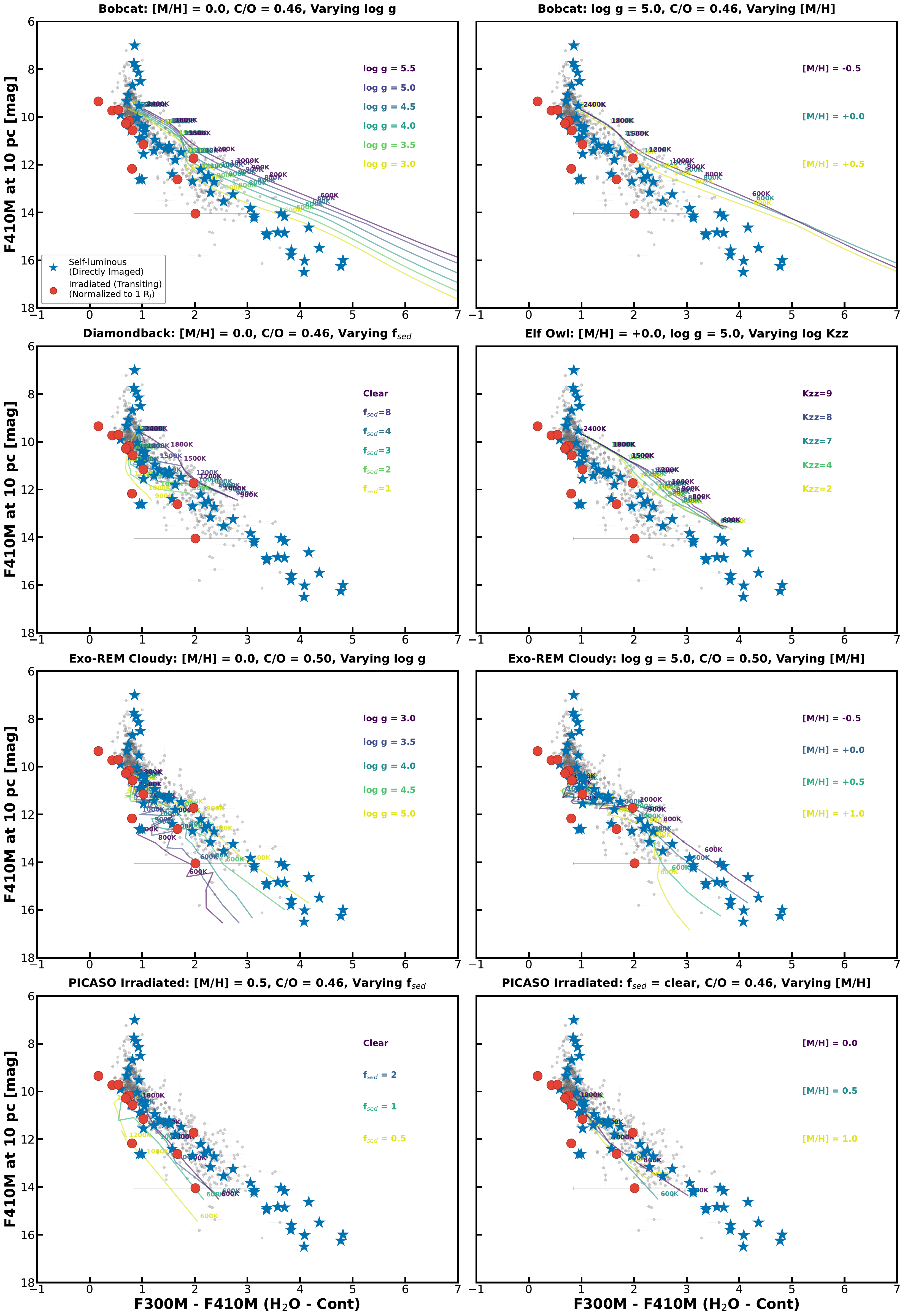}
    \caption{H$_2$O color-magnitude sequence. The observed water absorption strength is compared with multiple atmospheric model grids. The panels show predictions from Sonora Bobcat \citep{marleySonoraBrownDwarf2021}, Diamondback \citep{morleySonoraSubstellarAtmosphere2024}, Elf Owl \citep{mukherjeeSonoraSubstellarAtmosphere2024}, Exo-REM \citep{charnaySelfconsistentCloudModel2018}, and PICASO irradiated models (this work), illustrating the effects of gravity, metallicity, clouds, and vertical mixing on the H$_2$O feature. Small grey points show the quality-filtered SPHEREx ultracool dwarf sample \citep{gagneSPHERExPipelineSpectral2026}.}
\label{H2O_color_mag}   
\end{figure*}

\begin{figure*}[t]
    \centering
    \includegraphics[width=0.8\textwidth]{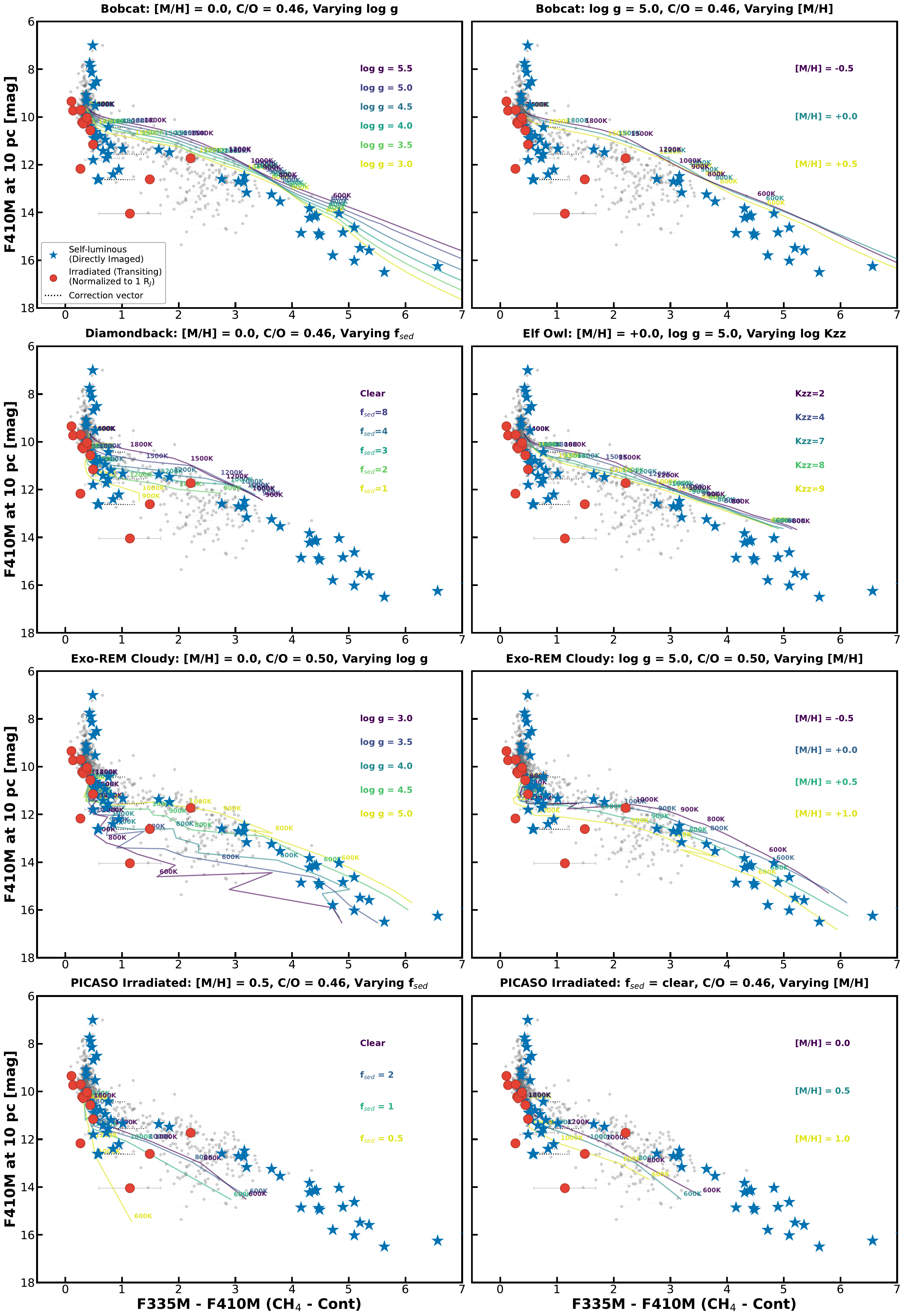}
    \caption{CH$_4$ color-magnitude sequence. The onset of methane absorption is compared with multiple atmospheric model grids. The panels show predictions from Sonora Bobcat \citep{marleySonoraBrownDwarf2021}, Diamondback \citep{morleySonoraSubstellarAtmosphere2024}, Elf Owl \citep{mukherjeeSonoraSubstellarAtmosphere2024}, Exo-REM \citep{charnaySelfconsistentCloudModel2018}, and PICASO irradiated models (this work), illustrating the effects of gravity, metallicity, clouds, and vertical mixing on the CH$_4$ feature. Small grey points show the quality-filtered SPHEREx ultracool dwarf sample \citep{gagneSPHERExPipelineSpectral2026}.}
\label{CH4_color_mag}   
\end{figure*}

\begin{figure*}[t]
    \centering
    \includegraphics[width=0.8\textwidth]{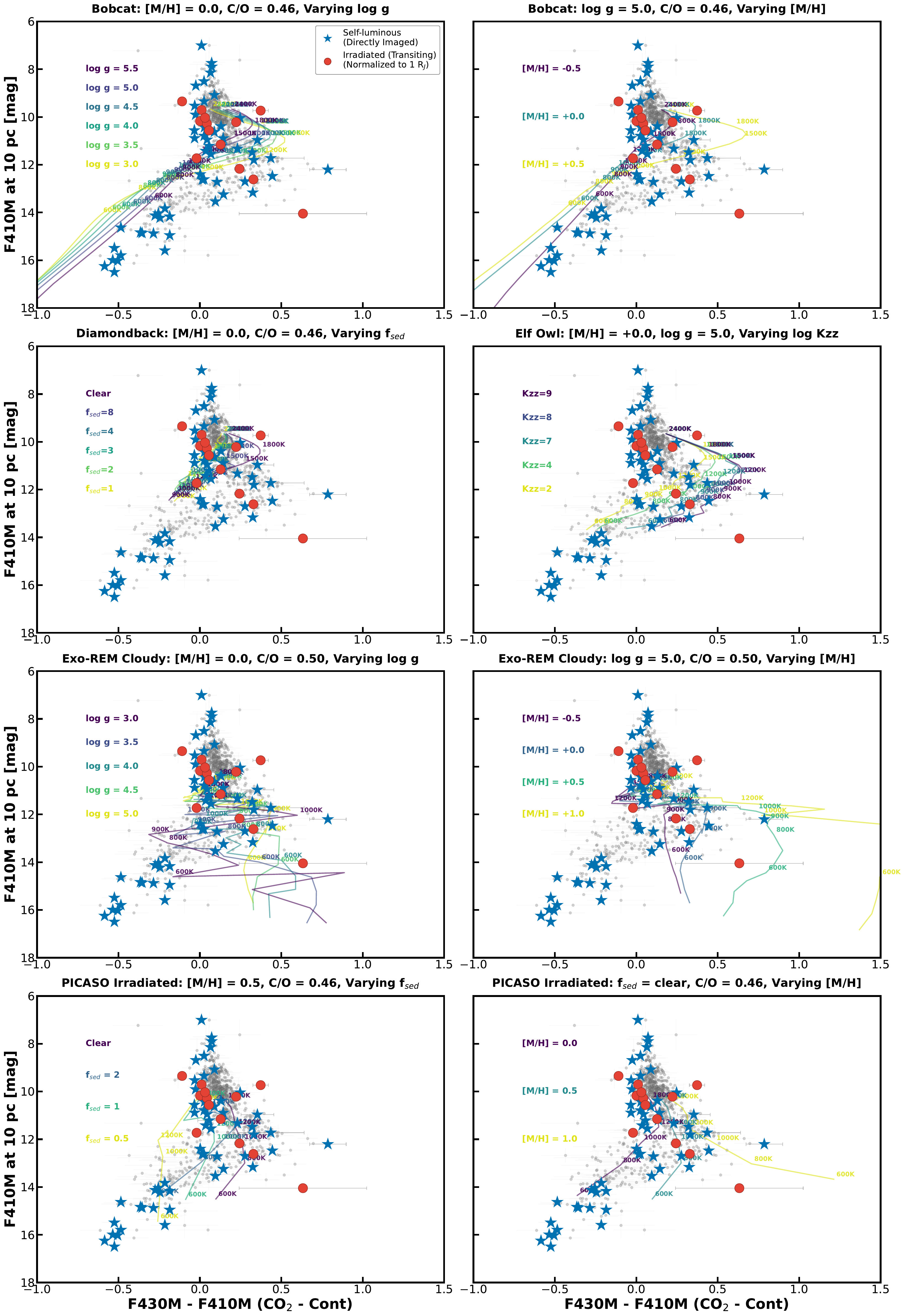}
    \caption{CO$_2$ color-magnitude sequence. The observed CO$_2$ absorption is compared with model grids. The panels display tracks from Sonora Bobcat \citep{marleySonoraBrownDwarf2021}, Diamondback \citep{morleySonoraSubstellarAtmosphere2024}, Elf Owl \citep{mukherjeeSonoraSubstellarAtmosphere2024}, Exo-REM \citep{charnaySelfconsistentCloudModel2018}, and PICASO irradiated models (this work), highlighting the sensitivity of CO$_2$ to metallicity and other atmospheric parameters. Small grey points show the quality-filtered SPHEREx ultracool dwarf sample \citep{gagneSPHERExPipelineSpectral2026}.}
\label{CO2_color_mag}   
\end{figure*}

\begin{figure*}[t]
    \centering
    \includegraphics[width=0.8\textwidth]{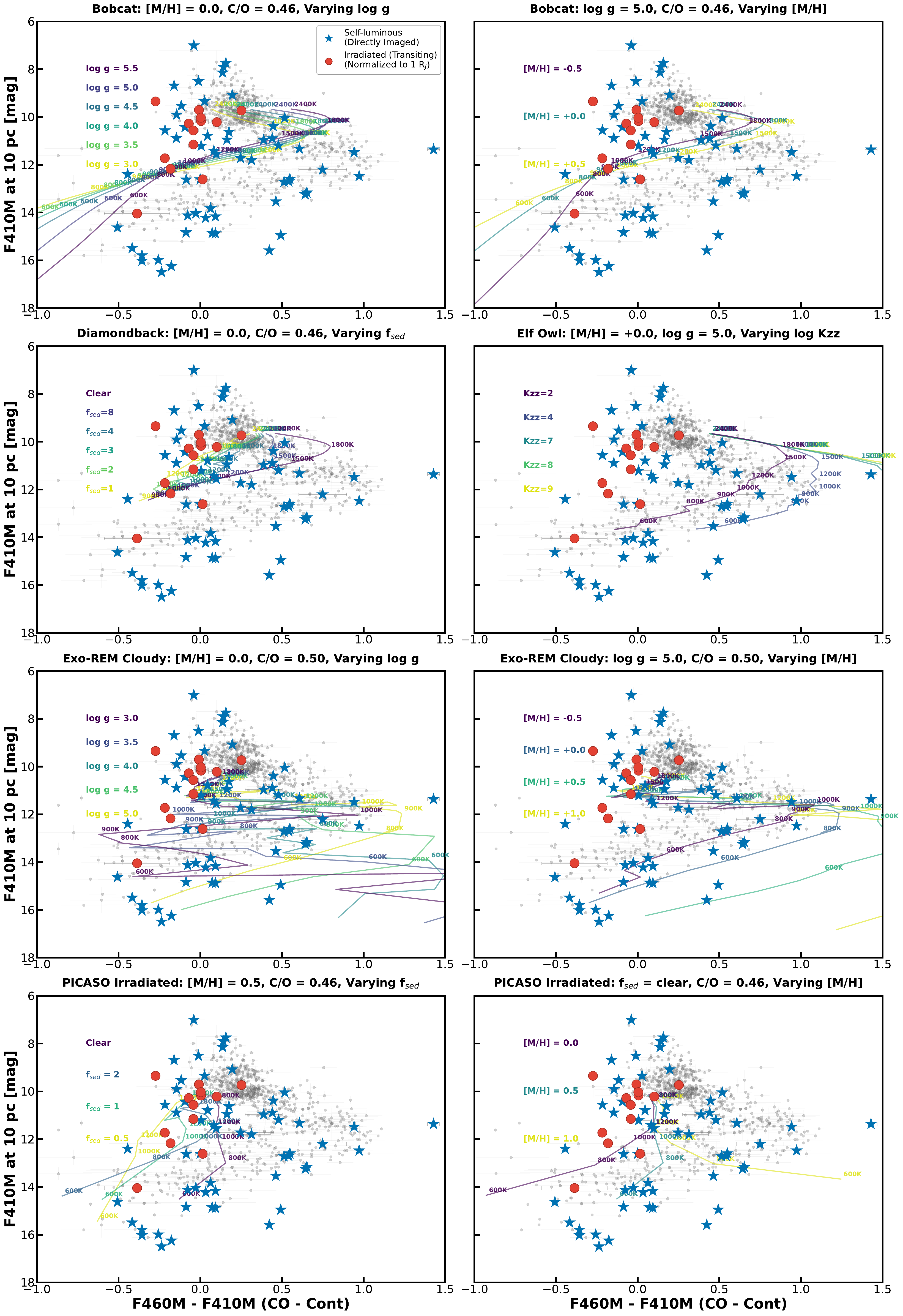}
    \caption{CO color-magnitude sequence. The CO absorption feature is compared with model predictions. The panels show tracks from Sonora Bobcat \citep{marleySonoraBrownDwarf2021}, Diamondback \citep{morleySonoraSubstellarAtmosphere2024}, Elf Owl \citep{mukherjeeSonoraSubstellarAtmosphere2024}, Exo-REM \citep{charnaySelfconsistentCloudModel2018}, and PICASO irradiated models (this work), demonstrating the persistence of CO across different atmospheric conditions. Small grey points show the quality-filtered SPHEREx ultracool dwarf sample \citep{gagneSPHERExPipelineSpectral2026}.}
\label{CO_color_mag}   
\end{figure*}

\begin{figure*}[t]
    \centering
    \includegraphics[width=0.8\textwidth]{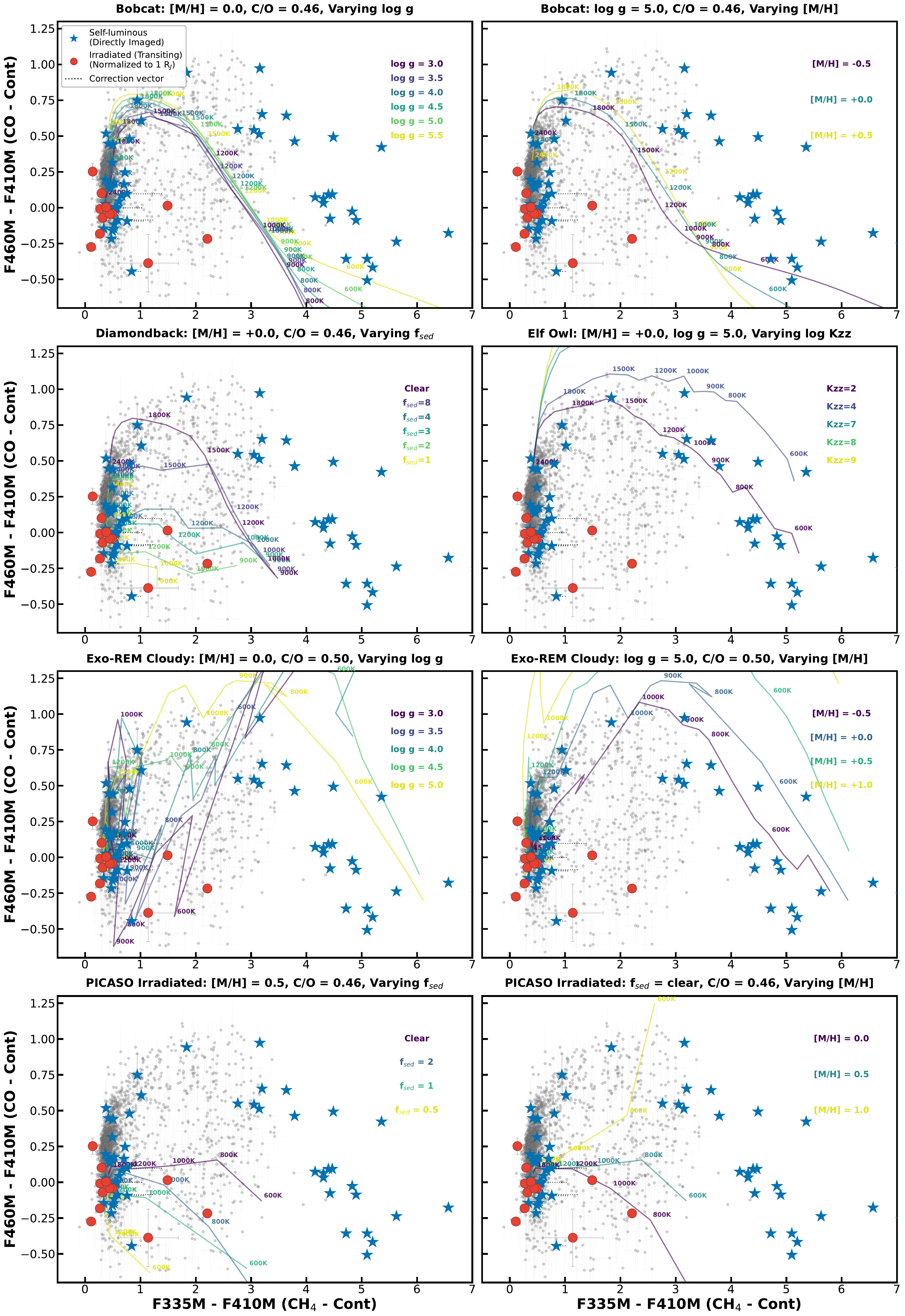}
    \caption{CH$_4$ versus CO color-color diagram. This diagnostic plot separates objects based on their carbon chemistry. The data are compared with model tracks from Sonora Bobcat \citep{marleySonoraBrownDwarf2021}, Diamondback \citep{morleySonoraSubstellarAtmosphere2024}, Elf Owl \citep{mukherjeeSonoraSubstellarAtmosphere2024}, Exo-REM \citep{charnaySelfconsistentCloudModel2018}, and PICASO irradiated models (this work) to probe the effects of disequilibrium chemistry and metallicity. Small grey points show the quality-filtered SPHEREx ultracool dwarf sample \citep{gagneSPHERExPipelineSpectral2026}.}
\label{CH4_CO_color_color}   
\end{figure*}

\begin{figure*}[t]
    \centering
    \includegraphics[width=0.8\textwidth]{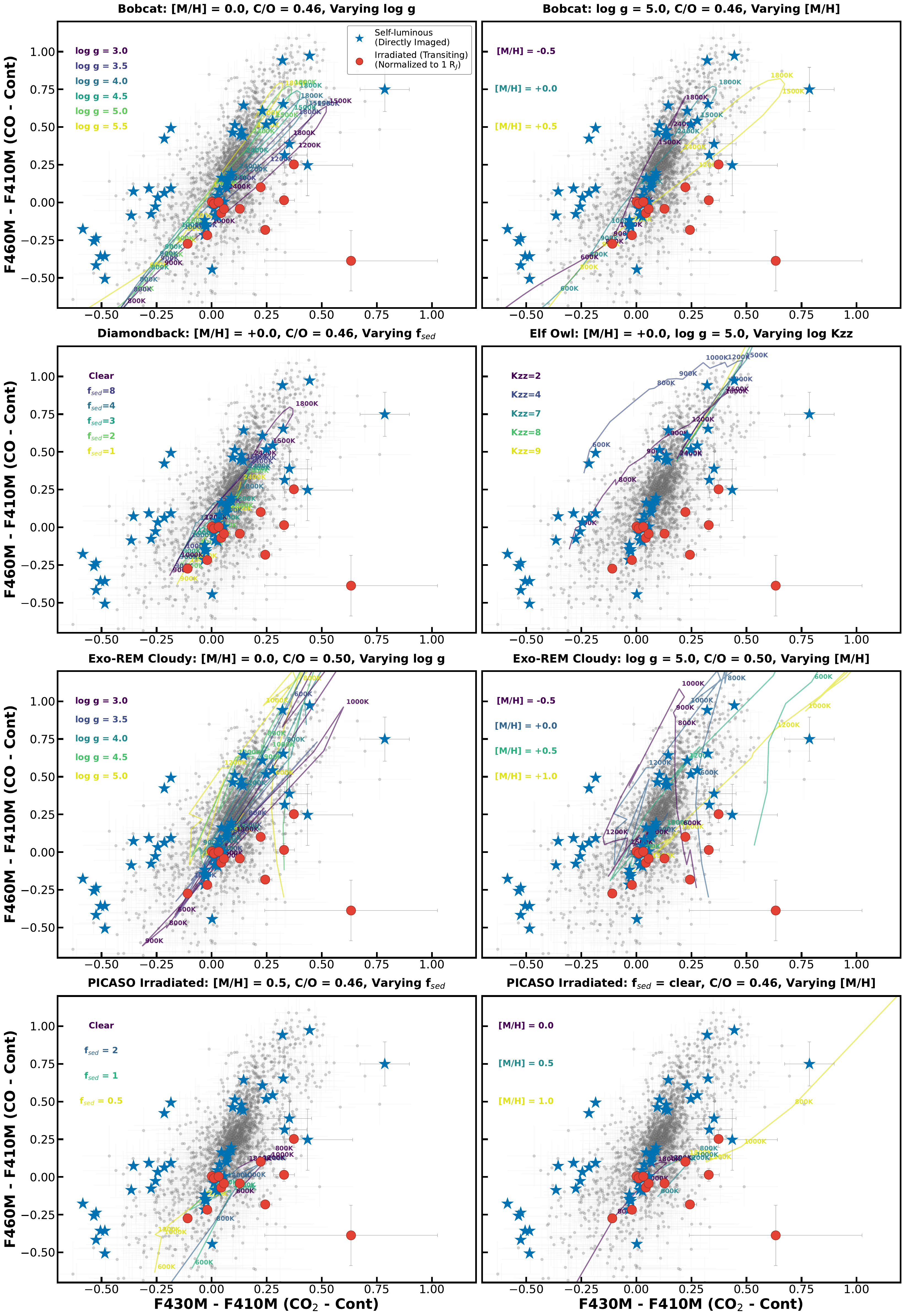}
    \caption{CO$_2$ versus CO color-color diagram. This plot probes the metallicity and C/O ratio of the atmospheres. The observed colors are compared with model predictions from Sonora Bobcat \citep{marleySonoraBrownDwarf2021}, Diamondback \citep{morleySonoraSubstellarAtmosphere2024}, Elf Owl \citep{mukherjeeSonoraSubstellarAtmosphere2024}, Exo-REM \citep{charnaySelfconsistentCloudModel2018}, and PICASO irradiated models (this work). Small grey points show the quality-filtered SPHEREx ultracool dwarf sample \citep{gagneSPHERExPipelineSpectral2026}.}
\label{CO2_CO_color_color}   
\end{figure*}

\clearpage
\begin{longrotatetable}
\begin{deluxetable}{lrrrrrrrl}
\tablecaption{Physical and Atmospheric Properties of Transiting Exoplanets and Substellar Objects \label{tab:all_objects}}
\tablehead{
\colhead{Object Name} & 
\colhead{Distance} & 
\colhead{$R_p$} & 
\colhead{$M_p$} & 
\colhead{log $g$} &
\colhead{Sp. Type} & 
\colhead{T$_{eq}$/T$_{eff}$} & 
\colhead{Age} & 
\colhead{Ref.} \\
\colhead{} & 
\colhead{(pc)} & 
\colhead{(R$_J$)} & 
\colhead{(M$_J$)} & 
\colhead{} &
\colhead{} & 
\colhead{(K)} & 
\colhead{(Gyr)} & 
\colhead{}
}
\startdata
\multicolumn{9}{c}{\textit{Hot Jupiters}} \\
GJ 436 b & $9.75 \pm 0.01$ & $0.37 \pm 0.02$ & $0.073 \pm 0.002$ & $3.120 \pm 0.049$ & — & $686 \pm 10$ & 5 & \citep{mukherjeeJWSTPanchromaticThermal2025} \\
HD 149026 b & $75.86 \pm 0.17$ & $0.74 \pm 0.02$ & $0.358 \pm 0.018$ & $3.210 \pm 0.032$ & — & $1706 \pm 50$ & 5 & \citep{beanHighAtmosphericMetal2023} \\
HD 189733 b & $19.76 \pm 0.01$ & $1.13 \pm 0.01$ & $1.15 \pm 0.039$ & $3.349 \pm 0.017$ & — & $1191 \pm 20$ & 5 & \citep{zhangRetrievalsNIRCamTransmission2024} \\
LTT 9779 b & $80.44 \pm 0.32$ & $0.42 \pm 0.02$ & $0.092 \pm 0.002$ & $3.113 \pm 0.043$ & — & $1978 \pm 19$ & 5 & \citep{jenkinsUltrahotNeptuneNeptune2020} \\
WASP-121 b & $269.9 \pm 1.58$ & $1.75 \pm 0.04$ & $1.157 \pm 0.070$ & $2.971 \pm 0.033$ & — & $2358 \pm 52$ & 5 & \citep{bourrierHotExoplanetAtmospheres2020} \\
WASP-17 b & $405.91 \pm 8.78$ & $1.99 \pm 0.08$ & $0.512 \pm 0.037$ & $2.506 \pm 0.047$ & — & $1755 \pm 28$ & 5 & \citep{lustig-yaegerJWSTTSTDREAMSNightside2025} \\
WASP-18 b & $123.48 \pm 0.37$ & $1.24 \pm 0.08$ & $10.2 \pm 0.35$ & $4.216 \pm 0.058$ & — & $2429 \pm 77$ & 5 & \citep{coulombeBroadbandThermalEmission2023} \\
WASP-19 b & $268.33 \pm 1.72$ & $1.42 \pm 0.05$ & $1.154 \pm 0.080$ & $3.152 \pm 0.043$ & — & $2113 \pm 29$ & 5 & \citep{sahaHighlyCarbonRichDayside2025} \\
WASP-52 b & $174.82 \pm 1.34$ & $1.27 \pm 0.03$ & $0.46 \pm 0.02$ & $2.849 \pm 0.028$ & — & $1315 \pm 35$ & 5 & This work \\
WASP-69 b & $49.96 \pm 0.13$ & $1.11 \pm 0.04$ & $0.26 \pm 0.02$ & $2.719 \pm 0.044$ & — & $963 \pm 18$ & 5 & \citep{schlawinMultipleCluesDayside2024} \\
WASP-77 A b & $105.17 \pm 1.21$ & $1.21 \pm 0.02$ & $1.667 \pm 0.068$ & $3.451 \pm 0.023$ & — & $1715 \pm 26$ & 5 & \citep{augustConfirmationSubsolarMetallicity2023} \\
WASP-80 b & $49.79 \pm 0.12$ & $1.00 \pm 0.03$ & $0.538 \pm 0.036$ & $3.125 \pm 0.039$ & — & $825 \pm 19$ & 5 & \citep{wiserPreciseMetallicityCarbontoOxygen2025, morelModerateAlbedoReflecting2025} \\
NGTS-10 b & $324.7 \pm 27.52$ & $1.241 \pm 0.006$ & $2.162 \pm 0.1$ & $3.542 \pm 0.021$ & — & $1508 \pm 12$ & 5 & \citep{parmentierHorizontalTransport2026} \\
\multicolumn{9}{c}{\textit{Irradiated Brown Dwarf}} \\
ZTF J0038+2030 B & $139 \pm 2$ & $0.759 \pm 0.011$ & $62.1 \pm 4.1$ & $5.425^{+0.020}_{-0.030}$ & T6 & $1049 \pm 6$ & 7.5--8.8 & \citep{broski-laingAsymmetricNightsideCO22026}; this work \\
\multicolumn{9}{c}{\textit{Substellar Companions and Free-floating Ultracool Dwarfs}} \\
TWA 27 A & $64.67 \pm 0.49$ & — & — & $4.0$ & M8 & $2600$ & 0.01 & \citep{manjavacasMediumresolution09753Mm2024} \\
Trappist-1 & $12.47 \pm 0.01$ & — & — & $5.24 \pm 0.007$ & M8V & $2566 \pm 26$ & 7.6 & \citep{gillonTemperateEarthsizedPlanets2016} \\
2MASS J1439+1929 & $14.33 \pm 0.09$ & — & — & $4.38 \pm 0.28$ & L1 & $1812 \pm 90$ & 5 & \citep{sanghiHawaiiInfraredParallax2023} \\
2MASS J0036+1821 & $8.74 \pm 0.01$ & — & — & $5.21 \pm 0.17$ & L4 & $1869 \pm 64$ & 5 & \citep{filippazzoFundamentalParametersSpectral2015} \\
J0624 & $12.19 \pm 0.05$ & — & — & $5.0$ & L5 & $1800$ & 5 & \citep{sanghiHawaiiInfraredParallax2023} \\
J2148 & $8.09 \pm 0.02$ & — & — & $5.13 \pm 0.28$ & L6 & $1446 \pm 72$ & 5 & \citep{filippazzoFundamentalParametersSpectral2015} \\
TWA 27 b & $64.67 \pm 0.49$ & — & $5 \pm 2$ & $3.5$ & L6 & $1300$ & 0.01 & \citep{manjavacasMediumresolution09753Mm2024} \\
VHS 1256 b & $21.15 \pm 0.21$ & — & $15 \pm 5$ & $4.212 \pm 0.005$ & L7 & $1137 \pm 1$ & 0.14 & \citep{milesJWSTEarlyreleaseScience2023, deregtNativeresolutionRetrievalsVHS2026} \\
PSO J318 & $22.2 \pm 0.8$ & — & $8.3 \pm 0.5$ & $4.0$ & L7.5 & $1114$ & 0.023 & \citep{molliereEvidenceSiOCloud2025} \\
W1049A & $1.996 \pm 0.0002$ & — & — & $4.82 \pm 0.07$ & L7.5 & $1242 \pm 16$ & 0.51 & \citep{billerJWSTWeatherReport2024} \\
W1049B & $1.996 \pm 0.0002$ & — & — & $4.78 \pm 0.06$ & T0.5 & $1188 \pm 16$ & 0.51 & \citep{billerJWSTWeatherReport2024} \\
SIMP 0136+0933 & $6.12 \pm 0.02$ & — & — & $4.31 \pm 0.03$ & T2 & $1098 \pm 6$ & 0.2 & \citep{mccarthyJWSTWeatherReport2025} \\
YSES-1b & $94.229 \pm 0.103$ & — & — & $4.56 \pm 0.41$ & L0 & $1607 \pm 24$ & 0.027 & \citep{hochSilicateCloudsCircumplanetary2025} \\
YSES-1c & $94.229 \pm 0.103$ & — & — & $3.69 \pm 0.19$ & L7.5 & $984 \pm 29$ & 0.027 & \citep{hochSilicateCloudsCircumplanetary2025} \\
Epsilon Indi Ba & $3.638 \pm 0.001$ & — & — & $5.5$ & T1 & $1370$ & 4 & \citep{kingIndiBaBb2010} \\
Epsilon Indi Bb & $3.638 \pm 0.001$ & — & — & $5.25$ & T6 & $1000$ & 4 & \citep{kingIndiBaBb2010} \\
HD 19467 b & $30.864 \pm 0.591$ & — & — & $4.5 \pm 0.5$ & T5.5 & $1103 \pm 100$ & 9.4 & \citep{hochJWSTTSTHighContrast2024} \\
HR 8799 b & $40.880 \pm 0.076$ & — & $6 \pm 0.3$ & $4.42$ & T2 & $1020$ & 0.023 & \citep{balmerJWSTTSTHighContrast2025} \\
HR 8799 c & $40.880 \pm 0.076$ & — & $8.5 \pm 0.4$ & $4.34$ & L7 & $1195$ & 0.023 & \citep{balmerJWSTTSTHighContrast2025} \\
HR 8799 d & $40.880 \pm 0.076$ & — & $9.2 \pm 0.1$ & $4.39$ & L7 & $1300$ & 0.023 & \citep{balmerJWSTTSTHighContrast2025} \\
HR 8799 e & $40.880 \pm 0.076$ & — & $9.6 \pm 1.9$ & $4.38$ & L7 & $1251$ & 0.023 & \citep{balmerJWSTTSTHighContrast2025} \\
29 Cygni b & $40.700 \pm 0.100$ & — & $13.1 \pm 5.2$ & $4.0$ & L8 & $1300$ & 0.0028 & Balmer et al. (in Prep) \\
LRL 2121 & 315 & — & — & 3.5 & H & — & 0.005 & \citep{luhmanNewSpectralClass2025} \\
LRL 11001 & 315 & — & — & 3.5 & H & — & 0.005 & \citep{luhmanNewSpectralClass2025} \\
LRL 11003 & 315 & — & — & 3.5 & H & — & 0.005 & \citep{luhmanNewSpectralClass2025} \\
LRL 11004 & 315 & — & — & 3.5 & H & — & 0.005 & \citep{luhmanNewSpectralClass2025} \\
LRL 11024 & 315 & — & — & 3.5 & H & — & 0.005 & \citep{luhmanNewSpectralClass2025} \\
LRL 11027 & 315 & — & — & 3.5 & L1 & — & 0.005 & \citep{luhmanNewSpectralClass2025} \\
LRL 11037 & 315 & — & — & 3.5 & H & — & 0.005 & \citep{luhmanNewSpectralClass2025} \\
LRL 11041 & 315 & — & — & 3.5 & H & — & 0.005 & \citep{luhmanNewSpectralClass2025} \\
LRL 11043 & 315 & — & — & 3.5 & H & — & 0.005 & \citep{luhmanNewSpectralClass2025} \\
LRL 11044 & 315 & — & — & 3.5 & H & — & 0.005 & \citep{luhmanNewSpectralClass2025} \\
LRL 1546 & 315 & — & — & 3.5 & L1 & — & 0.005 & \citep{luhmanNewSpectralClass2025} \\
LRL 22705 & 315 & — & — & 3.5 & L1 & — & 0.005 & \citep{luhmanNewSpectralClass2025} \\
LRL 40013 & 315 & — & — & 3.5 & H & — & 0.005 & \citep{luhmanNewSpectralClass2025} \\
WISE J0247+37 & $14.620 \pm 0.4275$ & — & — & $4.69 \pm 0.07$ & T8 & $689 \pm 4$ & 5 & \citep{beilerPreciseBolometricLuminosities2024, lueberCloudsChemistryAcross2026} \\
WISEPA J0313+78 & $7.375 \pm 0.1523$ & — & — & $5.20 \pm 0.07$ & T8.5 & $580 \pm 2$ & 5 & \citep{beilerPreciseBolometricLuminosities2024, lueberCloudsChemistryAcross2026} \\
WISE J0359-54 & $13.587 \pm 0.3692$ & — & — & $4.80 \pm 0.07$ & Y0 & $462 \pm 2$ & 5 & \citep{beilerPreciseBolometricLuminosities2024, lueberCloudsChemistryAcross2026} \\
WISE J0430+46 & $10.406 \pm 0.3140$ & — & — & $4.87 \pm 0.06$ & T8 & $567 \pm 2$ & 5 & \citep{beilerPreciseBolometricLuminosities2024, lueberCloudsChemistryAcross2026} \\
WISE J0535-75 & $14.556 \pm 0.4238$ & — & — & $4.30 \pm 0.09$ & Y1 & $399 \pm 3$ & 5 & \citep{beilerPreciseBolometricLuminosities2024, lueberCloudsChemistryAcross2026} \\
WISE J0734-71 & $13.423 \pm 0.3063$ & — & — & $4.34 \pm 0.04$ & Y0 & $468 \pm 2$ & 5 & \citep{beilerPreciseBolometricLuminosities2024, lueberCloudsChemistryAcross2026} \\
WISE J0825+28 & $6.553 \pm 0.0859$ & — & — & $4.32 \pm 0.09$ & Y0.5 & $378 \pm 35$ & 5 & \citep{beilerPreciseBolometricLuminosities2024, lueberCloudsChemistryAcross2026} \\
ULAS J1029+09 & $14.577 \pm 0.3612$ & — & — & $4.03 \pm 0.03$ & T8 & $742 \pm 7$ & 5 & \citep{beilerPreciseBolometricLuminosities2024, lueberCloudsChemistryAcross2026} \\
CWISEP J1047+54 & $14.684 \pm 1.0566$ & — & — & $5.34 \pm 0.08$ & Y1 & $430 \pm 6$ & 5 & \citep{beilerPreciseBolometricLuminosities2024, lueberCloudsChemistryAcross2026} \\
WISE J1206+84 & $11.806 \pm 0.2927$ & — & — & $5.33 \pm 0.07$ & Y0 & $456 \pm 2$ & 5 & \citep{beilerPreciseBolometricLuminosities2024, lueberCloudsChemistryAcross2026} \\
SDSSp J1346-00 & $14.451 \pm 0.4803$ & — & — & $4.84 \pm 0.12$ & T6.5 & $932 \pm 8$ & 5 & \citep{beilerPreciseBolometricLuminosities2024, lueberCloudsChemistryAcross2026} \\
WISEPC J1405+55 & $6.321 \pm 0.1039$ & — & — & 4.62 & Y0.5 & $394 \pm 3$ & 5 & \citep{beilerPreciseBolometricLuminosities2024, lueberCloudsChemistryAcross2026} \\
CWISEP J1446-23 & $9.634 \pm 0.4641$ & — & — & $4.88 \pm 0.11$ & Y1 & $372 \pm 5$ & 5 & \citep{beilerPreciseBolometricLuminosities2024, lueberCloudsChemistryAcross2026} \\
WISE J1501-40 & $13.736 \pm 0.4340$ & — & — & $4.24 \pm 0.07$ & T6 & $982 \pm 7$ & 5 & \citep{beilerPreciseBolometricLuminosities2024, lueberCloudsChemistryAcross2026} \\
WISEPA J1541-22 & $5.992 \pm 0.0718$ & — & — & $4.74 \pm 0.09$ & Y1 & $350 \pm 6$ & 5 & \citep{beilerPreciseBolometricLuminosities2024, lueberCloudsChemistryAcross2026} \\
SDSS J1624+00 & $10.893 \pm 0.1424$ & — & — & $4.90 \pm 0.10$ & T6 & $991 \pm 7$ & 5 & \citep{beilerPreciseBolometricLuminosities2024, lueberCloudsChemistryAcross2026} \\
WISEPA J1959-33 & $11.919 \pm 0.2841$ & — & — & $4.06 \pm 0.06$ & T8 & $736 \pm 6$ & 5 & \citep{beilerPreciseBolometricLuminosities2024, lueberCloudsChemistryAcross2026} \\
WISEPC J2056+14 & $7.102 \pm 0.1009$ & — & — & $4.96 \pm 0.12$ & Y0 & $444 \pm 4$ & 5 & \citep{beilerPreciseBolometricLuminosities2024, lueberCloudsChemistryAcross2026} \\
WISE J2102-44 & $10.764 \pm 0.2202$ & — & — & $4.70 \pm 0.08$ & T9 & $592 \pm 3$ & 5 & \citep{beilerPreciseBolometricLuminosities2024, lueberCloudsChemistryAcross2026} \\
WISEA J2159-48 & $13.532 \pm 0.4761$ & — & — & $4.96 \pm 0.05$ & T9 & $551 \pm 2$ & 5 & \citep{beilerPreciseBolometricLuminosities2024, lueberCloudsChemistryAcross2026} \\
WISE J2209+27 & $6.184 \pm 0.0765$ & — & — & $4.45 \pm 0.08$ & Y0 & $353 \pm 2$ & 5 & \citep{beilerPreciseBolometricLuminosities2024, lueberCloudsChemistryAcross2026} \\
WISEA J2354+02 & $7.657 \pm 0.1935$ & — & — & $4.76 \pm 0.07$ & Y1 & $381 \pm 5$ & 5 & \citep{beilerPreciseBolometricLuminosities2024, lueberCloudsChemistryAcross2026} \\
\enddata
\end{deluxetable}
\end{longrotatetable}

\begin{longtable}{lccccccc}
\caption{Synthetic photometry at 10 pc (1\,R$_J$ for transiting planets and ZTF J0038+2030 B)}
\label{tab:photometry_10pc}\\
\hline\hline
Object Name & F300M & F335M & F410M & F430M & F460M & J (2MASS) & Ks (2MASS) \\
\hline
\endfirsthead
\multicolumn{8}{c}{\tablename\ \thetable{} -- \textit{continued}}\\
\hline\hline
Object & F300M & F335M & F410M & F430M & F460M & J\_2MASS & Ks\_2MASS \\
\hline
\endhead
\hline
\endfoot
\hline
\endlastfoot
GJ 436 b & $16.06 \pm 1.16$ & $15.19 \pm 0.53$ & $14.05 \pm 0.14$ & $14.68 \pm 0.37$ & $13.66 \pm 0.14$ & \nodata & \nodata \\
HD 149026 b & $10.93 \pm 0.05$ & $10.51 \pm 0.03$ & $10.22 \pm 0.03$ & $10.44 \pm 0.04$ & $10.32 \pm 0.04$ & \nodata & \nodata \\
HD 189733 b & $12.17 \pm 0.03$ & $11.64 \pm 0.02$ & $11.15 \pm 0.01$ & $11.28 \pm 0.02$ & $11.11 \pm 0.02$ & \nodata & \nodata \\
LTT 9779 b & $10.16 \pm 0.03$ & $9.87 \pm 0.02$ & $9.73 \pm 0.02$ & $10.11 \pm 0.04$ & $9.98 \pm 0.05$ & $11.42 \pm 0.05$ & $10.44 \pm 0.06$ \\
WASP-121 b & $9.51 \pm 0.01$ & $9.46 \pm 0.01$ & $9.35 \pm 0.01$ & $9.24 \pm 0.01$ & $9.07 \pm 0.01$ & \nodata & \nodata \\
WASP-17 b & $10.97 \pm 0.02$ & $10.60 \pm 0.01$ & $10.28 \pm 0.01$ & $10.32 \pm 0.02$ & $10.21 \pm 0.02$ & $13.55 \pm 0.06$ & $11.71 \pm 0.04$ \\
WASP-18 b & \nodata & \nodata & \nodata & \nodata & \nodata & $10.92 \pm 0.01$ & $9.83 \pm 0.01$ \\
WASP-19 b & $10.25 \pm 0.01$ & $9.98 \pm 0.01$ & $9.71 \pm 0.01$ & $9.72 \pm 0.01$ & $9.70 \pm 0.02$ & $11.92 \pm 0.03$ & $10.57 \pm 0.01$ \\
WASP-52 b & $11.38 \pm 0.02$ & $11.01 \pm 0.02$ & $10.56 \pm 0.02$ & $10.62 \pm 0.02$ & $10.52 \pm 0.03$ & $14.29 \pm 0.11$ & $11.99 \pm 0.03$ \\
WASP-69 b & $12.98 \pm 0.04$ & $12.44 \pm 0.02$ & $12.17 \pm 0.02$ & $12.41 \pm 0.03$ & $11.99 \pm 0.02$ & \nodata & \nodata \\
WASP-77 A b & \nodata & $10.41 \pm 0.01$ & $10.03 \pm 0.01$ & $10.06 \pm 0.01$ & $10.04 \pm 0.01$ & \nodata & \nodata \\
WASP-80 b & $14.28 \pm 0.11$ & $14.10 \pm 0.08$ & $12.61 \pm 0.02$ & $12.94 \pm 0.04$ & $12.63 \pm 0.04$ & $16.83 \pm 0.23$ & $15.02 \pm 0.20$ \\
NGTS-10 b & $10.93 \pm 0.02$ & $10.55 \pm 0.02$ & $10.17 \pm 0.02$ & $10.17 \pm 0.03$ & $10.18 \pm 0.03$ & $13.46 \pm 0.07$ & $11.40 \pm 0.02$ \\
ZTF J0038+2030 B & $13.72 \pm 0.01$ & $13.96 \pm 0.02$ & $11.73 \pm 0.00$ & $11.71 \pm 0.01$ & $11.51 \pm 0.01$ & $14.89 \pm 0.01$ & $14.51 \pm 0.03$ \\
TWA 27 A & $7.86 \pm 0.00$ & $7.49 \pm 0.00$ & $7.00 \pm 0.00$ & $7.01 \pm 0.00$ & $6.96 \pm 0.00$ & $8.96 \pm 0.00$ & $7.93 \pm 0.00$ \\
Trappist-1 & $9.81 \pm 0.00$ & $9.45 \pm 0.00$ & $9.08 \pm 0.00$ & $9.17 \pm 0.00$ & $9.28 \pm 0.00$ & $10.84 \pm 0.00$ & $9.81 \pm 0.00$ \\
2MASS J1439+1929 & $10.95 \pm 0.01$ & $10.42 \pm 0.01$ & $10.04 \pm 0.01$ & $10.29 \pm 0.01$ & $10.56 \pm 0.02$ & \nodata & \nodata \\
2MASS J0036+1821 & $11.38 \pm 0.00$ & $10.81 \pm 0.00$ & $10.39 \pm 0.00$ & $10.52 \pm 0.01$ & $10.84 \pm 0.01$ & \nodata & \nodata \\
J0624 & $11.89 \pm 0.00$ & $11.40 \pm 0.00$ & $10.89 \pm 0.00$ & $11.04 \pm 0.00$ & $11.34 \pm 0.00$ & $14.08 \pm 0.00$ & $12.20 \pm 0.00$ \\
J2148 & $11.67 \pm 0.00$ & $11.15 \pm 0.00$ & $10.63 \pm 0.00$ & $10.72 \pm 0.00$ & $10.80 \pm 0.00$ & $14.63 \pm 0.00$ & $12.25 \pm 0.00$ \\
TWA 27 b & $11.84 \pm 0.00$ & $11.41 \pm 0.00$ & $10.89 \pm 0.00$ & $10.86 \pm 0.00$ & $10.74 \pm 0.00$ & $15.48 \pm 0.00$ & $12.72 \pm 0.00$ \\
VHS 1256 b & $12.19 \pm 0.00$ & $11.65 \pm 0.00$ & $10.95 \pm 0.00$ & $10.99 \pm 0.00$ & $11.11 \pm 0.00$ & $15.62 \pm 0.00$ & $13.00 \pm 0.00$ \\
PSO J318 & \nodata & $11.40 \pm 0.00$ & $10.79 \pm 0.00$ & $10.82 \pm 0.00$ & $10.84 \pm 0.00$ & $15.55 \pm 0.00$ & $12.76 \pm 0.00$ \\
W1049A & $12.52 \pm 0.00$ & $12.00 \pm 0.00$ & $11.20 \pm 0.00$ & $11.33 \pm 0.00$ & $11.68 \pm 0.01$ & $15.17 \pm 0.01$ & $13.07 \pm 0.00$ \\
W1049B & $12.78 \pm 0.00$ & $12.34 \pm 0.00$ & $11.33 \pm 0.00$ & $11.56 \pm 0.00$ & $11.94 \pm 0.00$ & $14.93 \pm 0.00$ & $13.26 \pm 0.00$ \\
SIMP 0136+0933 & $13.22 \pm 0.00$ & $13.32 \pm 0.00$ & $11.48 \pm 0.00$ & $11.80 \pm 0.00$ & $12.42 \pm 0.00$ & $14.48 \pm 0.00$ & $13.60 \pm 0.00$ \\
YSES-1b & $9.47 \pm 0.00$ & $9.06 \pm 0.00$ & $8.51 \pm 0.00$ & $8.53 \pm 0.00$ & $8.50 \pm 0.00$ & $11.27 \pm 0.00$ & $9.76 \pm 0.00$ \\
YSES-1c & $12.64 \pm 0.00$ & $12.09 \pm 0.00$ & $11.42 \pm 0.00$ & $11.46 \pm 0.00$ & $11.50 \pm 0.00$ & $16.61 \pm 0.00$ & $13.58 \pm 0.00$ \\
Epsilon Indi Ba & $12.90 \pm 0.00$ & $13.01 \pm 0.00$ & $11.36 \pm 0.01$ & \nodata & $12.79 \pm 0.04$ & $14.53 \pm 0.00$ & $13.53 \pm 0.00$ \\
Epsilon Indi Bb & $14.78 \pm 0.01$ & $15.36 \pm 0.00$ & $12.60 \pm 0.01$ & \nodata & $13.15 \pm 0.04$ & $15.40 \pm 0.00$ & $15.67 \pm 0.00$ \\
HD 19467 b & $14.66 \pm 0.01$ & $15.75 \pm 0.02$ & $12.70 \pm 0.00$ & $12.98 \pm 0.00$ & $13.24 \pm 0.01$ & \nodata & \nodata \\
HR 8799 b & $14.31 \pm 0.04$ & $13.15 \pm 0.05$ & $12.21 \pm 0.04$ & $12.99 \pm 0.10$ & $12.95 \pm 0.14$ & \nodata & \nodata \\
HR 8799 c & $13.43 \pm 0.09$ & $12.29 \pm 0.12$ & $11.80 \pm 0.09$ & $12.13 \pm 0.05$ & $12.12 \pm 0.08$ & \nodata & \nodata \\
HR 8799 d & $12.72 \pm 0.10$ & $11.76 \pm 0.06$ & $11.21 \pm 0.04$ & $11.27 \pm 0.05$ & $11.22 \pm 0.06$ & \nodata & \nodata \\
HR 8799 e & \nodata & $12.44 \pm 0.16$ & $11.72 \pm 0.16$ & $12.16 \pm 0.12$ & $11.97 \pm 0.12$ & \nodata & \nodata \\
29 Cygni b & \nodata & \nodata & $10.97 \pm 0.06$ & $11.32 \pm 0.08$ & $11.35 \pm 0.10$ & \nodata & \nodata \\
LRL 2121 & 10.49 & 10.31 & 9.91 & 9.88 & 9.76 & 12.31 & 10.78 \\
LRL 11001 & 11.48 & 11.48 & 10.43 & 10.46 & 10.34 & 13.37 & 11.68 \\
LRL 11003 & 12.57 & 12.94 & 11.55 & 11.62 & 11.65 & 14.76 & 12.85 \\
LRL 11004 & 10.47 & 10.07 & 9.53 & 9.50 & 9.42 & 13.42 & 11.14 \\
LRL 11024 & 13.97 & 13.41 & 12.40 & 12.40 & 11.96 & 15.43 & 13.76 \\
LRL 11027 & 10.06 & 9.72 & 9.35 & 9.37 & 9.37 & 12.41 & 10.48 \\
LRL 11037 & 13.57 & 13.85 & 12.62 & 12.64 & 12.54 & 15.65 & 13.95 \\
LRL 11041 & 13.61 & 13.64 & 12.61 & 12.64 & 12.61 & 16.15 & 13.99 \\
LRL 11043 & 11.35 & 11.16 & 10.56 & 10.53 & 10.34 & 13.33 & 11.47 \\
LRL 11044 & 9.50 & 9.28 & 8.69 & 8.66 & 8.53 & 11.57 & 9.70 \\
LRL 1546 & 8.58 & 8.17 & 7.74 & 7.81 & 7.90 & 10.31 & 8.67 \\
LRL 22705 & 8.78 & 8.35 & 7.90 & 7.97 & 8.04 & 10.68 & 8.92 \\
LRL 40013 & 9.07 & 8.64 & 8.14 & 8.21 & 8.28 & 11.30 & 9.36 \\
WISE J0247+37 & 16.88 & 18.14 & 13.83 & 13.62 & 13.89 & $17.49 \pm 0.00$ & $17.71 \pm 0.02$ \\
WISEPA J0313+78 & 17.25 & 18.55 & 14.13 & 13.85 & 14.05 & \nodata & \nodata \\
WISE J0359-54 & 18.56 & 19.02 & 14.86 & 14.50 & 14.93 & \nodata & \nodata \\
WISE J0430+46 & 17.87 & 18.57 & 14.17 & 13.98 & 14.26 & \nodata & \nodata \\
WISE J0535-75 & 18.79 & 19.73 & 14.63 & 14.14 & 14.12 & \nodata & \nodata \\
WISE J0734-71 & 18.41 & 19.73 & 14.84 & 14.47 & 14.75 & \nodata & \nodata \\
WISE J0825+28 & 20.11 & 21.12 & 16.02 & 15.52 & 15.67 & \nodata & \nodata \\
ULAS J1029+09 & 16.08 & 17.32 & 13.54 & 13.63 & 14.00 & \nodata & \nodata \\
CWISEP J1047+54 & 19.43 & 20.95 & 15.59 & 15.38 & 16.01 & \nodata & \nodata \\
WISE J1206+84 & 18.32 & 19.44 & 14.95 & 14.77 & 15.45 & \nodata & \nodata \\
SDSSp J1346-00 & 14.73 & 15.63 & 12.48 & 12.92 & 13.45 & $15.14 \pm 0.00$ & $15.02 \pm 0.00$ \\
WISEPC J1405+55 & 19.86 & 20.69 & 15.49 & 14.97 & 15.07 & \nodata & \nodata \\
CWISEP J1446-23 & 20.57 & 22.12 & 16.50 & 15.97 & 16.26 & \nodata & \nodata \\
WISE J1501-40 & 15.46 & 16.36 & 13.17 & 13.49 & 13.82 & $15.65 \pm 0.00$ & $15.67 \pm 0.00$ \\
WISEPA J1541-22 & 19.62 & 20.51 & 15.80 & 15.31 & 15.44 & \nodata & \nodata \\
SDSS J1624+00 & 15.08 & 15.87 & 12.72 & 12.82 & 13.23 & $15.30 \pm 0.00$ & $15.40 \pm 0.00$ \\
WISEPA J1959-33 & 15.97 & 16.88 & 13.24 & 13.39 & 13.89 & $16.52 \pm 0.00$ & $16.42 \pm 0.01$ \\
WISEPC J2056+14 & 18.24 & 19.35 & 14.88 & 14.59 & 14.97 & \nodata & \nodata \\
WISE J2102-44 & 17.35 & 18.53 & 14.23 & 13.98 & 14.26 & \nodata & \nodata \\
WISEA J2159-48 & 17.67 & 18.87 & 14.04 & 13.78 & 14.01 & \nodata & \nodata \\
WISE J2209+27 & 21.02 & 22.82 & 16.25 & 15.66 & 16.07 & \nodata & \nodata \\
WISEA J2354+02 & 20.81 & 23.07 & 15.99 & 15.46 & 15.74 & \nodata & \nodata \\
\end{longtable}


\begin{thebibliography}{}
\expandafter\ifx\csname natexlab\endcsname\relax\def\natexlab#1{#1}\fi
\providecommand{\url}[1]{\href{#1}{#1}}
\providecommand{\dodoi}[1]{doi:~\href{http://doi.org/#1}{\nolinkurl{#1}}}
\providecommand{\doeprint}[1]{\href{http://ascl.net/#1}{\nolinkurl{http://ascl.net/#1}}}
\providecommand{\doarXiv}[1]{\href{https://arxiv.org/abs/#1}{\nolinkurl{https://arxiv.org/abs/#1}}}

\bibitem[{A.~S. Ackerman \& M.~S. Marley(2001)Ackerman \&
  Marley}]{ackermanPrecipitatingCondensationClouds2001}
Ackerman, A.~S., \& Marley, M.~S. 2001, \bibinfo{title}{Precipitating
  {Condensation} {Clouds} in {Substellar} {Atmospheres},} The Astrophysical
  Journal, 556, 872, \dodoi{10.1086/321540}

\bibitem[{F. Allard {et~al.}(2001)Allard, Hauschildt, Alexander, Tamanai, \&
  Schweitzer}]{allardLimitingEffectsDust2001}
Allard, F., Hauschildt, P.~H., Alexander, D.~R., Tamanai, A., \& Schweitzer, A.
  2001, \bibinfo{title}{The {Limiting} {Effects} of {Dust} in {Brown} {Dwarf}
  {Model} {Atmospheres},} The Astrophysical Journal, 556, 357,
  \dodoi{10.1086/321547}

\bibitem[{P.~C. August {et~al.}(2023)August, Bean, Zhang, Lunine, Xue, Line, \&
  Smith}]{augustConfirmationSubsolarMetallicity2023}
August, P.~C., Bean, J.~L., Zhang, M., {et~al.} 2023,
  \bibinfo{title}{Confirmation of sub-solar metallicity for {WASP}-{77Ab} from
  {JWST} thermal emission spectroscopy,} arXiv.
\newblock \url{http://arxiv.org/abs/2305.07753}

\bibitem[{W.~O. Balmer {et~al.}(2025)Balmer, Kammerer, Pueyo, Perrin, Girard,
  Leisenring, Lawson, Dennen, van~der Marel, Beichman, Bryden, Llop-Sayson,
  Valenti, Lothringer, Lewis, Mâlin, Rebollido, Rickman, Hoch, Soummer,
  Clampin, \& Mountain}]{balmerJWSTTSTHighContrast2025}
Balmer, W.~O., Kammerer, J., Pueyo, L., {et~al.} 2025,
  \bibinfo{title}{{JWST}-{TST} {High} {Contrast}: {Living} on the {Wedge}, or,
  {NIRCam} {Bar} {Coronagraphy} {Reveals} {CO2} in the {HR} 8799 and 51 {Eri}
  {Exoplanets}’ {Atmospheres},} The Astronomical Journal, 169, 209,
  \dodoi{10.3847/1538-3881/adb1c6}

\bibitem[{N.~E. Batalha {et~al.}(2019)Batalha, Marley, Lewis, \&
  Fortney}]{batalhaExoplanetReflectedlightSpectroscopy2019}
Batalha, N.~E., Marley, M.~S., Lewis, N.~K., \& Fortney, J.~J. 2019,
  \bibinfo{title}{Exoplanet {Reflected}-light {Spectroscopy} with {PICASO},}
  The Astrophysical Journal, 878, 70, \dodoi{10.3847/1538-4357/ab1b51}

\bibitem[{N.~E. Batalha {et~al.}(2026)Batalha, Rooney, Visscher, Moran, Marley,
  Sengupta, Kiefer, Lodge, Mang, Morley, Mukherjee, Fortney, Gao, Lewis,
  Mayorga, Pearce, \& Wakeford}]{batalha_condensation_2026}
Batalha, N.~E., Rooney, C.~M., Visscher, C., {et~al.} 2026,
  \bibinfo{title}{Condensation {Clouds} in {Substellar} {Atmospheres} with
  {Virga},} The Astronomical Journal, 171, 98, \dodoi{10.3847/1538-3881/ae29e5}

\bibitem[{J.~L. Bean {et~al.}(2023)Bean, Xue, August, Lunine, Zhang, Thorngren,
  Tsai, Stassun, Schlawin, Ahrer, Ih, \&
  Mansfield}]{beanHighAtmosphericMetal2023}
Bean, J.~L., Xue, Q., August, P.~C., {et~al.} 2023, \bibinfo{title}{High
  atmospheric metal enrichment for a {Saturn}-mass planet,} Nature, 1,
  \dodoi{10.1038/s41586-023-05984-y}

\bibitem[{S.~A. Beiler {et~al.}(2024)Beiler, Cushing, Kirkpatrick, Schneider,
  Mukherjee, Marley, Marocco, \&
  Smart}]{beilerPreciseBolometricLuminosities2024}
Beiler, S.~A., Cushing, M.~C., Kirkpatrick, J.~D., {et~al.} 2024,
  \bibinfo{title}{Precise {Bolometric} {Luminosities} and {Effective}
  {Temperatures} of 23 {Late}-{T} and {Y} {Dwarfs} {Obtained} with {JWST},} The
  Astrophysical Journal, 973, 107, \dodoi{10.3847/1538-4357/ad6301}

\bibitem[{B. Benneke {et~al.}(2024)Benneke, Roy, Coulombe, Radica, Piaulet,
  Ahrer, Pierrehumbert, Krissansen-Totton, Schlichting, Hu, Yang, Christie,
  Thorngren, Young, Pelletier, Knutson, Miguel, Evans-Soma, Dorn, Gagnebin,
  Fortney, Komacek, MacDonald, Raul, Cloutier, Acuna, Lafrenière, Cadieux,
  Doyon, Welbanks, \& Allart}]{bennekeJWSTRevealsCH$_4$2024}
Benneke, B., Roy, P.-A., Coulombe, L.-P., {et~al.} 2024, \bibinfo{title}{{JWST}
  {Reveals} {CH}\$\_4\$, {CO}\$\_2\$, and {H}\$\_2\${O} in a {Metal}-rich
  {Miscible} {Atmosphere} on a {Two}-{Earth}-{Radius} {Exoplanet},} arXiv.
\newblock \url{http://arxiv.org/abs/2403.03325}

\bibitem[{W.~M.~J. Best {et~al.}(2025)Best, Dupuy, Liu, Sanghi, Siverd, \&
  Zhang}]{bestUltracoolSheetPhotometryAstrometry2025}
Best, W. M.~J., Dupuy, T.~J., Liu, M.~C., {et~al.} 2025, \bibinfo{title}{The
  {UltracoolSheet}: {Photometry}, {Astrometry}, {Spectroscopy}, and
  {Multiplicity} for 4000+ {Ultracool} {Dwarfs} and {Imaged} {Exoplanets},},
  Version 2.1.0 Zenodo, \dodoi{10.5281/zenodo.15802304}

\bibitem[{B.~A. Biller {et~al.}(2024)Biller, Vos, Zhou, McCarthy, Tan,
  Crossfield, Whiteford, Suarez, Faherty, Manjavacas, Chen, Liu, Sutlieff,
  Limbach, Molliere, Dupuy, Oliveros-Gomez, Muirhead, Henning, Mace, Crouzet,
  Karalidi, Morley, Tremblin, \& Kataria}]{billerJWSTWeatherReport2024}
Biller, B.~A., Vos, J.~M., Zhou, Y., {et~al.} 2024, \bibinfo{title}{The {JWST}
  weather report from the nearest brown dwarfs {I}: multiperiod {JWST}
  {NIRSpec} + {MIRI} monitoring of the benchmark binary brown dwarf {WISE}
  {1049AB},} Monthly Notices of the Royal Astronomical Society, 532, 2207,
  \dodoi{10.1093/mnras/stae1602}

\bibitem[{V. Bourrier {et~al.}(2020)Bourrier, Ehrenreich, Lendl, Cretignier,
  Allart, Dumusque, Cegla, Suarez-Mascareno, Wyttenbach, Hoeijmakers, Melo,
  Kuntzer, Astudillo-Defru, Giles, Heng, Kitzmann, Lavie, Lovis, Murgas,
  Nascimbeni, Pepe, Pino, Segransan, \&
  Udry}]{bourrierHotExoplanetAtmospheres2020}
Bourrier, V., Ehrenreich, D., Lendl, M., {et~al.} 2020, \bibinfo{title}{Hot
  {Exoplanet} {Atmospheres} {Resolved} with {Transit} {Spectroscopy} ({HEARTS})
  {III}. {Atmospheric} structure of the misaligned ultra-hot {Jupiter}
  {WASP}-121b,} Astronomy \& Astrophysics, 635, A205,
  \dodoi{10.1051/0004-6361/201936640}

\bibitem[{D. Broski-Laing {et~al.}(2026)Broski-Laing, Zhou, Lothringer, Apai,
  French, Casewell, Mayorga, Shin, Lew, Tan, Parmentier, Xu, \&
  Marley}]{broski-laingAsymmetricNightsideCO22026}
Broski-Laing, D., Zhou, Y., Lothringer, J.~D., {et~al.} 2026,
  \bibinfo{title}{Asymmetric nightside {CO2} features, inefficient heat
  transport, and precise evolutionary constraints: {Spectroscopic} phase curves
  reveal the past and present of a white dwarf-brown dwarf binary,} arXiv
  e-prints, arXiv:2606.30112, \dodoi{10.48550/arXiv.2606.30112}

\bibitem[{A.~J. Burgasser {et~al.}(2006)Burgasser, Geballe, Leggett,
  Kirkpatrick, \& Golimowski}]{burgasserUnifiedNearInfraredSpectral2006}
Burgasser, A.~J., Geballe, T.~R., Leggett, S.~K., Kirkpatrick, J.~D., \&
  Golimowski, D.~A. 2006, \bibinfo{title}{A {Unified} {Near}-{Infrared}
  {Spectral} {Classification} {Scheme} for {T} {Dwarfs},} The Astrophysical
  Journal, 637, 1067, \dodoi{10.1086/498563}

\bibitem[{A.~J. Burgasser {et~al.}(2025)Burgasser, Gonzales, Beiler, Visscher,
  Burningham, Mace, Faherty, Zhang, Sousa-Silva, Lodieu, Metchev, Meisner,
  Cushing, Schneider, Suarez, Hsu, Gerasimov, Aganze, \&
  Theissen}]{burgasserObservationUndepletedPhosphine2025}
Burgasser, A.~J., Gonzales, E.~C., Beiler, S.~A., {et~al.} 2025,
  \bibinfo{title}{Observation of undepleted phosphine in the atmosphere of a
  low-temperature brown dwarf,} Science, eadu0401,
  \dodoi{10.1126/science.adu0401}

\bibitem[{H. Bushouse {et~al.}(2025)Bushouse, Eisenhamer, Dencheva, Davies,
  Greenfield, Morrison, Hodge, Simon, Grumm, Droettboom, \&
  et~al.}]{bushouseJWSTCalibrationPipeline2025}
Bushouse, H., Eisenhamer, J., Dencheva, N., {et~al.} 2025,
  \bibinfo{title}{{JWST} {Calibration} {Pipeline},}, 1.19.1 Zenodo,
  \dodoi{10.5281/zenodo.16280965}

\bibitem[{G. Chabrier {et~al.}(2000)Chabrier, Baraffe, Allard, \&
  Hauschildt}]{chabrierEvolutionaryModelsVery2000}
Chabrier, G., Baraffe, I., Allard, F., \& Hauschildt, P. 2000,
  \bibinfo{title}{Evolutionary {Models} for {Very} {Low}-{Mass} {Stars} and
  {Brown} {Dwarfswith} {Dusty} {Atmospheres},} The Astrophysical Journal, 542,
  464, \dodoi{10.1086/309513}

\bibitem[{D. Charbonneau {et~al.}(2005)Charbonneau, Allen, Megeath, Torres,
  Alonso, Brown, Gilliland, Latham, Mandushev, O’Donovan, \&
  Sozzetti}]{charbonneauDetectionThermalEmission2005}
Charbonneau, D., Allen, L.~E., Megeath, S.~T., {et~al.} 2005,
  \bibinfo{title}{Detection of {Thermal} {Emission} from an {Extrasolar}
  {Planet},} The Astrophysical Journal, 626, 523, \dodoi{10.1086/429991}

\bibitem[{B. Charnay {et~al.}(2018)Charnay, Bézard, Baudino, Bonnefoy,
  Boccaletti, \& Galicher}]{charnaySelfconsistentCloudModel2018}
Charnay, B., Bézard, B., Baudino, J.-L., {et~al.} 2018, \bibinfo{title}{A
  {Self}-consistent {Cloud} {Model} for {Brown} {Dwarfs} and {Young} {Giant}
  {Exoplanets}: {Comparison} with {Photometric} and {Spectroscopic}
  {Observations},} The Astrophysical Journal, 854, 172,
  \dodoi{10.3847/1538-4357/aaac7d}

\bibitem[{M. Cohen {et~al.}(2003)Cohen, Wheaton, \&
  Megeath}]{cohenSpectralIrradianceCalibration2003}
Cohen, M., Wheaton, W.~A., \& Megeath, S.~T. 2003, \bibinfo{title}{Spectral
  {Irradiance} {Calibration} in the {Infrared}. {XIV}. {The} {Absolute}
  {Calibration} of {2MASS},} The Astronomical Journal, 126, 1090,
  \dodoi{10.1086/376474}

\bibitem[{L.-P. Coulombe {et~al.}(2023)Coulombe, Benneke, Challener, Piette,
  Wiser, Mansfield, MacDonald, Beltz, Feinstein, Radica, Savel, {Dos Santos},
  Bean, Parmentier, Wong, Rauscher, Komacek, Kempton, Tan, Hammond, Lewis,
  Line, Lee, Shivkumar, Crossfield, Nixon, Rackham, Wakeford, Welbanks, Zhang,
  Batalha, Berta-Thompson, Changeat, Désert, Espinoza, Goyal, Harrington,
  Knutson, Kreidberg, López-Morales, Shporer, Sing, Stevenson, Aggarwal,
  Ahrer, Alam, Bell, Blecic, Caceres, Carter, Casewell, Crouzet, Cubillos,
  Decin, Fortney, Gibson, Heng, Henning, Iro, Kendrew, Lagage, Leconte, Lendl,
  Lothringer, Mancini, Mikal-Evans, Molaverdikhani, Nikolov, Ohno, Palle,
  Piaulet, Redfield, Roy, Tsai, Venot, \&
  Wheatley}]{coulombeBroadbandThermalEmission2023}
Coulombe, L.-P., Benneke, B., Challener, R., {et~al.} 2023, \bibinfo{title}{A
  broadband thermal emission spectrum of the ultra-hot {Jupiter} {WASP}-18b,}
  Nature, 620, 292, \dodoi{10.1038/s41586-023-06230-1}

\bibitem[{N.~B. Cowan \& E. Agol(2011)Cowan \&
  Agol}]{cowanStatisticsAlbedoHeat2011}
Cowan, N.~B., \& Agol, E. 2011, \bibinfo{title}{The statistics of albedo and
  heat recirculation on hot exoplanets,} The Astrophysical Journal, 729, 54,
  \dodoi{10.1088/0004-637X/729/1/54}

\bibitem[{M.~C. Cushing {et~al.}(2005)Cushing, Rayner, \&
  Vacca}]{cushing_infrared_2005}
Cushing, M.~C., Rayner, J.~T., \& Vacca, W.~D. 2005, \bibinfo{title}{An
  {Infrared} {Spectroscopic} {Sequence} of {M}, {L}, and {T} {Dwarfs}*,} The
  Astrophysical Journal, 623, 1115, \dodoi{10.1086/428040}

\bibitem[{S. {de Regt} {et~al.}(2026){de Regt}, Whiteford, Miles,
  {et~al.}}]{deregtNativeresolutionRetrievalsVHS2026}
{de Regt}, S., Whiteford, N., Miles, B.~E., {et~al.} 2026,
  \bibinfo{title}{Native-resolution retrievals of {VHS} 1256-1257 b spanning
  the {JWST}/{NIRSpec} wavelength range: {Chemical} composition of a partially
  cloudy atmosphere,} arXiv e-prints, arXiv:2607.00952

\bibitem[{D. Deming {et~al.}(2006)Deming, Harrington, Seager, \&
  Richardson}]{demingStrongInfraredEmission2006}
Deming, D., Harrington, J., Seager, S., \& Richardson, L.~J. 2006,
  \bibinfo{title}{Strong {Infrared} {Emission} from the {Extrasolar} {Planet}
  {HD} 189733b,} The Astrophysical Journal, 644, 560, \dodoi{10.1086/503358}

\bibitem[{T.~J. Dupuy \& M.~C. Liu(2017)Dupuy \&
  Liu}]{dupuyIndividualDynamicalMasses2017}
Dupuy, T.~J., \& Liu, M.~C. 2017, \bibinfo{title}{Individual {Dynamical}
  {Masses} of {Ultracool} {Dwarfs},} The Astrophysical Journal Supplement
  Series, 231, 15, \dodoi{10.3847/1538-4365/aa5e4c}

\bibitem[{T.~M. Evans-Soma {et~al.}(2025)Evans-Soma, Sing, Barstow, Piette,
  Taylor, Lothringer, Reggiani, Goyal, Ahrer, Mayne, Rustamkulov, Kataria,
  Christie, Gapp, Dong, Foreman-Mackey, Hattori, \&
  Marley}]{evans-somaSiOSuperstellarRatio2025}
Evans-Soma, T.~M., Sing, D.~K., Barstow, J.~K., {et~al.} 2025,
  \bibinfo{title}{{SiO} and a super-stellar {C}/{O} ratio in the atmosphere of
  the giant exoplanet {WASP}-121 b,} Nature Astronomy, 9, 845,
  \dodoi{10.1038/s41550-025-02513-x}

\bibitem[{J.~K. Faherty {et~al.}(2016)Faherty, Riedel, Cruz, Gagne, Filippazzo,
  Lambrides, Fica, Weinberger, Thorstensen, Tinney, Baldassare, Lemonier, \&
  Rice}]{faherty_population_2016}
Faherty, J.~K., Riedel, A.~R., Cruz, K.~L., {et~al.} 2016,
  \bibinfo{title}{Population {Properties} of {Brown} {Dwarf} {Analogs} to
  {Exoplanets},} The Astrophysical Journal Supplement Series, 225, 10,
  \dodoi{10.3847/0067-0049/225/1/10}

\bibitem[{J.~C. Filippazzo {et~al.}(2015)Filippazzo, Rice, Faherty, Cruz,
  Van~Gordon, \& Looper}]{filippazzoFundamentalParametersSpectral2015}
Filippazzo, J.~C., Rice, E.~L., Faherty, J., {et~al.} 2015,
  \bibinfo{title}{Fundamental {Parameters} and {Spectral} {Energy}
  {Distributions} of {Young} and {Field} {Age} {Objects} with {Masses}
  {Spanning} the {Stellar} to {Planetary} {Regime},} The Astrophysical Journal,
  810, 158, \dodoi{10.1088/0004-637X/810/2/158}

\bibitem[{J. Fortney {et~al.}(2008)Fortney, Lodders, Marley, \&
  Freedman}]{fortneyUnifiedTheoryAtmospheres2008}
Fortney, J., Lodders, K., Marley, M., \& Freedman, R. 2008, \bibinfo{title}{A
  {Unified} {Theory} for the {Atmospheres} of the {Hot} and {Very} {Hot}
  {Jupiters}: {Two} {Classes} of {Irradiated} {Atmospheres},} The Astrophysical
  Journal, 678, 1419, \dodoi{10.1086/528370}

\bibitem[{J.~J. Fortney {et~al.}(2020)Fortney, Visscher, Marley, Hood, Line,
  Thorngren, Freedman, \& Lupu}]{fortneyEquilibriumTemperatureHow2020}
Fortney, J.~J., Visscher, C., Marley, M.~S., {et~al.} 2020,
  \bibinfo{title}{Beyond {Equilibrium} {Temperature}: {How} the
  {Atmosphere}/{Interior} {Connection} {Affects} the {Onset} of {Methane},
  {Ammonia}, and {Clouds} in {Warm} {Transiting} {Giant} {Planets},} The
  Astronomical Journal, 160, 288, \dodoi{10.3847/1538-3881/abc5bd}

\bibitem[{G. Fu {et~al.}(2017)Fu, Deming, Knutson, Madhusudhan, Mandell, \&
  Fraine}]{fuStatisticalAnalysisHubble2017}
Fu, G., Deming, D., Knutson, H., {et~al.} 2017, \bibinfo{title}{Statistical
  {Analysis} of {Hubble}/{WFC3} {Transit} {Spectroscopy} of {Extrasolar}
  {Planets},} The Astrophysical Journal Letters, 847, L22,
  \dodoi{10.3847/2041-8213/aa8e40}

\bibitem[{G. Fu {et~al.}(2024)Fu, Welbanks, Deming, Inglis, Zhang, Lothringer,
  Ih, Moses, Schlawin, Knutson, Henry, Greene, Sing, Savel, Kempton, Louie,
  Line, \& Nixon}]{fuHydrogenSulfideMetalenriched2024}
Fu, G., Welbanks, L., Deming, D., {et~al.} 2024, \bibinfo{title}{Hydrogen
  sulfide and metal-enriched atmosphere for a {Jupiter}-mass exoplanet,}
  Nature, 632, 752, \dodoi{10.1038/s41586-024-07760-y}

\bibitem[{J. Gagn{\'e} {et~al.}(2026)Gagn{\'e}, Faherty, {Ruiz Diaz},
  {et~al.}}]{gagneSPHERExPipelineSpectral2026}
Gagn{\'e}, J., Faherty, J.~K., {Ruiz Diaz}, A., {et~al.} 2026,
  \bibinfo{title}{A {SPHEREx} {Pipeline} and {Spectral} {Library} for
  {Ultracool} {Dwarfs},} arXiv e-prints, arXiv:2604.22012

\bibitem[{P. Gao {et~al.}(2020)Gao, Thorngren, Lee, Fortney, Morley, Wakeford,
  Powell, Stevenson, \& Zhang}]{gaoAerosolCompositionHot2020}
Gao, P., Thorngren, D.~P., Lee, G. K.~H., {et~al.} 2020,
  \bibinfo{title}{Aerosol {Composition} of {Hot} {Giant} {Exoplanets}
  {Dominated} by {Silicates} and {Hydrocarbon} {Hazes},} Nature Astronomy, 4,
  951, \dodoi{10.1038/s41550-020-1114-3}

\bibitem[{M. Gillon {et~al.}(2016)Gillon, Jehin, Lederer, Delrez, de~Wit,
  Burdanov, Van~Grootel, Burgasser, Triaud, Opitom, Demory, Sahu,
  Bardalez~Gagliuffi, Magain, \& Queloz}]{gillonTemperateEarthsizedPlanets2016}
Gillon, M., Jehin, E., Lederer, S.~M., {et~al.} 2016, \bibinfo{title}{Temperate
  {Earth}-sized planets transiting a nearby ultracool dwarf star,} Nature, 533,
  221, \dodoi{10.1038/nature17448}

\bibitem[{J.~M. Goyal {et~al.}(2018)Goyal, Mayne, Sing, Drummond, Tremblin,
  Amundsen, Evans, Carter, Spake, Baraffe, Nikolov, Manners, Chabrier, \&
  Hebrard}]{goyalLibraryATMOForward2018}
Goyal, J.~M., Mayne, N., Sing, D.~K., {et~al.} 2018, \bibinfo{title}{A library
  of {ATMO} forward model transmission spectra for hot {Jupiter} exoplanets,}
  Monthly Notices of the Royal Astronomical Society, 474, 5158,
  \dodoi{10.1093/mnras/stx3015}

\bibitem[{K.~K.~W. Hoch {et~al.}(2024)Hoch, Theissen, Barman, Perrin, Ruffio,
  Rickman, Konopacky, Manjavacas, Balmer, Pueyo, Kammerer, van~der Marel,
  Lewis, Girard, Seager, Clampin, \& Mountain}]{hochJWSTTSTHighContrast2024}
Hoch, K. K.~W., Theissen, C.~A., Barman, T.~S., {et~al.} 2024,
  \bibinfo{title}{{JWST}-{TST} {High} {Contrast}: {Spectroscopic}
  {Characterization} of the {Benchmark} {Brown} {Dwarf} {HD} 19467 {B} with the
  {NIRSpec} {Integral} {Field} {Spectrograph},} The Astronomical Journal, 168,
  187, \dodoi{10.3847/1538-3881/ad6cd3}

\bibitem[{K.~K.~W. Hoch {et~al.}(2025)Hoch, Rowland, Petrus, Nasedkin,
  Ingebretsen, Kammerer, Perrin, D’Orazi, Balmer, Barman, Bonnefoy, Chauvin,
  Chen, De~Rosa, Girard, Gonzales, Kenworthy, Konopacky, Macintosh, Moran,
  Morley, Palma-Bifani, Pueyo, Ren, Rickman, Ruffio, Theissen, Ward-Duong, \&
  Zhang}]{hochSilicateCloudsCircumplanetary2025}
Hoch, K. K.~W., Rowland, M., Petrus, S., {et~al.} 2025,
  \bibinfo{title}{Silicate clouds and a circumplanetary disk in the {YSES}-1
  exoplanet system,} Nature, 643, 938, \dodoi{10.1038/s41586-025-09174-w}

\bibitem[{M. Ikoma \& H. Kobayashi(2025)Ikoma \&
  Kobayashi}]{ikomaFormationGiantPlanets2025}
Ikoma, M., \& Kobayashi, H. 2025, \bibinfo{title}{Formation of {Giant}
  {Planets},} arXiv, \dodoi{10.48550/arXiv.2504.04090}

\bibitem[{P. Jakobsen {et~al.}(2022)Jakobsen, Ferruit, {Alves de Oliveira},
  Arribas, Bagnasco, Barho, Beck, Birkmann, B{\"o}ker, Bunker, Charlot, \&
  et~al.}]{jakobsenNearInfraredSpectrographNIRSpec2022}
Jakobsen, P., Ferruit, P., {Alves de Oliveira}, C., {et~al.} 2022,
  \bibinfo{title}{The {Near-Infrared} {Spectrograph} ({NIRSpec}) on the {James}
  {Webb} {Space} {Telescope}. {I}. {Overview} of the instrument and its
  capabilities,} Astronomy \& Astrophysics, 661, A80,
  \dodoi{10.1051/0004-6361/202142663}

\bibitem[{J.~S. Jenkins {et~al.}(2020)Jenkins, Díaz, Kurtovic, Espinoza,
  Vines, Rojas, Brahm, Torres, Cortés-Zuleta, Soto, Lopez, King, Wheatley,
  Winn, Ciardi, Ricker, Vanderspek, Latham, Seager, Jenkins, Beichman, Bieryla,
  Burke, Christiansen, Henze, Klaus, McCauliff, Mori, Narita, Nishiumi, Tamura,
  de~Leon, Quinn, Villaseñor, Vezie, Lissauer, Collins, Collins, Isopi,
  Mallia, Ercolino, Petrovich, Jordán, Acton, Armstrong, Bayliss, Bouchy,
  Belardi, Bryant, Burleigh, Cabrera, Casewell, Chaushev, Cooke, Eigmüller,
  Erikson, Foxell, Gänsicke, Gill, Gillen, Günther, Goad, Hooton, Jackman,
  Louden, McCormac, Moyano, Nielsen, Pollacco, Queloz, Rauer, Raynard, Smith,
  Tilbrook, Titz-Weider, Turner, Udry, Walker, Watson, West, Palle, Ziegler,
  Law, \& Mann}]{jenkinsUltrahotNeptuneNeptune2020}
Jenkins, J.~S., Díaz, M.~R., Kurtovic, N.~T., {et~al.} 2020,
  \bibinfo{title}{An ultrahot {Neptune} in the {Neptune} desert,} Nature
  Astronomy, 4, 1148, \dodoi{10.1038/s41550-020-1142-z}

\bibitem[{T.~D. Kennedy {et~al.}(2024)Kennedy, Rauscher, Malsky, Roman, \&
  Beltz}]{kennedy_radiatively_2024}
Kennedy, T.~D., Rauscher, E., Malsky, I., Roman, M.~T., \& Beltz, H. 2024,
  \bibinfo{title}{Radiatively {Active} {Clouds} and {Magnetic} {Effects}
  {Explored} in a {Grid} of {Hot} {Jupiter} {GCMs},} The Astrophysical Journal,
  978, 82, \dodoi{10.3847/1538-4357/ad9394}

\bibitem[{R.~R. King {et~al.}(2010)King, McCaughrean, Homeier, Allard, Scholz,
  \& Lodieu}]{kingIndiBaBb2010}
King, R.~R., McCaughrean, M.~J., Homeier, D., {et~al.} 2010,
  \bibinfo{title}{Indi {Ba}, {Bb}: a detailed study of the nearest known brown
  dwarfs,} Astronomy \& Astrophysics, 510, A99,
  \dodoi{10.1051/0004-6361/200912981}

\bibitem[{J.~D. Kirkpatrick {et~al.}(2006)Kirkpatrick, Barman, Burgasser,
  McGovern, McLean, Tinney, \& Lowrance}]{kirkpatrick_discovery_2006}
Kirkpatrick, J.~D., Barman, T.~S., Burgasser, A.~J., {et~al.} 2006,
  \bibinfo{title}{Discovery of a {Very} {Young} {Field} {L} {Dwarf}, {2MASS}
  {J01415823}–4633574*,} The Astrophysical Journal, 639, 1120,
  \dodoi{10.1086/499622}

\bibitem[{J.~D. Kirkpatrick {et~al.}(2012)Kirkpatrick, Gelino, Cushing, Mace,
  Griffith, Skrutskie, Marsh, Wright, Eisenhardt, McLean, Mainzer, Burgasser,
  Tinney, Parker, \& Salter}]{kirkpatrick_further_2012}
Kirkpatrick, J.~D., Gelino, C.~R., Cushing, M.~C., {et~al.} 2012,
  \bibinfo{title}{Further {Defining} {Spectral} {Type} "{Y}" and {Exploring}
  the {Low}-mass {End} of the {Field} {Brown} {Dwarf} {Mass} {Function},} The
  Astrophysical Journal, 753, 156, \dodoi{10.1088/0004-637X/753/2/156}

\bibitem[{T.~D. Komacek \& A.~P. Showman(2016)Komacek \&
  Showman}]{komacekAtmosphericCirculationHot2016}
Komacek, T.~D., \& Showman, A.~P. 2016, \bibinfo{title}{Atmospheric
  {Circulation} of {Hot} {Jupiters}: {Dayside}-{Nightside} {Temperature}
  {Differences},} The Astrophysical Journal, 821, 16,
  \dodoi{10.3847/0004-637X/821/1/16}

\bibitem[{T.~D. Komacek {et~al.}(2017)Komacek, Showman, \&
  Tan}]{komacekAtmosphericCirculationHot2017}
Komacek, T.~D., Showman, A.~P., \& Tan, X. 2017, \bibinfo{title}{Atmospheric
  {Circulation} of {Hot} {Jupiters}: {Dayside}–{Nightside} {Temperature}
  {Differences}. {II}. {Comparison} with {Observations},} The Astrophysical
  Journal, 835, 198, \dodoi{10.3847/1538-4357/835/2/198}

\bibitem[{L. Kreidberg(2015)Kreidberg}]{kreidbergBatmanBAsicTransit2015}
Kreidberg, L. 2015, \bibinfo{title}{batman: {BAsic} {Transit} {Model}
  {cAlculatioN} in {Python},} Publications of the Astronomical Society of the
  Pacific, 127, 1161, \dodoi{10.1086/683602}

\bibitem[{K. Lodders \& B. Fegley(2002)Lodders \&
  Fegley}]{loddersAtmosphericChemistryGiant2002}
Lodders, K., \& Fegley, B. 2002, \bibinfo{title}{Atmospheric {Chemistry} in
  {Giant} {Planets}, {Brown} {Dwarfs}, and {Low}-{Mass} {Dwarf} {Stars}: {I}.
  {Carbon}, {Nitrogen}, and {Oxygen},} Icarus, 155, 393,
  \dodoi{10.1006/icar.2001.6740}

\bibitem[{K. Lodders {et~al.}(2009)Lodders, Palme, \&
  Gail}]{lodders44AbundancesElements2009}
Lodders, K., Palme, H., \& Gail, H.-P. 2009, in Solar {System}, ed.
  J.~Trümper, Vol.~4B (Berlin, Heidelberg: Springer Berlin Heidelberg),
  712--770, \dodoi{10.1007/978-3-540-88055-4_34}

\bibitem[{J.~D. Lothringer {et~al.}(2018)Lothringer, Barman, \&
  Koskinen}]{lothringerExtremelyIrradiatedHot2018}
Lothringer, J.~D., Barman, T., \& Koskinen, T. 2018, \bibinfo{title}{Extremely
  {Irradiated} {Hot} {Jupiters}: {Non}-oxide {Inversions}, {H}
  $^{\textrm{−}}$ {Opacity}, and {Thermal} {Dissociation} of {Molecules},}
  The Astrophysical Journal, 866, 27, \dodoi{10.3847/1538-4357/aadd9e}

\bibitem[{J.~D. Lothringer {et~al.}(2025)Lothringer, Lowson, \&
  Fu}]{lothringer_library_2025}
Lothringer, J.~D., Lowson, N., \& Fu, G. 2025, \bibinfo{title}{The {Library} of
  {Exoplanet} {Atmospheric} {Composition} {Measurements}: {Population}-level
  {Trends} in {Exoplanet} {Composition} with {ExoComp},} The Astronomical
  Journal, 171, 31, \dodoi{10.3847/1538-3881/ae1b8e}

\bibitem[{J.~D. Lothringer {et~al.}(2022)Lothringer, Sing, Rustamkulov,
  Wakeford, Stevenson, Nikolov, Lavvas, Spake, \&
  Winch}]{lothringerUVAbsorptionSilicate2022}
Lothringer, J.~D., Sing, D.~K., Rustamkulov, Z., {et~al.} 2022,
  \bibinfo{title}{{UV} absorption by silicate cloud precursors in ultra-hot
  {Jupiter} {WASP}-178b,} Nature, 604, 49, \dodoi{10.1038/s41586-022-04453-2}

\bibitem[{A. Lueber {et~al.}(2026)Lueber, Kitzmann, \&
  Heng}]{lueberCloudsChemistryAcross2026}
Lueber, A., Kitzmann, D., \& Heng, K. 2026, \bibinfo{title}{Clouds and
  chemistry across the brown dwarf {T}--{Y} sequence: {Insights} from {JWST}
  atmospheric retrievals,} Astronomy \& Astrophysics, 707, A92,
  \dodoi{10.1051/0004-6361/202556207}

\bibitem[{K.~L. Luhman \& C. Alves~de Oliveira(2025)Luhman \& Alves~de
  Oliveira}]{luhmanNewSpectralClass2025}
Luhman, K.~L., \& Alves~de Oliveira, C. 2025, \bibinfo{title}{A {New}
  {Spectral} {Class} of {Brown} {Dwarfs} at the {Bottom} of the {IMF} in {IC}
  348,} The Astrophysical Journal Letters, 986, L14,
  \dodoi{10.3847/2041-8213/addc55}

\bibitem[{J. Lustig-Yaeger {et~al.}(2025)Lustig-Yaeger, Sotzen, Stevenson,
  Tsai, Challener, Goyal, Lewis, Louie, Mayorga, Valentine, Wakeford, Alderson,
  Allen, Fauchez, Glidden, Gressier, Hörst, Huang, Lin, Mandell, Mullens,
  Peacock, Schwieterman, Valenti, Mountain, Perrin, \& van~der
  Marel}]{lustig-yaegerJWSTTSTDREAMSNightside2025}
Lustig-Yaeger, J., Sotzen, K.~S., Stevenson, K.~B., {et~al.} 2025,
  \bibinfo{title}{{JWST}-{TST} {DREAMS}: {The} {Nightside} {Emission} and
  {Chemistry} of {WASP}-17b,} arXiv, \dodoi{10.48550/arXiv.2510.06169}

\bibitem[{J. Mang {et~al.}(2026)Mang, Batalha, Morley, Wogan, Mukherjee,
  Visscher, Marley, Fortney, Chubb, Gao, \& Malsky}]{mang_picaso_2026}
Mang, J., Batalha, N.~E., Morley, C.~V., {et~al.} 2026,
  \bibinfo{title}{{PICASO} 4.0: {Clouds} and {Photochemistry} in {Climate}
  {Models} of {Brown} {Dwarfs} and {Exoplanets},} The Astrophysical Journal,
  1000, 98, \dodoi{10.3847/1538-4357/ae47ff}

\bibitem[{E. Manjavacas {et~al.}(2019)Manjavacas, Apai, Zhou, Lew, Schneider,
  Metchev, Miles-Páez, Radigan, Marley, Cowan, Karalidi, Burgasser, Bedin,
  Lowrance, \& Kauffmann}]{manjavacasCloudAtlasHubble2019}
Manjavacas, E., Apai, D., Zhou, Y., {et~al.} 2019, \bibinfo{title}{Cloud
  {Atlas}: {Hubble} {Space} {Telescope} {Near}-infrared {Spectral} {Library} of
  {Brown} {Dwarfs}, {Planetary}-mass {Companions}, and {Hot} {Jupiters},} The
  Astronomical Journal, 157, 101, \dodoi{10.3847/1538-3881/aaf88f}

\bibitem[{E. Manjavacas {et~al.}(2024)Manjavacas, Tremblin, Birkmann, Valenti,
  Alves~de Oliveira, Beck, Giardino, Lützgendorf, Rauscher, \&
  Sirianni}]{manjavacasMediumresolution09753Mm2024}
Manjavacas, E., Tremblin, P., Birkmann, S., {et~al.} 2024,
  \bibinfo{title}{Medium-resolution 0.97–5.3 μm {Spectra} of {Very} {Young}
  {Benchmark} {Brown} {Dwarfs} with {NIRSpec} on {Board} the {James} {Webb}
  {Space} {Telescope},} The Astronomical Journal, 167, 168,
  \dodoi{10.3847/1538-3881/ad2938}

\bibitem[{M. Mansfield {et~al.}(2018)Mansfield, Bean, Line, Parmentier,
  Kreidberg, Desert, Fortney, Stevenson, Arcangeli, \&
  Dragomir}]{mansfieldHSTWFC3Thermal2018}
Mansfield, M., Bean, J.~L., Line, M.~R., {et~al.} 2018, \bibinfo{title}{An
  {HST}/{WFC3} {Thermal} {Emission} {Spectrum} of the {Hot} {Jupiter}
  {HAT}-{P}-7b,} The Astronomical Journal, 156, 10,
  \dodoi{10.3847/1538-3881/aac497}

\bibitem[{M.~S. Marley {et~al.}(2010)Marley, Saumon, \&
  Goldblatt}]{marleyPATCHYCLOUDMODEL2010}
Marley, M.~S., Saumon, D., \& Goldblatt, C. 2010, \bibinfo{title}{A {PATCHY}
  {CLOUD} {MODEL} {FOR} {THE} {L} {TO} {T} {DWARF} {TRANSITION},} The
  Astrophysical Journal Letters, 723, L117,
  \dodoi{10.1088/2041-8205/723/1/L117}

\bibitem[{M.~S. Marley {et~al.}(2021)Marley, Saumon, Visscher, Lupu, Freedman,
  Morley, Fortney, Seay, Smith, Teal, \& Wang}]{marleySonoraBrownDwarf2021}
Marley, M.~S., Saumon, D., Visscher, C., {et~al.} 2021, \bibinfo{title}{The
  {Sonora} {Brown} {Dwarf} {Atmosphere} and {Evolution} {Models}. {I}. {Model}
  {Description} and {Application} to {Cloudless} {Atmospheres} in {Rainout}
  {Chemical} {Equilibrium},} The Astrophysical Journal, 920, 85,
  \dodoi{10.3847/1538-4357/ac141d}

\bibitem[{C. Marois {et~al.}(2008)Marois, Macintosh, Barman, Zuckerman, Song,
  Patience, Lafrenière, \& Doyon}]{marois_direct_2008}
Marois, C., Macintosh, B., Barman, T., {et~al.} 2008, \bibinfo{title}{Direct
  {Imaging} of {Multiple} {Planets} {Orbiting} the {Star} {HR} 8799,} Science,
  322, 1348, \dodoi{10.1126/science.1166585}

\bibitem[{A.~M. McCarthy {et~al.}(2025)McCarthy, Vos, Muirhead, Biller, Morley,
  Faherty, Burningham, Calamari, Cowan, Cruz, Gonzales, Limbach, Liu, Nasedkin,
  Suárez, Tan, O’Toole, Visscher, Whiteford, \&
  Zhou}]{mccarthyJWSTWeatherReport2025}
McCarthy, A.~M., Vos, J.~M., Muirhead, P.~S., {et~al.} 2025,
  \bibinfo{title}{The {JWST} {Weather} {Report} from the {Isolated} {Exoplanet}
  {Analog} {SIMP} 0136+0933: {Pressure}-dependent {Variability} {Driven} by
  {Multiple} {Mechanisms},} The Astrophysical Journal Letters, 981, L22,
  \dodoi{10.3847/2041-8213/ad9eaf}

\bibitem[{B.~E. Miles {et~al.}(2023)Miles, Biller, Patapis, Worthen, Rickman,
  Hoch, Skemer, Perrin, Whiteford, Chen, Sargent, Mukherjee, Morley, Moran,
  Bonnefoy, Petrus, Carter, Choquet, Hinkley, Ward-Duong, Leisenring,
  Millar-Blanchaer, Pueyo, Ray, Sallum, Stapelfeldt, Stone, Wang, Absil,
  Balmer, Boccaletti, Bonavita, Booth, Bowler, Chauvin, Christiaens, Currie,
  Danielski, Fortney, Girard, Grady, Greenbaum, Henning, Hines, Janson, Kalas,
  Kammerer, Kennedy, Kenworthy, Kervella, Lagage, Lew, Liu, Macintosh, Marino,
  Marley, Marois, Matthews, Matthews, Mawet, McElwain, Metchev, Meyer,
  Molliere, Pantin, Quirrenbach, Rebollido, Ren, Schneider, Vasist, Wyatt,
  Zhou, Briesemeister, Bryan, Calissendorff, Cantalloube, Cugno, De~Furio,
  Dupuy, Factor, Faherty, Fitzgerald, Franson, Gonzales, Hood, Howe, Kraus,
  Kuzuhara, Lagrange, Lawson, Lazzoni, Liu, Llop-Sayson, Lloyd, Martinez,
  Mazoyer, Quanz, Redai, Samland, Schlieder, Tamura, Tan, Uyama, Vigan, Vos,
  Wagner, Wolff, Ygouf, Zhang, Zhang, \&
  Zhang}]{milesJWSTEarlyreleaseScience2023}
Miles, B.~E., Biller, B.~A., Patapis, P., {et~al.} 2023, \bibinfo{title}{The
  {JWST} {Early}-release {Science} {Program} for {Direct} {Observations} of
  {Exoplanetary} {Systems} {II}: {A} 1 to 20 μm {Spectrum} of the
  {Planetary}-mass {Companion} {VHS} 1256–1257 b,} The Astrophysical Journal
  Letters, 946, L6, \dodoi{10.3847/2041-8213/acb04a}

\bibitem[{P. Mollière {et~al.}(2025)Mollière, Kühnle, Matthews, Henning,
  Min, Patapis, Lagage, Waters, Güdel, Jäger, Zhang, Decin, Biller, Absil,
  Argyriou, Barrado, Cossou, Glasse, Olofsson, Pye, Rouan, Samland,
  Scheithauer, Tremblin, Whiteford, van Dishoeck, Östlin, \&
  Ray}]{molliereEvidenceSiOCloud2025}
Mollière, P., Kühnle, H., Matthews, E.~C., {et~al.} 2025,
  \bibinfo{title}{Evidence for {SiO} cloud nucleation in the rogue planet {PSO}
  {J318},} Astronomy \& Astrophysics, 703, A79,
  \dodoi{10.1051/0004-6361/202555732}

\bibitem[{K. Morel {et~al.}(2025)Morel, Coulombe, Rowe, Lafrenière, Albert,
  Artigau, Cowan, Dang, Radica, Taylor, Piaulet-Ghorayeb, Roy, Benneke,
  Darveau-Bernier, Pelletier, Doyon, Johnstone, Langeveld, Allart, Flagg, \&
  Turner}]{morelModerateAlbedoReflecting2025}
Morel, K., Coulombe, L.-P., Rowe, J.~F., {et~al.} 2025, \bibinfo{title}{A
  {Moderate} {Albedo} from {Reflecting} {Aerosols} on the {Dayside} of
  {WASP}-80 b {Revealed} by {JWST}/{NIRISS} {Eclipse} {Spectroscopy},} The
  Astronomical Journal, 169, 277, \dodoi{10.3847/1538-3881/adc43f}

\bibitem[{C.~V. Morley {et~al.}(2024)Morley, Mukherjee, Marley, Fortney,
  Visscher, Lupu, Gharib-Nezhad, Thorngren, Freedman, \&
  Batalha}]{morleySonoraSubstellarAtmosphere2024}
Morley, C.~V., Mukherjee, S., Marley, M.~S., {et~al.} 2024, \bibinfo{title}{The
  {Sonora} {Substellar} {Atmosphere} {Models}. {III}. {Diamondback}:
  {Atmospheric} {Properties}, {Spectra}, and {Evolution} for {Warm} {Cloudy}
  {Substellar} {Objects},} The Astrophysical Journal, 975, 59,
  \dodoi{10.3847/1538-4357/ad71d5}

\bibitem[{S. Mukherjee {et~al.}(2023)Mukherjee, Batalha, Fortney, \&
  Marley}]{mukherjeePICASO30Onedimensional2023}
Mukherjee, S., Batalha, N.~E., Fortney, J.~J., \& Marley, M.~S. 2023,
  \bibinfo{title}{{PICASO} 3.0: {A} {One}-dimensional {Climate} {Model} for
  {Giant} {Planets} and {Brown} {Dwarfs},} The Astrophysical Journal, 942, 71,
  \dodoi{10.3847/1538-4357/ac9f48}

\bibitem[{S. Mukherjee {et~al.}(2022)Mukherjee, Fortney, Batalha, Karalidi,
  Marley, Visscher, Miles, \& Skemer}]{mukherjeeProbingExtentVertical2022}
Mukherjee, S., Fortney, J.~J., Batalha, N.~E., {et~al.} 2022,
  \bibinfo{title}{Probing the {Extent} of {Vertical} {Mixing} in {Brown}
  {Dwarf} {Atmospheres} with {Disequilibrium} {Chemistry},} The Astrophysical
  Journal, 938, 107, \dodoi{10.3847/1538-4357/ac8dfb}

\bibitem[{S. Mukherjee {et~al.}(2024)Mukherjee, Fortney, Morley, Batalha,
  Marley, Karalidi, Visscher, Lupu, Freedman, \&
  Gharib-Nezhad}]{mukherjeeSonoraSubstellarAtmosphere2024}
Mukherjee, S., Fortney, J.~J., Morley, C.~V., {et~al.} 2024,
  \bibinfo{title}{The {Sonora} {Substellar} {Atmosphere} {Models}. {IV}. {Elf}
  {Owl}: {Atmospheric} {Mixing} and {Chemical} {Disequilibrium} with {Varying}
  {Metallicity} and {C}/{O} {Ratios},} The Astrophysical Journal, 963, 73,
  \dodoi{10.3847/1538-4357/ad18c2}

\bibitem[{S. Mukherjee {et~al.}(2025)Mukherjee, Schlawin, Bell, Fortney,
  Beatty, Greene, Ohno, Murphy, Parmentier, Line, Welbanks, Wiser, \&
  Rieke}]{mukherjeeJWSTPanchromaticThermal2025}
Mukherjee, S., Schlawin, E., Bell, T.~J., {et~al.} 2025, \bibinfo{title}{A
  {JWST} {Panchromatic} {Thermal} {Emission} {Spectrum} of the {Warm} {Neptune}
  {Archetype} {GJ} 436b,} The Astrophysical Journal Letters, 982, L39,
  \dodoi{10.3847/2041-8213/adba46}

\bibitem[{T. Nakajima {et~al.}(1995)Nakajima, Oppenheimer, Kulkarni,
  Golimowski, Matthews, \& Durrance}]{nakajimaDiscoveryCoolBrown1995}
Nakajima, T., Oppenheimer, B.~R., Kulkarni, S.~R., {et~al.} 1995,
  \bibinfo{title}{Discovery of a cool brown dwarf,} Nature, 378, 463,
  \dodoi{10.1038/378463a0}

\bibitem[{V. Parmentier {et~al.}(2026)Parmentier, Stevenson, Welbanks, Taylor,
  Schlawin, Coulombe, Tang, Line, Shivkumar, Tan, Bean, D{\'e}sert, Fortney,
  Gao, Hammond, Kempton, Komacek, \&
  Weiner~Mansfield}]{parmentierHorizontalTransport2026}
Parmentier, V., Stevenson, K.~B., Welbanks, L., {et~al.} 2026,
  \bibinfo{title}{Horizontal transport as a source of disequilibrium chemistry
  on the nightside of a hot exoplanet,} Nature Astronomy,
  \dodoi{10.1038/s41550-026-02845-2}

\bibitem[{D. P{\'e}rez-Becker \& A.~P. Showman(2013)P{\'e}rez-Becker \&
  Showman}]{perezbeckerAtmosphericHeatRedistribution2013}
P{\'e}rez-Becker, D., \& Showman, A.~P. 2013, \bibinfo{title}{Atmospheric
  {Heat} {Redistribution} on {Hot} {Jupiters},} The Astrophysical Journal, 776,
  134, \dodoi{10.1088/0004-637X/776/2/134}

\bibitem[{M.~W. Phillips {et~al.}(2020)Phillips, Tremblin, Baraffe, Chabrier,
  Allard, Spiegelman, Goyal, Drummond, \&
  H{\'e}brard}]{phillipsNewSetAtmosphere2020}
Phillips, M.~W., Tremblin, P., Baraffe, I., {et~al.} 2020, \bibinfo{title}{A
  new set of atmosphere and evolution models for cool {T}--{Y} brown dwarfs and
  giant exoplanets,} Astronomy \& Astrophysics, 637, A38,
  \dodoi{10.1051/0004-6361/201937381}

\bibitem[{A. Pr{\v s}a {et~al.}(2016)Pr{\v s}a, Harmanec, Torres, Mamajek,
  Asplund, \& et~al.}]{prsaNominalValuesSelected2016}
Pr{\v s}a, A., Harmanec, P., Torres, G., {et~al.} 2016, \bibinfo{title}{Nominal
  {Values} for {Selected} {Solar} and {Planetary} {Quantities}: {IAU} 2015
  {Resolution} {B3},} The Astronomical Journal, 152, 41,
  \dodoi{10.3847/0004-6256/152/2/41}

\bibitem[{R. Rebolo {et~al.}(1995)Rebolo, {Zapatero Osorio}, \&
  Mart{\'i}n}]{reboloDiscoveryBrownDwarf1995}
Rebolo, R., {Zapatero Osorio}, M.~R., \& Mart{\'i}n, E.~L. 1995,
  \bibinfo{title}{Discovery of a brown dwarf in the {Pleiades} star cluster,}
  Nature, 377, 129, \dodoi{10.1038/377129a0}

\bibitem[{H.~N. Russell(1914)Russell}]{russell_relations_1914}
Russell, H.~N. 1914, \bibinfo{title}{Relations {Between} the {Spectra} and
  {Other} {Characteristics} of the {Stars},} Popular Astronomy, 22, 275.
\newblock \url{https://ui.adsabs.harvard.edu/abs/1914PA.....22..275R}

\bibitem[{S. Saha \& J.~S. Jenkins(2025)Saha \&
  Jenkins}]{sahaHighlyCarbonRichDayside2025}
Saha, S., \& Jenkins, J.~S. 2025, \bibinfo{title}{Dayside {Clouds} and an
  {Elevated} {C}/{O} {Ratio} in the {Atmosphere} of the {Ultra}-hot {Jupiter}
  {WASP}-19b,} arXiv, \dodoi{10.48550/arXiv.2507.02797}

\bibitem[{A. Sanghi {et~al.}(2023)Sanghi, Liu, Best, Dupuy, Siverd, Zhang,
  Hurt, Magnier, Aller, \& Deacon}]{sanghiHawaiiInfraredParallax2023}
Sanghi, A., Liu, M.~C., Best, W. M.~J., {et~al.} 2023, \bibinfo{title}{The
  {Hawaii} {Infrared} {Parallax} {Program}. {VI}. {The} {Fundamental}
  {Properties} of 1000+ {Ultracool} {Dwarfs} and {Planetary}-mass {Objects}
  {Using} {Optical} to {Mid}-infrared {Spectral} {Energy} {Distributions} and
  {Comparison} to {BT}-{Settl} and {ATMO} 2020 {Model} {Atmospheres},} The
  Astrophysical Journal, 959, 63, \dodoi{10.3847/1538-4357/acff66}

\bibitem[{D. Saumon \& M.~S. Marley(2008)Saumon \&
  Marley}]{saumonEvolutionDwarfsColorMagnitude2008}
Saumon, D., \& Marley, M.~S. 2008, \bibinfo{title}{The {Evolution} of {L} and
  {T} {Dwarfs} in {Color}-{Magnitude} {Diagrams},} The Astrophysical Journal,
  689, 1327, \dodoi{10.1086/592734}

\bibitem[{E. Schlawin {et~al.}(2024)Schlawin, Mukherjee, Ohno, Bell, Beatty,
  Greene, Line, Challener, Parmentier, Fortney, Rauscher, Wiser, Welbanks,
  Murphy, Edelman, Batalha, Moran, Mehta, \&
  Rieke}]{schlawinMultipleCluesDayside2024}
Schlawin, E., Mukherjee, S., Ohno, K., {et~al.} 2024, \bibinfo{title}{Multiple
  {Clues} for {Dayside} {Aerosols} and {Temperature} {Gradients} in {WASP}-69 b
  from a {Panchromatic} {JWST} {Emission} {Spectrum},} The Astronomical
  Journal, 168, 104, \dodoi{10.3847/1538-3881/ad58e0}

\bibitem[{S. Sorahana \& I. Yamamura(2012)Sorahana \&
  Yamamura}]{sorahanaAKARIOBSERVATIONSBROWN2012}
Sorahana, S., \& Yamamura, I. 2012, \bibinfo{title}{{AKARI} {OBSERVATIONS} {OF}
  {BROWN} {DWARFS}. {III}. {CO}, {CO2}, {AND} {CH4} {FUNDAMENTAL} {BANDS} {AND}
  {PHYSICAL} {PARAMETERS},} The Astrophysical Journal, 760, 151,
  \dodoi{10.1088/0004-637X/760/2/151}

\bibitem[{G. Suárez \& S. Metchev(2022)Suárez \&
  Metchev}]{suarezUltracoolDwarfsObserved2022}
Suárez, G., \& Metchev, S. 2022, \bibinfo{title}{Ultracool dwarfs observed
  with the \textit{{Spitzer}} infrared spectrograph – {II}. {Emergence} and
  sedimentation of silicate clouds in {L} dwarfs, and analysis of the full
  {M5}–{T9} field dwarf spectroscopic sample,} Monthly Notices of the Royal
  Astronomical Society, 513, 5701, \dodoi{10.1093/mnras/stac1205}

\bibitem[{D. Thorngren {et~al.}(2019)Thorngren, Gao, \&
  Fortney}]{thorngrenIntrinsicTemperatureRadiative2019}
Thorngren, D., Gao, P., \& Fortney, J.~J. 2019, \bibinfo{title}{The {Intrinsic}
  {Temperature} and {Radiative}–{Convective} {Boundary} {Depth} in the
  {Atmospheres} of {Hot} {Jupiters},} The Astrophysical Journal Letters, 884,
  L6, \dodoi{10.3847/2041-8213/ab43d0}

\bibitem[{S.-M. Tsai {et~al.}(2023)Tsai, Lee, Powell, Gao, Zhang, Moses,
  Hébrard, Venot, Parmentier, Jordan, Hu, Alam, Alderson, Batalha, Bean,
  Benneke, Bierson, Brady, Carone, Carter, Chubb, Inglis, Leconte, Line,
  López-Morales, Miguel, Molaverdikhani, Rustamkulov, Sing, Stevenson,
  Wakeford, Yang, Aggarwal, Baeyens, Barat, de~Val-Borro, Daylan, Fortney,
  France, Goyal, Grant, Kirk, Kreidberg, Louca, Moran, Mukherjee, Nasedkin,
  Ohno, Rackham, Redfield, Taylor, Tremblin, Visscher, Wallack, Welbanks,
  Youngblood, Ahrer, Batalha, Behr, Berta-Thompson, Blecic, Casewell,
  Crossfield, Crouzet, Cubillos, Decin, Désert, Feinstein, Gibson, Harrington,
  Heng, Henning, Kempton, Krick, Lagage, Lendl, Lothringer, Mansfield, Mayne,
  Mikal-Evans, Palle, Schlawin, Shorttle, Wheatley, \&
  Yurchenko}]{tsaiPhotochemicallyProducedSO22023}
Tsai, S.-M., Lee, E. K.~H., Powell, D., {et~al.} 2023,
  \bibinfo{title}{Photochemically produced {SO2} in the atmosphere of
  {WASP}-39b,} Nature, 617, 483, \dodoi{10.1038/s41586-023-05902-2}

\bibitem[{F. Usui {et~al.}(2018)Usui, Onaka, \& {The AKARI/IRC
  team}}]{usuiAKARIIRCNearInfrared2018}
Usui, F., Onaka, T., \& {The AKARI/IRC team}. 2018, in The {Cosmic} {Wheel} and
  the {Legacy} of the {AKARI} {Archive}: {From} {Galaxies} and {Stars} to
  {Planets} and {Life}, ed. T.~Ootsubo, I.~Yamamura, K.~Murata, \& T.~Onaka
  (ISAS/JAXA), 237--240.
\newblock \url{https://jaxa.repo.nii.ac.jp/records/3115}

\bibitem[{C. Visscher \& J.~I. Moses(2011)Visscher \&
  Moses}]{visscherQUENCHINGCARBONMONOXIDE2011}
Visscher, C., \& Moses, J.~I. 2011, \bibinfo{title}{{QUENCHING} {OF} {CARBON}
  {MONOXIDE} {AND} {METHANE} {IN} {THE} {ATMOSPHERES} {OF} {COOL} {BROWN}
  {DWARFS} {AND} {HOT} {JUPITERS},} The Astrophysical Journal, 738, 72,
  \dodoi{10.1088/0004-637X/738/1/72}

\bibitem[{C. Visscher {et~al.}(2010)Visscher, Moses, \&
  Saslow}]{visscherDeepWaterAbundance2010}
Visscher, C., Moses, J.~I., \& Saslow, S.~A. 2010, \bibinfo{title}{The deep
  water abundance on {Jupiter}: {New} constraints from thermochemical kinetics
  and diffusion modeling,} Icarus, 209, 602,
  \dodoi{10.1016/j.icarus.2010.03.029}

\bibitem[{L. Welbanks {et~al.}(2025)Welbanks, Nixon, McGill, Tilke, Wiser,
  Rotman, Mukherjee, Feinstein, Line, Benneke, Seager, Beatty, Seligman,
  Parmentier, \& Sing}]{welbanksChallengesDetectingGases2025}
Welbanks, L., Nixon, M.~C., McGill, P., {et~al.} 2025,
  \bibinfo{title}{Challenges in the detection of gases in exoplanet
  atmospheres,} Nature Astronomy, 10, 234, \dodoi{10.1038/s41550-025-02730-4}

\bibitem[{L.~S. Wiser {et~al.}(2026)Wiser, Roth, Parmentier, \&
  Line}]{wiser_comparison_2026}
Wiser, L.~S., Roth, A., Parmentier, V., \& Line, M.~R. 2026, \bibinfo{title}{A
  {Comparison} of {One}-dimensional and {Three}-dimensional {Exoplanet}
  {Atmosphere} {Model} {Grids}: {ScCHIMERA} and the {SPARC}/{MiTgcm},} The
  Astrophysical Journal, 997, 365, \dodoi{10.3847/1538-4357/ae2b6c}

\bibitem[{L.~S. Wiser {et~al.}(2025)Wiser, Bell, Line, Schlawin, Beatty,
  Welbanks, Greene, Parmentier, Murphy, Fortney, Arnold, Mehta, Ohno, \&
  Mukherjee}]{wiserPreciseMetallicityCarbontoOxygen2025}
Wiser, L.~S., Bell, T.~J., Line, M.~R., {et~al.} 2025, \bibinfo{title}{A
  {Precise} {Metallicity} and {Carbon}-to-{Oxygen} {Ratio} for a {Warm} {Giant}
  {Exoplanet} from its {Panchromatic} {JWST} {Emission} {Spectrum},} arXiv,
  \dodoi{10.48550/arXiv.2506.01800}

\bibitem[{J.~W. Xuan {et~al.}(2026)Xuan, Ruffio, Chachan, Ohno, Kesseli,
  Murray-Clay, Lee, Moses, Balmer, Baburaj, Blake, Johnstone, Zhang, Knutson,
  Mawet, Beichman, Hodapp, Perrin, Konopacky, Meyer, Bryden, Greene,
  Leisenring, Ygouf, Benneke, Inglis, \& Wallack}]{xuanCompositionsHR87992026}
Xuan, J.~W., Ruffio, J.-B., Chachan, Y., {et~al.} 2026, \bibinfo{title}{The
  compositions of the {HR} 8799 planets reflect accretion of both solids and
  metal-enriched gas,} The Astrophysical Journal, 1000, 27,
  \dodoi{10.3847/1538-4357/ae448f}

\bibitem[{K.~J. Zahnle \& M.~S. Marley(2014)Zahnle \&
  Marley}]{zahnleMETHANECARBONMONOXIDE2014}
Zahnle, K.~J., \& Marley, M.~S. 2014, \bibinfo{title}{{METHANE}, {CARBON}
  {MONOXIDE}, {AND} {AMMONIA} {IN} {BROWN} {DWARFS} {AND} {SELF}-{LUMINOUS}
  {GIANT} {PLANETS},} The Astrophysical Journal, 797, 41,
  \dodoi{10.1088/0004-637X/797/1/41}

\bibitem[{M. Zhang {et~al.}(2024)Zhang, Paragas, Bean, Yeung, Chachan, Greene,
  Lunine, \& Deming}]{zhangRetrievalsNIRCamTransmission2024}
Zhang, M., Paragas, K., Bean, J.~L., {et~al.} 2024, \bibinfo{title}{Retrievals
  on {NIRCam} transmission and emission spectra of {HD} 189733b with {PLATON}
  6, a {GPU} code for the {JWST} era,} arXiv, \dodoi{10.48550/arXiv.2410.22398}

\end{thebibliography}


\begin{references}
\bibliographystyle{aasjournal}
\bibliography{references,references-2}
\end{references}
\end{document}